\documentclass[aps,prd,showpacs,preprintnumbers]{revtex4}
\usepackage{graphics}
\def \beq{\begin{equation}}
\def \eeq{\end{equation}}
\def \beqa{\begin{eqnarray}}
\def \eeqa{\end{eqnarray}}
\def \tr{{\rm Tr}\,}
\def \det{{\rm Det}\,}
\def \osum{\,{\oplus}\,}
\def \lb{\lfloor}
\def \rb{\rceil}
\def \ppbar{\langle\overline\psi\psi\rangle}
\def \lamms{\Lambda_{\overline{{\scriptscriptstyle MS}}}}
\def \O{{\cal O}}
\def \etc{{\sl etc.\/}}
\def \eg{{\sl e.g.\/}}
\def \ie{{\sl i.e.\/}}
\def \etal{{\sl et al.\/}}
\def \jhep{{\sl J.\ H.\ E.\ P.\/}}
\def \np{{\sl Nucl.\ Phys.\/}}
\def \pl{{\sl Phys.\ Lett.\/}}
\def \pr{{\sl Phys.\ Rev.\/}}
\def \prl{{\sl Phys.\ Rev.\ Lett.\/}}

\begin{document}
 
\title{On the critical end point of QCD}
\author{R.\ V.\ \surname{Gavai}}
\email{gavai@tifr.res.in}
\affiliation{Department of Theoretical Physics, Tata Institute of Fundamental
         Research,\\ Homi Bhabha Road, Mumbai 400005, India.}
\author{Sourendu \surname{Gupta}}
\email{sgupta@tifr.res.in}
\affiliation{Department of Theoretical Physics, Tata Institute of Fundamental
         Research,\\ Homi Bhabha Road, Mumbai 400005, India.}

\begin{abstract}
We investigate the critical end point of QCD with two flavours of
light dynamical quarks at finite lattice cutoff $a=1/4T$ using a
Taylor expansion of the baryon number susceptibility. We find a
strong volume dependence of its radius of convergence. In the large
volume limit we obtain $\mu_B/T\approx1.1$ as the radius of convergence
at $T/T_c\approx0.95$ where $T_c$ is the cross over temperature at
zero chemical potential.  Since this estimate is a lower bound on
the critical end point of QCD, the above small value of $\mu$ may place
it in the range of observability in energy scans at the RHIC.
\end{abstract}
\pacs{12.38.Aw, 11.15.Ha, 05.70.Fh}
\preprint{TIFR/TH/04-32, hep-lat/0412035}
\maketitle

\section{Introduction}\label{sc.intro}

\begin{figure}
\begin{center}
   \scalebox{0.8}{\includegraphics{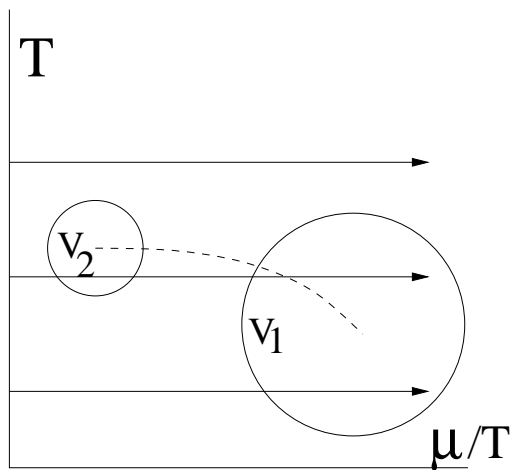}}
\end{center}
\caption{The method followed in this work seeks to Taylor expand
   \cite{pressure} results obtained at vanishing $\mu$ along lines of constant $T$;
   thereby bracketing the critical region, shown by circles. This region
   can vary in location (as shown by the dashed line) and size (indicated by
   the circles) as the volume changes, growing smaller with increasing volume.
   Measurements of these changes can, in principle, be used to identify the
   critical exponents and thereby pin down the universality class at the critical
   end point.}
\label{fg.method}\end{figure}

The phase diagram of QCD contains two experimentally tunable couplings---
the temperature, $T$, and the baryon chemical potential, $\mu_B$.
For physical values of the quark masses, it is expected to have a line
of first order phase transitions in the $T$-$\mu_B$ plane. This line
rises from the $\mu_B$ axis to terminate at the critical end point,
which is specified by its coordinates, $(T^E,\mu_B^E)$. There is
no critical point on the $T$ axis, but there is a rapid cross over
in quantities such as the Wilson line or the quark condensate at a
temperature $T_c$. This temperature is therefore identifiable through
peaks in the susceptibilities, \ie, the temperature derivatives, of
these quantities. In the limit of vanishing quark mass, $T_c$ becomes a
critical point and the critical end point becomes a tricritical point
\cite{pettini,kr,ms}.  Model estimates of the location of the end point vary
widely, thus calling for a first principles determination through a
lattice computation.  If $\mu_B^E$ is small enough that it can be reached
in present day experiments, then it may be detectable in a variety of
ways \cite{shuryak}.

Direct simulations of QCD at finite chemical potential are not possible
due to the fermion-sign problem. However, various methods have been
developed to continue lattice simulations at computable points in the
phase diagram to the more interesting and uncomputable regimes. These
methods include reweighting and its variants \cite{fk,biswaold},
analytic continuation of computations at imaginary chemical potential
\cite{maria,owe} and Taylor series expansions of the free energy
\cite{pressure,biswa}.

In this paper we make the first attempt to locate the critical end point
on large volumes and small pion masses through a Taylor expansion of the
quark number susceptibility in the variable $\mu_B$ at several different
$T$. This Taylor expansion allows us to extrapolate out in $\mu_B$
along lines of constant $T$ (see Figure \ref{fg.method}). Computing the
extrapolation at several $T$ allows us to bracket the critical region
and track its change as the lattice volume is changed.  Systematic
uncertainties in setting the scale of $T$ (see Section \ref{sc.lattice})
determine how closely these extrapolation lines can be spaced. Typically,
these uncertainties decrease with lattice spacing. In the future, as
one approaches the continuum limit with increased computing power, it
should become possible to obtain the critical exponents using this method.

At this time however, computational resources are not sufficient
to complete this program. In this work we concentrate on bracketing
the critical region and investigating gross changes with volume.
All finite volume estimates should technically be called pseudo-critical
values, which tend to the critical values on infinite lattices. On
large enough lattices the distinction is often immaterial. Our larger
lattices are the largest ever used in investigating this problem,
and the estimates of the radius of convergence that we obtain here
are compatible with the infinite volume results, as
we discuss later.

Operating within the context of the present understanding of the phase
diagram, we scan downwards in temperature from $T>T_c$. The first sign
of a phase transition that one meets is then an estimate of the critical
end point.

Our results could be compared with earlier work on the estimation
of the end point with either two flavours (up and down) of light staggered
quarks ($N_f=2$) or two light (up and down) and a third heavier (strange)
staggered quark ($N_f=2+1$). All
such computations have been performed on lattices with temporal extent
$N_t=4$, \ie, with lattice spacing $a=1/4T$. All have comparable lattice
spacings in physical units, since $m_\rho/T_c=5$--$5.5$. Nevertheless,
the quark masses
and lattice volumes ($N_s^3$, where $N_s$ is the spatial extent)
differ widely (a detailed comparison is given later in Table \ref{tb.summ}). The $N_f=2$ computation of \cite{biswa} used a very
high up/down quark mass \cite{biel}, 
although the lattice size was large in units of the pion's Compton
wavelength. Significantly lighter up/down quark masses were used on
much smaller lattices
in the $N_f=2+1$ study of \cite{fk,fktwo} and the $N_f=2$ work of \cite{owe}.
The only errors reported for these
estimates are statistical (see Table \ref{tb.summ} later).

In fact, the largest errors are likely to be systematic. The two
computations of \cite{fk,fktwo} indicate that quark mass effects are
large.  Finite lattice spacing effects have also been estimated to
be as large as the statistical errors \cite{pressure}. Moreover, one
should expect two kinds of finite volume effects. A strong ``small''
volume effect must arise from distortions of the spectrum of the
Dirac operator when the spatial size is as small as 2--3 pion Compton
wavelengths. These must be removed before any physics is extracted. The more
benign ``large'' volume effects are welcome, since it is the analysis of
such effects which would eventually verify whether one is indeed studying
a critical point. In view of this we decided to work with $N_f=2$ QCD
with intermediate quark masses giving $m_\pi/m_\rho=0.31\pm0.01$ at a
lattice spacing of $1/4T$ where $m_\rho/T_c=5.4\pm0.2$.  We expect our results to be
comparable with those of \cite{fk} because 
the scales determined at
$\mu=0$ are comparable. We differ from earlier studies with similiar
quark masses in that our lattices are significantly larger.
From our estimate of the radius of convergence on lattices with spatial sizes
in the range $N_sm_\pi\approx3$--10, we found strong ``small'' volume
effects at the lower end of $N_s$, and a more stable physical value for
larger $N_s$.

The plan of our paper is as follows--- the Taylor expansion is introduced
and the quantities to be evaluated on the lattice are set out in detail in
Section \ref{sc.taylor}. Efficient numerical techniques for performing
the trace computations are given in Section \ref{sc.traces}, along
with details of the performance of the algorithms. The determination of
simulation parameters, the details of the simulations and the numerical
results are given in Section \ref{sc.lattice}. The main results are
collected and discussed in Section \ref{sc.disc}. Detailed formul\ae{} for
the non-linear susceptibilities are given in the apppendix. The paper is organized
such that a perusal of Section \ref{sc.disc} alone would satisfy a reader
who is familiar with the context and needs to extract our main results.
If such a reader finds the language to be unfamiliar, then referring to
Section \ref{sc.taylor} would suffice.

\section{The Taylor expansion}\label{sc.taylor}

The partition function for QCD at temperature $T$ and chemical potentials
$\mu_f$ for each of $N_f$ flavours, can be written in the form
\beq
   Z(T,\{\mu_f\}) = \int{\cal D}U\,{\rm e}^{-S_G(T)}\,\prod_f\det M_f(m_f,T,\mu_f),
\label{part}\eeq
where $S_G$ is the gluon part of the action and $M$ denotes the Dirac
operator.  We employ the standard Wilson action for $S_G$ and staggered
fermions to define $M$.  The fermion-sign problem, which prevents direct
Monte Carlo simulations of QCD at finite chemical potential, is due to the
fact that the determinants have arbitrary complex phases for non-vanishing
$\mu_f$.  One  possible solution to is to recognize that the pressure,
\beq
   P(T,\{\mu_f\})=-\,\frac FV = \left(\frac TV\right)\log Z(T,\{\mu_f\}),
\label{pres}\eeq
can be expanded in a Taylor series about the point where all the
$\mu_f=0$ \cite{pressure}. The leading term, independent of $\mu_f$,
has been obtained in previous computations \cite{qcdpax}. The pressure
is a convex function of the intensive thermodynamic variables, $T$ and
$\{\mu_f\}$. The first derivative of the pressure with respect to $\mu_f$
is the quark number density for flavour $f$. The second derivative is
the corresponding quark number susceptibility (QNS) \cite{gott}. As one
approaches the critical end point, the diagonal QNS \cite{first} diverges
as a power-law determined by the critical exponents. The location of the
divergence gives $(T^E,\mu_B^E)$, and a determination of the critical
exponent could verify the arguments that the QCD critical end point is
in the same universality class as the 3-d Ising model \cite{shuryak}.

In this paper we deal with $N_f=2$ and mainly with the baryon chemical
potential $\mu_B=3\mu_u=3\mu_d$. The quarks have a small but non-vanishing
mass, $m_u=m_d=m$. The $N$-th order derivatives in the Taylor expansion
then can be taken $n_u$ times with respect to $\mu_u$ and $n_d=N-n_u$
times with respect to $\mu_d$. We denote this non-linear quark number
susceptibility (NLS) as $\chi_{n_u,n_d}$. Explicit operator expressions
for the NLS are given in the appendix, along with an exposition of the
methods and operators, $\O_n$, used to compute them. Flavour symmetry implies
that $\chi_{n_u,n_d}=\chi_{n_d,n_u}$. The Taylor expansion
of the chemical potential dependent part of the pressure,
\beq
   \Delta P(T,\mu_u,\mu_d) \equiv
   P(T,\mu_u,\mu_d)-P(T,0,0) = \sum_{n_u,n_d} \chi_{n_u,n_d}\;
        \frac{\mu_u^{n_u}}{n_u!}\, \frac{\mu_d^{n_d}}{n_d!}
\label{presst}\eeq
can be translated into a joint Taylor expansion in the baryon
chemical potential $\mu_B=3(\mu_u+\mu_d)/2$ and the iso-vector
chemical potential $\mu_I=(\mu_u-\mu_d)/4$. This double expansion
can be specialized to one in $\mu_B/3=\mu_u=\mu_d$, when $\mu_I=0$,
or one in $2\mu_I = \mu_u = -\mu_d$, when $\mu_B=0$.  Due to CP
symmetry, the terms for odd $N=n_u+n_d$ in eq.\ (\ref{presst})
vanish. In the remaining terms, the two expansions in $\mu_B/3$ and
$2\mu_I$ differ by a sum of susceptibilities with odd $n_u$ and
$n_d$. The difference has been taken to quantify the seriousness
of the Fermion sign problem \cite{biswa}. We will use a somewhat
different method here, as we explain below.

The non-linear susceptibilities defined above can be written down in
terms of the derivatives of $Z$. From the expression above it is clear
that the derivatives with respect to the $\mu_f$ land entirely on the
determinants. Now, since $\det M=\exp{\tr\log M}$, the first derivative
gives $(\det M)'=\tr(M^{-1} M')\det M\equiv\O_1\det M$. Higher derivatives can be found
systematically using the additional relation $M M^{-1}=1$, which yields
$(M^{-1})'=-M^{-1} M' M^{-1}$. Clearly, therefore, the derivatives of
$Z$ can be written in terms of expectation values of certain operators
involving traces of inverses and derivatives of the Dirac operator
(see the appendix for details).

A Taylor expansion of the pressure immediately yields one for the QNS.
For comparison of the notation of this paper to our earlier works, note
that the flavour susceptibilities $\chi_u=\chi_d$ used earlier correspond
to $\chi_{20}$ in the present notation; $\chi_{ud}$, to
$\chi_{11}$; the isovector susceptibility, $\chi_3$, to $\chi_{20}-\chi_{11}$;
and the baryon susceptibility, $\chi_B=2\chi_0/9$, to $2(\chi_{20}+\chi_{11})/9$.
The radius of convergence of the series for $\mu_B$ gives the location of the
nearest critical point. The Taylor coefficients of $\chi_{20}$ up to 6th order in
$\mu_B/3$ are
\beqa
\nonumber
   \chi^0_B = \chi_{20},
\hfill&&\qquad
   \chi^2_B = \frac1{2!}\left[\chi_{40}+2\chi_{31}+\chi_{22}\right],\\
   \chi^4_B = \frac1{4!}\left[\chi_{60}+4\chi_{51}
                  +7\chi_{42}+4\chi_{33}\right],
\hfill&&\qquad
   \chi^6_B = \frac1{6!}\left[\chi_{80}+6\chi_{71}
                  +16\chi_{62}+26\chi_{53}+15\chi_{44}\right].
\label{taylord}\eeqa
The Taylor coefficients for the off-diagonal QNS, $\chi_{11}$, up to the
same order in $\mu_B/3$ are
\beqa
\nonumber
   \underline\chi^0_B = \chi_{11},
\hfill&&\qquad
   \underline\chi^2_B = \frac1{2!}\left[2\chi_{31}+2\chi_{22}\right],\\
   \underline\chi^4_B = \frac1{4!}\left[2\chi_{51}
                  +8\chi_{42}+6\chi_{33}\right],
\hfill&&\qquad
   \underline\chi^6_B = \frac1{6!}\left[2\chi_{71}
                  +12\chi_{62}+30\chi_{53}+20\chi_{44}\right].
\label{taylorn}\eeqa
The coefficients of the Taylor series in $2\mu_I$ can be obtained from
eqs.\ (\ref{taylord}, \ref{taylorn}) by flipping the sign of every NLS
which has odd $n_d$.  The two QNS above are in principle observable, being
connected to many interesting pieces of physics such as fluctuations and
chemical composition in heavy-ion collisions.

In addition, the QNS can be used to quantify the magnitude of the fermion-sign
problem.  Recall that the sign problem in the measure for the partition
function in eq.\ (\ref{part}) comes entirely from the determinant. Thus,
we can Taylor expand the determinant to get
\beq
   \det M(\mu) = \det M(\mu_0)\left[ 1 + (\mu-\mu_0)\O_1 + \frac1{2!}
     (\mu-\mu_0)^2\O_2 +\cdots\right],
\label{dexpn}\eeq
where $\O_1$ is anti-Hermitean and $\O_2=\O_1'$
is Hermitean.
We show later that
$(T/V) \langle\O_1\rangle$ is the number density and $(T/V) \langle\O_2
+ \O_1\O_1\rangle$ is the susceptibility $\chi_{20}$. If we perform the
expansion around $\mu_0=0$, then by exponentiating the bracket, we can see
that $\O_2$ determines the width of the real part of the measure. Also,
$\langle\O_1\O_1\rangle = (V/T) \chi_{11}$ is the width of the imaginary
part of the measure, up to a sign. Thus, $\chi_{11}/\chi_{20}$ gives the
ratio of the widths of the measure in the imaginary and real directions,
and hence quantifies the magnitude of the fermion-sign problem. With a
little care, the same argument can be extended to any $\mu_0$. However,
in that case care has to be taken to subtract the number density.

We use the NLS to analyze the radius of convergence of the Taylor expansion
of $\chi_B$. Analysis of series expansions in order to extract critical
behaviour is a method of long standing in statistical mechanics and lattice
gauge theory \cite{gaunt,zuber}. The caveats about such analysis have clear
physical meaning. First, the fact that in any infinite series, a few terms
can be changed without changing the radius of convergence implies that one
must check the series for signs of irregularity. In many cases these can be
attributed to special features of the model (see \cite{gaunt}) and can be
seen as interference from ``critical'' points off the real axis. We make
such tests and find that the radius of convergence can indeed be attributed
to a point on the real axis at which a divergence begins to build up.

\subsection{Volume dependence}\label{sc.connected}

Since $P$, $T$ and $\{\mu_f\}$ are intensive variables, it is clear
that each Taylor coefficient should be independent of volume in the
thermodynamic limit. In a lattice computation, this is a non-trivial
check since an arbitrary sub-expression for each Taylor coefficient
could diverge with volume. This divergence is canceled between terms,
and with increasing volume, calls for ever more careful study of possible
numerical inaccuracies and their elimination. A method for doing this
was set out in \cite{pressure}.

In order to do this systematically, we need to construct all possible
sub-expressions which are independent of volume. At the second order,
there are two sub-expressions, $\langle \O_2 \rangle$ and $\langle
\O_{11} \rangle$, which can be expressed as linear combinations of the
two observables $\chi_{20}$ and $\chi_{11}$. It is clear therefore that
the two operator expectation values are the basic volume independent
expressions. The individual traces which constitute them can, and do,
diverge, and it is a delicate, but controllable, numerical task to get
a sensible answer with increasing volume.

In \cite{pressure} we had suggested that this procedure could be
generalized by notionally taking a version of QCD with larger number
of degenerate flavours.  This allows us to define a larger number of
observable NLS, the maximum number possible at a given order $N$ is
for $N_f\ge N$. This shows that the connected part of every fermion
line disconnected operator must be volume independent, in agreement
with other proofs, for example, in perturbation theory.

We outline the argument at the fourth order.
With four degenerate flavours at the fourth order we get---
\beqa
\nonumber &&
   \chi_{4000} = \left(\frac TV\right)\left[\frac{Z_{4000}}Z
        -3\left(\frac{Z_{2000}}Z\right)^2\right],\qquad
   \chi_{3100} = \left(\frac TV\right)\left[\frac{Z_{3100}}Z
        -3\left(\frac{Z_{2000}}Z\right)\left(\frac{Z_{1100}}Z\right)\right],
\\ \nonumber &&
   \chi_{2200} = \left(\frac TV\right)\left[\frac{Z_{2200}}Z
        -\left(\frac{Z_{2000}}Z\right)^2-2\left(\frac{Z_{1100}}Z\right)^2\right],\qquad
   \chi_{1111} = \left(\frac TV\right)\left[\frac{Z_{1111}}Z
        -3\left(\frac{Z_{1100}}Z\right)^2\right],
\\ &&
   \chi_{2110} = \left(\frac TV\right)\left[\frac{Z_{2110}}Z
        -\left(\frac{Z_{2000}}Z\right) \left(\frac{Z_{1100}}Z\right)
        -2\left(\frac{Z_{1100}}Z\right)^2\right].
\label{nf4sus4}\eeqa
The derivatives can be written easily in terms of the traces using the methods
explained in the appendix---
\beqa
\nonumber &&
    Z_{4000} = Z\biggl\langle\O_{1111}+6\O_{112}+4\O_{13}+3\O_{22}+\O_4\biggr\rangle,\qquad
    Z_{3100} = Z\biggl\langle\O_{1111}+3\O_{112}+\O_{13}\biggr\rangle,
\\ &&
    Z_{2200} = Z\biggl\langle\O_{1111}+2\O_{112}+\O_{22}\biggr\rangle,\qquad
    Z_{2110} = Z\biggl\langle\O_{1111}+\O_{112}\biggr\rangle,\qquad
    Z_{1111} = Z\biggl\langle\O_{1111}\biggr\rangle.
\label{nf4ord4}\eeqa
Finally this gives precisely the connected parts of each expectation value
as the invariant---
\beqa
\nonumber &&
   \left(\frac TV\right)\biggl\langle\O_{1111}\biggr\rangle_c=
   \left(\frac TV\right)\left[\biggl\langle\O_{1111}\biggr\rangle
        -3\biggl\langle\O_{11}\biggr\rangle^2\right] = \chi_{1111},
\\ \nonumber &&
   \left(\frac TV\right)\biggl\langle\O_{112}\biggr\rangle_c=
   \left(\frac TV\right)\left[\biggl\langle\O_{112}\biggr\rangle
        -\biggl\langle\O_{11}\biggr\rangle\biggl\langle\O_2\biggr\rangle\right]
   = \chi_{2110}-\chi_{1111},
\\ \nonumber &&
   \left(\frac TV\right)\biggl\langle\O_{22}\biggr\rangle_c=
   \left(\frac TV\right)\left[\biggl\langle\O_{22}\biggr\rangle -\biggl\langle\O_{22}\biggr\rangle^2\right]
   = \chi_{2200}-2\chi_{2110}+\chi_{1111},
\\ \nonumber &&
   \left(\frac TV\right)\biggl\langle\O_{31}\biggr\rangle
   = \chi_{3100}-3\chi_{2110}+2\chi_{1111},
\\ &&
   \left(\frac TV\right)\biggl\langle\O_4\biggr\rangle
   = \chi_{4000}-4\chi_{3100}-3\chi_{2200}+12\chi_{2110}-6\chi_{1111}.
\label{nf4inv}\eeqa
Recall that each expectation value of order two in the above expressions
scales as a single power of the volume, so that their squares scale
with two powers.  This is therefore the leading divergence in the 4-th
order terms, which are canceled to leave a result linear in volume.
This first power is canceled by the factor $T/V$ in the definition of the
susceptibilities, giving the correct scaling of the Taylor coefficients.
Such a construction generalizes.  Terms of order $2N$ grow as $N$-th
powers of the volume. The leading $N-1$ powers are canceled by taking
connected parts, leaving a result linear in volume. Thus, the need for
controlling numerical inaccuracies grows more severe with increasing
order.

\subsection{Finite size scaling}\label{sc.fss}

Consider the free energy, $F(T,\mu_B;V,a,m) = F(T,\mu_B; V, N_t,
m_\rho, m_\pi)$, where the second expression is obtained by trading
the bare parameters for physical quantities.  Instead of the lattice
spacing, $a$, we can specify the rho meson mass, $m_\rho$. Then the
quark mass, $m$, can be traded for the ratio $m_\pi/m_\rho$ since
the ratio can be tuned to its observed value by adjusting the bare
quark mass, $m$, all else being fixed.  This process of regularizing
the ultraviolet divergences of the theory is carried out at
$T=\mu_B=0$.  In other words, $m_\rho$ and $m_\pi$ are computed at
$T=\mu_B=0$. $T$ is then specified by $N_t$ and appropriate boundary
conditions.

``Small volume'' finite size scaling (FSS), which we perform,
establishes the minimum lattice sizes required
to obtain the physics appropriate to the thermodynamic
limit.  This uses the fact that the simulations are actually carried
out at $\mu_B=0$, below the crossover temperature for chiral symmetry
breaking, $T_c$, by measuring the Taylor coefficients of a
susceptibility which is expected to diverge at the critical end
point. These Taylor coefficients can be written in the eigenbasis
of the Dirac operator; for example---
\beq
   \frac{\chi_{20}}{m_\rho^2} = \frac{T}{m_\rho^2V}\sum_{ij} \left[
      \frac{\left|\langle i|\gamma_0|j\rangle\right|^2}{\lambda_i\lambda_j}
     +\left(\sum_i\frac{\langle i|\gamma_0|i\rangle}{\lambda_i}\right)^2
       \right],
\eeq
where $|i\rangle$ denotes the eigenstate of the Dirac operator with
eigenvalue $\lambda_i$. Rewriting the coefficient, $T/m_\rho^2V$, as
$Tm_\pi(m_\pi/m_\rho)^2/ (Vm_\pi^3)$, one can take the factor
$Tm_\pi$ inside the sum, to form dimensionless combinations such
as $Tm_\pi/\lambda^2$. Then the factors $m_\pi/m_\rho$ and
$Tm_\pi/\lambda^2$ are quark mass and lattice spacing dependent
quantities, leaving $Vm_\pi^3$ to be the FSS variable at fixed $T$
and $m$. Below $T_c$,
at small values of $m_\pi V^{1/3}$, one is in the so-called
$\epsilon$-region of physics which is dominated by the pion only.
At larger values of this variable one enters the regime where the
thermodynamic limit is approached. Experience in $T=0$ physics has
shown that $m_\pi V^{1/3}\approx$4--5 is the dividing line between
these two regimes of physics. This is the FSS which
we explore for the first time.

Once one reaches the thermodynamic regime of sizes, further finite
size scaling at the critical end point is determined by the divergent
correlation length at the critical point.  The results of \cite{ray}
indicate that this is not related to the pion. However, as mentioned
earlier, detailed ``large volume'' FSS of this kind, which requires
resources far beyond those available today, will eventually furnish
direct measurements of the critical indices. This we leave to the
future.  In this work we present a restricted form of this ``large
volume'' FSS. We show that the finite size movement in the
radius of convergence on even bigger lattices than the ones we use
here is smaller than our statistical errors.

\section{Numerical methods for traces}\label{sc.traces}

This section is devoted to technical matters involving the numerical evaluation
of the operators $\O_n$. We describe a technique which minimizes the number of
matrix inversions required to determine all these operators up to a given maximum
$n$. This systematic technique also allows us to identify all possible numerical
tests of accuracy which come at negligible extra cost. We also discuss the
optimization of the conjugate gradient matrix inversion. Readers with no interest
in these details may skip this section.

\subsection{Evaluating traces}
\begin{figure}
\begin{center}
   \scalebox{0.7}{\includegraphics{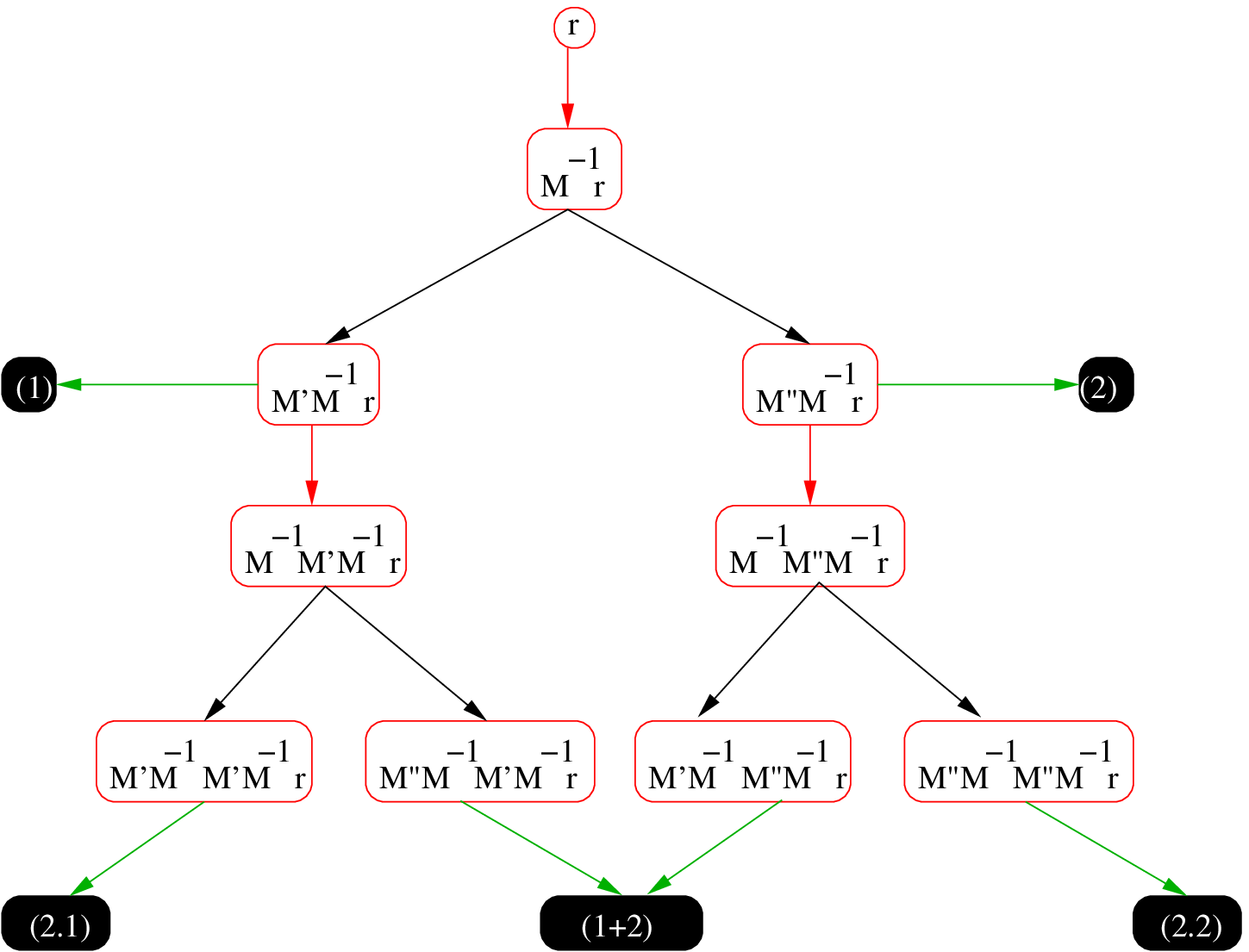}}\hfill
   \scalebox{0.5}{\includegraphics{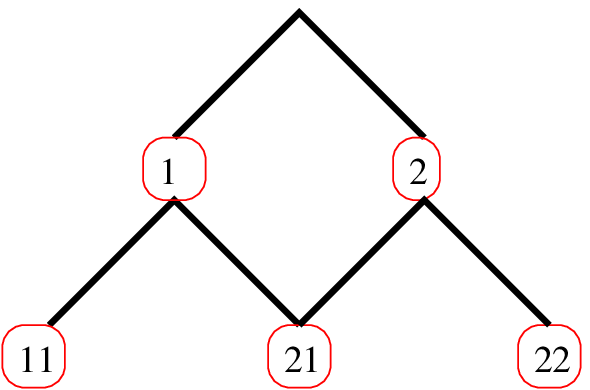}}
\end{center}
\caption{Representing the evaluation tree for NLS. The root node is
a random vector $r$, internal nodes stand for vectors formed
by application of a matrix on a vector, and leaves (in black) stand for
scalars.  Here every vertical edge represents the application of $M^{-1}$
on a vector (which is the costliest part of the computation), every
horizontal edge stands for dotting with $r^\dag$ to give a scalar, and
other edges for application of $M'$ (left slanting) or $M^{\prime\prime}$
(right slanting). The Steiner problem corresponds to finding the minimum
cost of evaluating a given set of scalars. In the compact representation
at the right, the horizontal edges have been contracted to bring the leaf
nodes into the body of the tree and vertical edges have been contracted to
a point.  Note that the redundancy in the computation of $\lb1\oplus2\rb$
(also written as 21) can be used as a check of the program.}
\label{fg.eval}\end{figure}

\begin{figure}
\begin{center}\scalebox{0.7}{\includegraphics{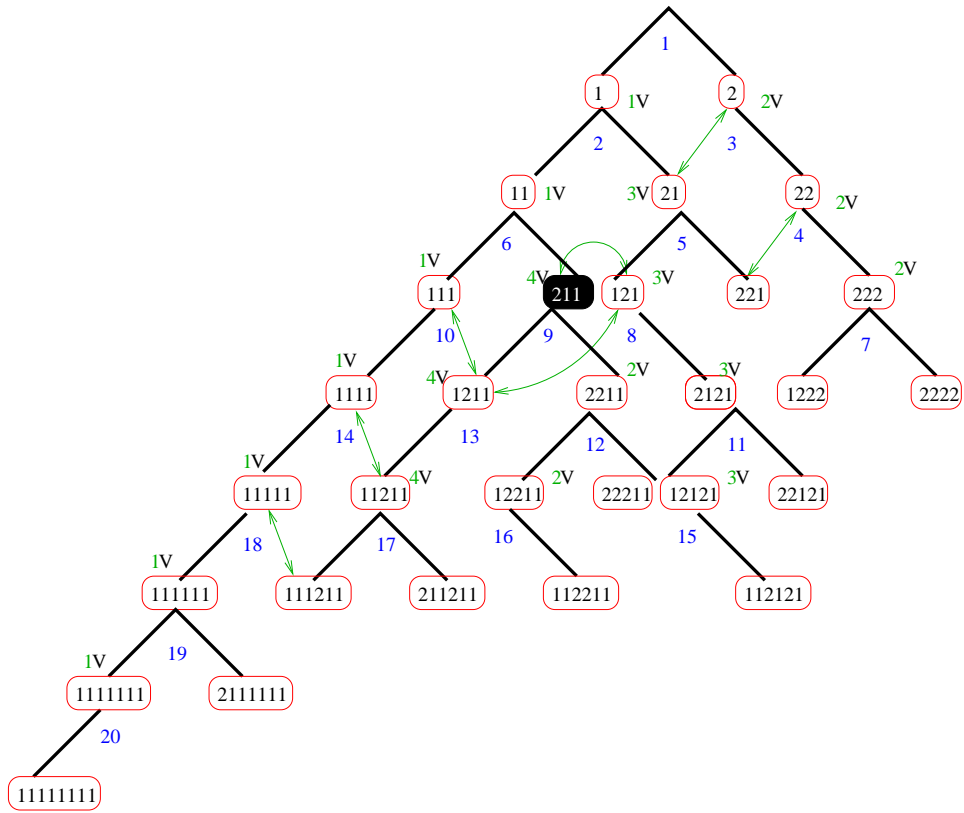}}\end{center}
\caption{The evaluation tree for 8th order NLS requires 20 matrix
inversions (numbered at the branches in the tree) and one extra node
(in black). Since leaves are denser on the right of the tree, evaluating
the inverses from right to left results in the use of less memory in
the form of vectors to be stored.  An assignment of vectors is shown
(1V, 2V \etc),
which uses only 4 intermediate storage vectors. Bidirectional arrows
connect quantities that serve as checks of the program.}
\label{fg.num8}\end{figure}

The traces are evaluated through a noisy technique---
\beq
   \tr A = \overline{r^\dag A r}/\overline{r^\dag r},
\label{trace}\eeq
where the bar denotes averaging over random vectors $r$. An unbiased
measurement is provided for ensembles of random vectors which satisfy
the conditions---
\beq
   \overline{(r^\alpha_i)^* r^\beta_j} = N \delta_{\alpha\beta}
     \delta_{ij}
\label{noise}\eeq
where $\alpha$, $\beta$ label the vector, $i$, $j$ label the
components, and $N$ may depend on the ensemble but is independent of
either index.  The matrix $A$ consists of a product of $M^{-1}$, $M'$
and $M^{\prime\prime}$ such that there is one $M^{-1}$ for each of the
other two. The costliest part of the evaluation is the computation of
$M^{-1}$ acting on a vector. While evaluating a certain (fixed) set of
traces, we need to minimise the number of matrix inversions.

We represent this problem through a directed graph without cycles.
Each internal node of the graph contains a vector, and each leaf contains
a scalar. The root node is the random vector $r$, and every other
internal node is obtained by the action of some matrix on $r$. Since each
vector is obtained from another by the action of either $M'M^{-1}$ or
$M^{\prime\prime}M^{-1}$, the internal nodes form a binary tree. To count
the cost of any operation we separate the action into a single $M^{-1}$
followed by either $M'$ or $M^{\prime\prime}$. Each internal node (except
the root) gives rise to a leaf by dotting it with $r^\dag$. Since this
action is the stochastic evaluation of a trace, several internal nodes
(differing by cyclic permutations of the operations) may connect to each
leaf. A representation of the evaluation tree up to level 2 is given in
Figure \ref{fg.eval}. A compressed representation, also shown in Figure
\ref{fg.eval}, is obtained by collapsing together the nodes containing
a vector $v$ and the vector $M^{-1}v$, by writing $1$ for each $M'$,
$2$ for each $M^{\prime\prime}$ and never writing $M^{-1}$.

Our problem is--- given a set of target leaf nodes, find the path on
the tree using the minimum number of $M^{-1}$ which evaluates these
leaves. This is one version of a problem known in the computer science
literature as the group Steiner problem on directed graphs, for which a
solution has been given recently \cite{steiner}.
The computer science interest arises from the fact that
the general problem is known to be in the class of NP complete problems.

Since our problem size is small, we do not use the general algorithm, but
a heuristic which shares the idea of ``bunching'' with that algorithm,
uses the structure specific to this problem, and is easily implemented in
wetware.  Given the set of matrix elements, write down all strings which
have to be generated, and group each with all its cyclic permutations.
For each set of strings of length $\ell$, starting from the largest,
enumerate the suffixes of length $\ell-1$. Choose that representation
in which the suffixes are common at the nearest possible level. This is
the central heuristic--- bunch the strings by largest suffixes. Build
this back all the way to the empty string, backtracking only if there
is insufficient bunching close to the root. This gives an evaluation
tree. Run over all such trees and find the ones with lowest cost.

For the 4th order problem this yields:
\beqa
\nonumber (1)&:& 1,\\
\nonumber (2)&:& 2,\\
\nonumber (2\cdot1)&:& 1\underline1,\\
\nonumber (1\osum2)&:& 12,\;2\underline1,\\
\nonumber (2\cdot2)&:& 2\underline2,\\
\nonumber (3\cdot1)&:& 1\underline{11},\\
\nonumber (2\cdot1\osum2)&:& 112,\;121,\;2\underline{11}\\
\nonumber (4\cdot1)&:& 1\underline{111},\\
\eeqa
The underlined suffixes are used in the evaluation. The tree is identified
from bottom up, but evaluated top down.  A solution for the 8th order
problem is given in Figure \ref{fg.num8}.

There seems to be a large degree of redundancy in the most efficient
evaluation. This can be used to minimize the number of internal nodes
not connected to the target leaf set. Even after this is done, there
seems to be further degeneracy, and this can be utilised to minimize the
breadth of the tree, since that determines the number of vectors needed
in the evaluation. 

Sometimes only one of the two branches is needed in the evaluation. Since
both branches can be generated at nearly the same cost as one, the other
branch can then be used to check the numerical accuracy of the procedure.
Such possible checks are marked by bi-directional
arrows in Figure \ref{fg.num8}.

\subsection{Inverting the fermion matrix}

\begin{figure}
\begin{center}\scalebox{1.0}{\includegraphics{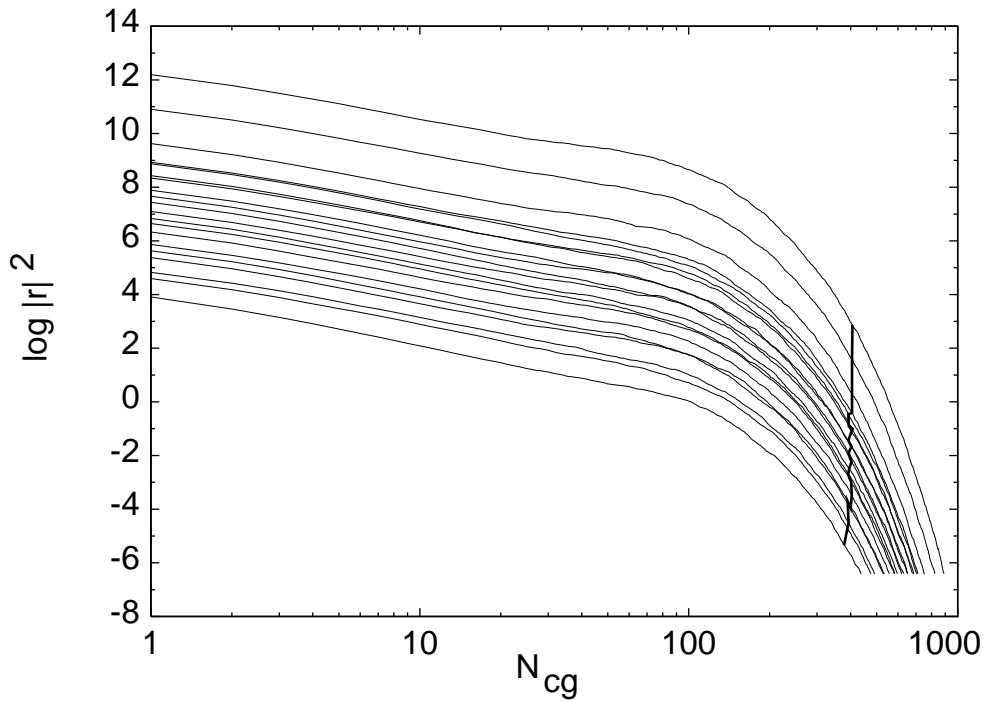}}\end{center}
\caption{The evolution of the norm of the residual, $|r_n|^2$ against
   the number of conjugate gradient iterations, $n$. This figure shows
   the computation on a typical gauge configuration at $T=1.05T_c$ using
   one random vector for the full evaluation tree of the 8th order NLS.
   In Method 1 $n$ increases as one proceeds down the tree. In Method 2
   $n$ remains almost constant (see the thick line).}
\label{fg.cgstop}\end{figure}

\begin{figure}
\begin{center}
\scalebox{0.65}{\includegraphics{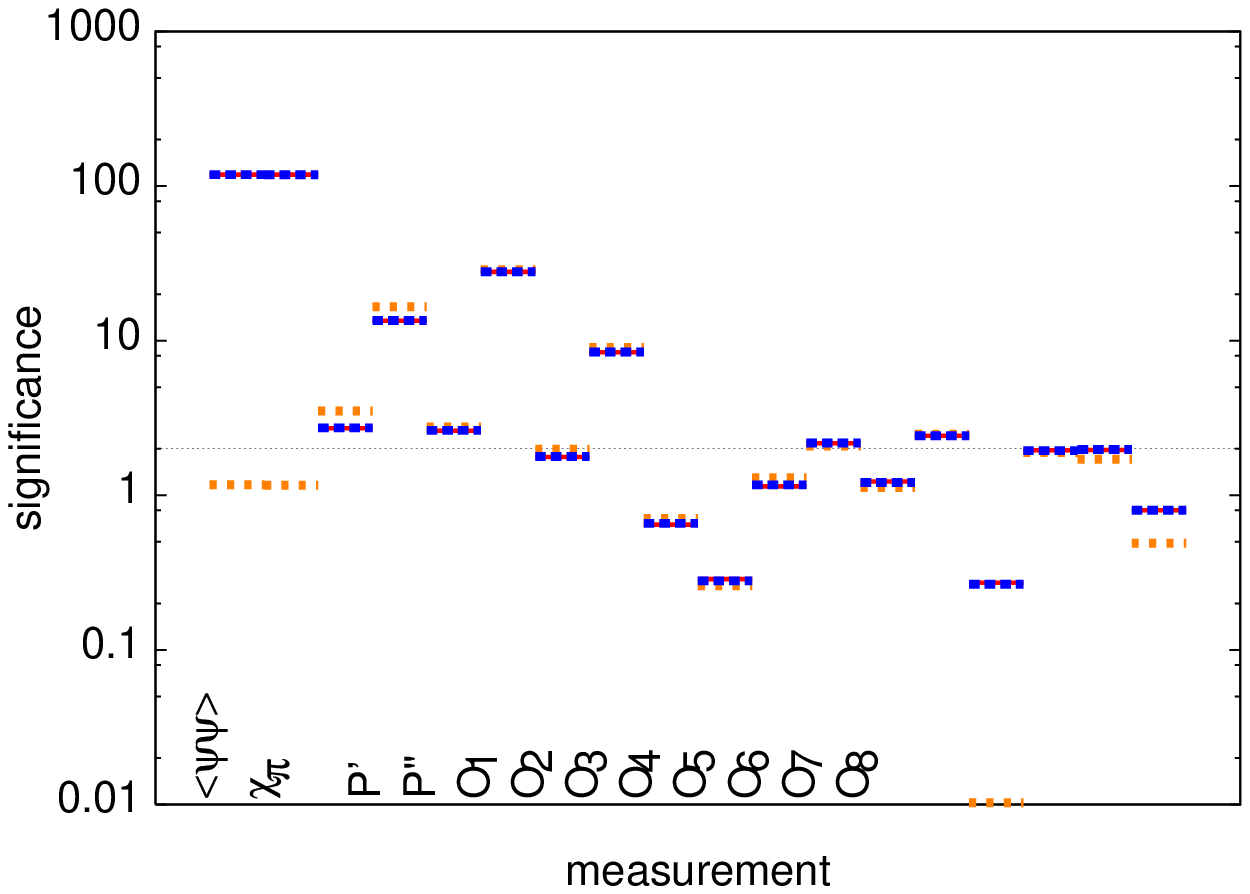}}
\scalebox{0.65}{\includegraphics{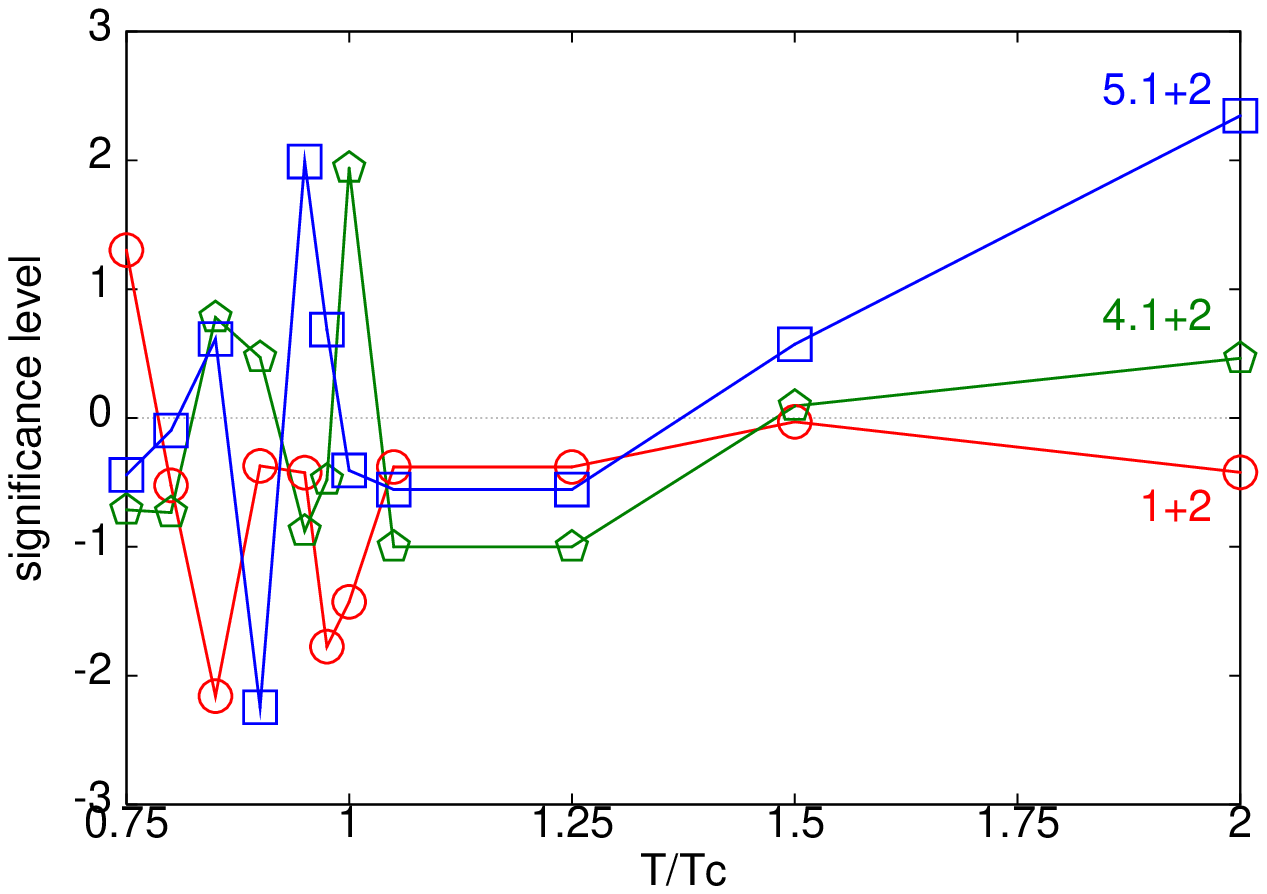}}
\end{center}
\caption{Results of inversion with $\epsilon=10^{-6}$ with method 1 (full
   lines) and with $\epsilon=10^{-3}$ with method 2 (dotted lines). The
   latter are 
   significantly different for $\ppbar$ and $\chi_\pi$, but not for the
   traces $\O_n$.  Inversions with
$\epsilon=10^{-4}$ with method 2 are indistinguishable from
   the full lines on the scale of this figure. The panel on the right displays the accuracy of the inversions
   as one proceeds down the evaluation tree. The significance level is defined as
   the difference of two evaluations of the same trace divided by the error in the
   difference.}
\label{fg.howgood}\end{figure}

The fermion matrix inversion is done through a conjugate gradient
method.  The most time consuming part of this problem is the matrix
multiplication $M^\dag M x$. A huge gain could be obtained if the
multiplication could be speeded up in any way. Each matrix
multiplication involves $V$ diagonal terms and $2dV$ off-diagonal
terms (where $V$ is the number of sites, and $d$ the number of
dimensions). Each diagonal term involves $N_c$ multiplications.
Each off-diagonal term involves $2N_c^2$ operations ($N_c^2$
multiplications and $N_c^2$ additions). Hence the total operation
count for the double matrix multiplication is $VN_c(1+4dN_c)$.

One consequence of this is that `improved' methods such as
preconditioning or deflation are not of much use unless the number of
matrix multiplications can be cut down drastically. For example, in our
application where the same matrix is applied repeatedly on many different
vectors, one might expect deflation to work wonders \cite{saad}.
However, this requires two double matrix multiplications per step
and hence actually performs worse.

The simplest solution, and the one that seems to work best, is to tune
the stopping criterion. In Figure \ref{fg.cgstop} we show the performance
of two different criteria---
\begin{enumerate}
\item Method 1 is to require the norm of the residual, $|r_n|^2<\epsilon V$.
   Since each inversion increases the norm of the right-hand side, through
   the 20 inversions required for evaluating all the traces, the CG has to
   take increasingly larger number of steps, $n$, for convergence.
\item Method 2 is to require the norm of the residual, $|r_n|^2<\epsilon|r_0|^2 $.
   As shown in the figure, this seems to keep $n$ almost fixed as one proceeds
   through the evaluation tree.
\end{enumerate}

There is really no {\sl a priori} reason for the two stopping criteria to
give equally accurate results with the same value of $\epsilon$. However,
we see that for $\epsilon\le10^{-4}$ the results do not differ for the
$\O_n$'s (they do for $\ppbar$), while
method 2 takes less CPU time. To have some understanding of this, we note
that since any matrix is diagonalised by a bi-unitary transformation,
we can write $M=U^\dag DV$, where $D$ is diagonal, and $U$ and $V$ are
unitary matrices.
Similarly, $M^{\prime\prime}=W^\dag D^{\prime\prime}X$, giving
\beq
   \tr M^{-1} M^{\prime\prime} = \tr D^{-1} U W^\dag D^{\prime\prime} XV^\dag
     = \sum_{ij} \frac{d_{jj}^{\prime\prime}}{d_{ii}} (UW^\dag)_{ij} (XV^\dag)_{ji}
\label{dtrc}\eeq
The triangle inequality can then be used to bound 
\beq
   |\tr M^{-1}M^{\prime\prime} |\le
     \sum_{ij} \left|\frac{d_{jj}^{\prime\prime}}{d{ii}}(UW^\dag)_{ij} (XV^\dag)_{ji}\right|
     \le \left\{\sum_i |1/d_{ii}|\right\}\,\left\{\sum_i |d_{ii}^{\prime\prime}|\right\}.
\label{bnd}\eeq
Since every element of an unitary matrix is bounded by unity, we get the
second bound above by inserting unity for these matrix elements. However,
this bound is too rough. If it is (nearly) saturated then the trace would
depend on the stopping criterion as strongly as $\ppbar=\tr M^{-1}$,
contrary to observation. Therefore it seems likely that the structure of
the unitary matrices plays a role in reducing the bound. In particular,
we note that the low eigenvalues of $M$, $d_{ii}\le\rho$ can be killed
only by eigenvalues of $M^{\prime\prime}$. This can happen if the unitary matrices
$WU^\dag$ and $XV^\dag$ are nearly block-diagonal in this subspace
(they are exactly block-diagonal in the vacuum or in the presence of
stationary gauge fields, when the Dirac operator becomes separable). In
other words, the observed lack of sensitivity to small eigenvalues of
$M$ can be understood if every near-zero mode of $M$ is also a near-zero
mode of $M^{\prime\prime}$.

\section{Lattice simulations and results}\label{sc.lattice}

\subsection{Parameters and simulations}\label{sc.setup}

\begin{table}[htb]
\begin{tabular}{ll|l|rr|rr|rr|rr|rr}
\hline
$\beta$ & $\;m$ & $T/T_c$ & \multicolumn{2}{|c}{$4\times8^3$}
                        & \multicolumn{2}{|c}{$4\times10^3$}
                        & \multicolumn{2}{|c}{$4\times12^3$}
                        & \multicolumn{2}{|c}{$4\times16^3$}
                        & \multicolumn{2}{|c}{$4\times24^3$} \\
& & & $N_{\rm stat}$ & $\tau_{\rm max}$ 
    & $N_{\rm stat}$ & $\tau_{\rm max}$ 
    & $N_{\rm stat}$ & $\tau_{\rm max}$ 
    & $N_{\rm stat}$ & $\tau_{\rm max}$ 
    & $N_{\rm stat}$ & $\tau_{\rm max}$ \\
\hline
$5.20$ & $\;0.033$ & $0.75\pm0.02$ &
  86 & 25 & 28 & 20 & 172&14  & 89 & 15 & --- & --- \\
$5.22$ & $\;0.03125$ & $0.8\pm0.02$ &
  206 & 19 & 193 & 14 & 94 &14  & 84 & 15 & --- & --- \\
$5.24$ & $\;0.0298$ & $0.85\pm0.01$ &
  89 & 14 & 43 & 14 &271 & 15 & 210 & 15 & --- & --- \\
$5.26$ & $\;0.02778$ & $0.9\pm0.01$ &
  144 & 29 & 70 & 59 & 50 & 21 & 65 & 17 & 63 & 12 \\
$5.275$ & $\;0.02631$ & $0.951\pm0.01$ &
  115 & 34 & 110 & 112 & 98 & 42 & 26 & 35 & 56 & 17 \\
$5.2875$ & $\;0.025$ & $1.00\pm0.01$ &
  44 & 207 & 22 & 362 & 56 &217 & 51 & 131 & 123 & 43 \\
$5.30$ & $\;0.0238$ & $1.048\pm0.01$ &
  4 & 279 & 178 & 114 &110 &41  & 111 & 42 & --- & --- \\
$5.35$ & $\;0.02$ & $1.25\pm0.02$ &
  139 & 6 &--- &--- &--- &--- & 73 & 9 & --- & --- \\
$5.425$ & $\;0.01667$ & $1.65\pm0.06$ &
  1146 & 6 &--- &--- &--- &--- & 112 & 7 & --- & --- \\
$5.54$ & $\;0.0125$ & $2.15\pm0.10$ &
  1648 & 6 &--- &--- & 49 & 6 & 313 & 4 & --- & --- \\
\hline
\end{tabular}
\caption{The couplings $\beta$ and bare quark masses $m$ at which
the simulations were performed. The temperature scale, $T/T_c$, is given
in the $\overline{\rm MS}$ scheme, with error assignment as discussed in
the text. The bare quark mass at
all temperatures correspond to $m=0.1T_c$.  The statistics collected
in each simulation, $N_{\rm stat}$, is given in terms of the longest
autocorrelation time, $\tau_{\rm max}$, measured in that simulation in MD
time units. The trajectory lengths are 1 MD time unit on the $4\times8^3$
lattice and scaled up in proportion to the length of the spatial lattice,
\ie, growing to 3 MD time units on the $4\times24^3$ lattice.}
\label{tb.runs}\end{table}

\begin{figure}
\begin{center}
\scalebox{0.65}{\includegraphics{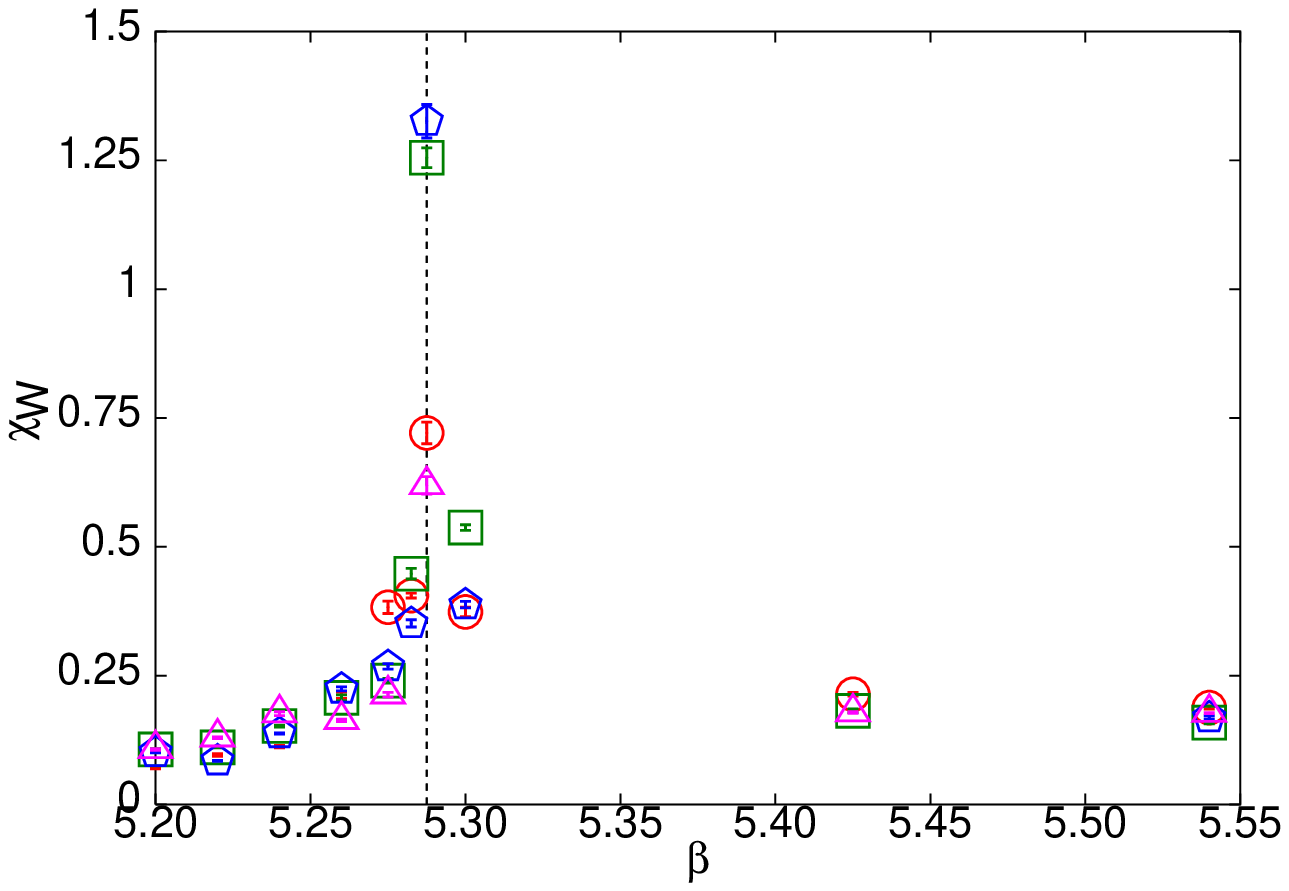}}
\scalebox{0.65}{\includegraphics{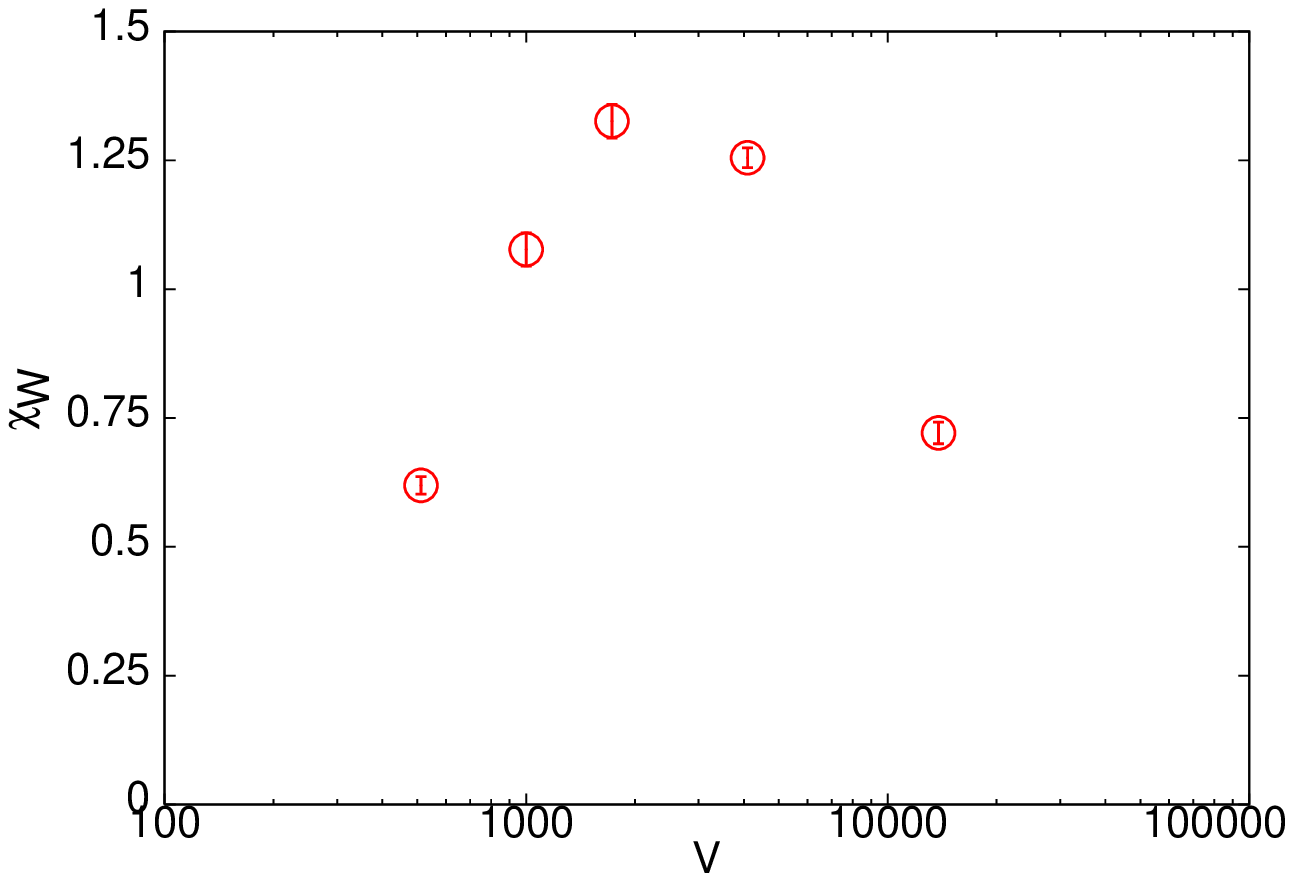}}
\end{center}
\caption{The Wilson line susceptibility, $\chi_L$, shown on the left as a function
   of the bare coupling, $\beta$ on lattice sizes of $8^3$ (triangles), $12^3$
   (pentagons), $16^3$ (boxes) and $24^3$ (circles). It is clear that $\beta=5.2875$
   does not locate the exact position of the peak on the larger volumes. This is
   also indicated in the figure on the right, which depicts $\chi_L$
   at a fixed $\beta=5.2875$ as a function of $V$.}
\label{fg.vdep}\end{figure}

\begin{figure}
\begin{center}
\scalebox{0.65}{\includegraphics{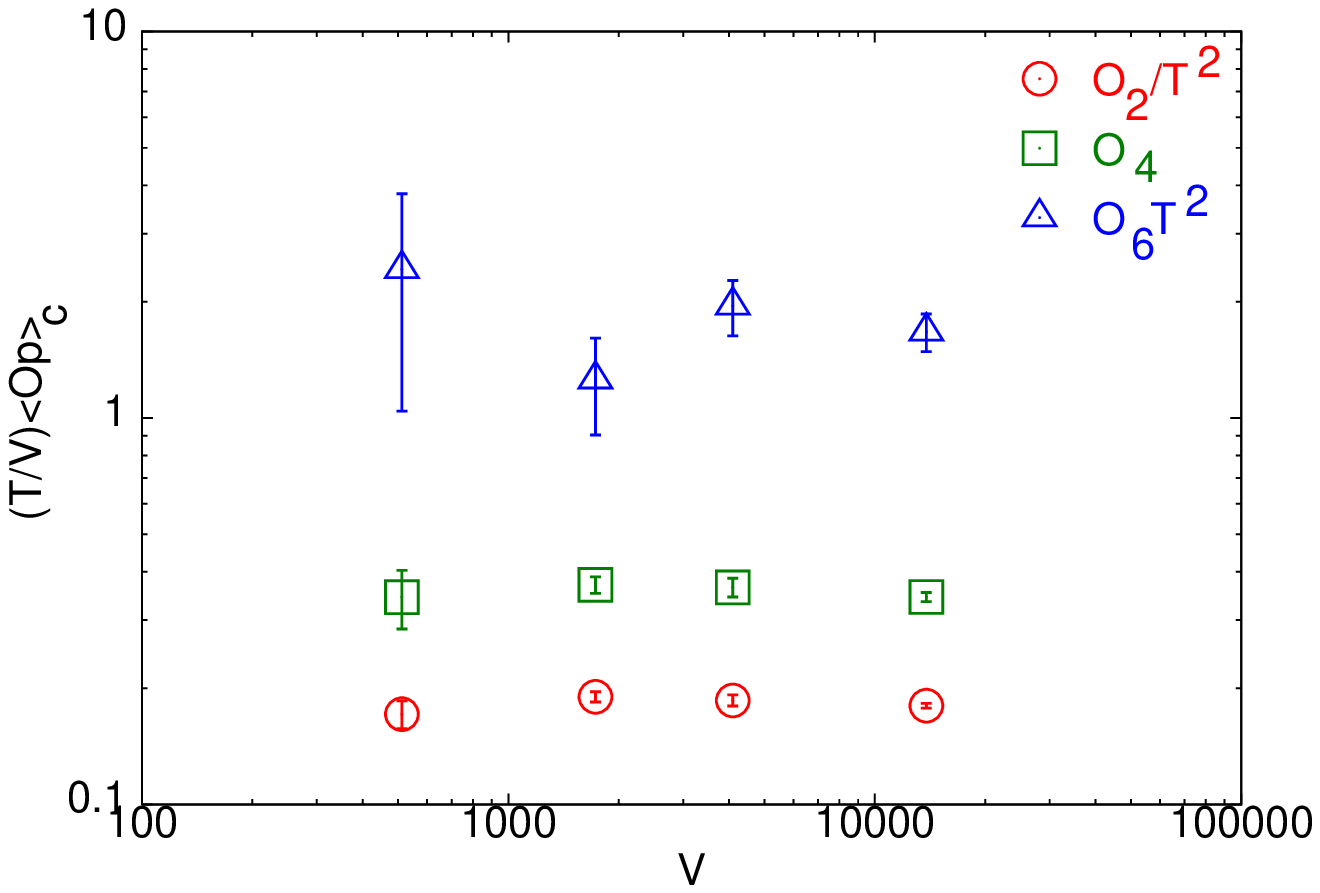}}
\scalebox{0.65}{\includegraphics{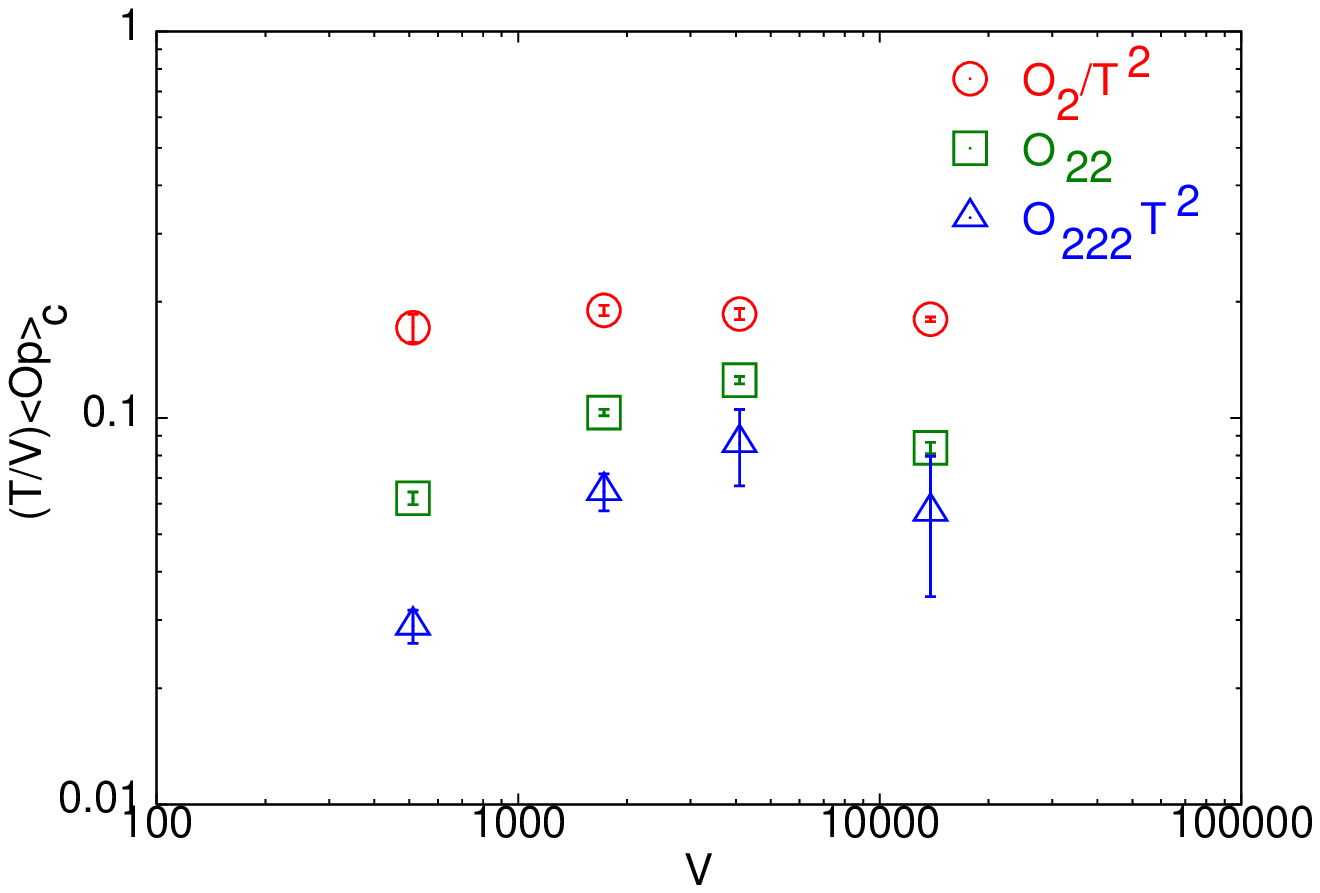}}
\end{center}
\caption{The panel on the left shows operator expectation values
   $\langle\O_n\rangle$ as functions of the volume at $T=0.75T_c$.
   The panel on the right shows that the connected parts $\langle O_{22}\rangle_c$,
   and $\langle O_{222}\rangle_c$ at $0.75T_c$
   scale with the correct power of $V$.}
\label{fg.volloss}\end{figure}

\begin{figure}
\begin{center}
\scalebox{0.65}{\includegraphics{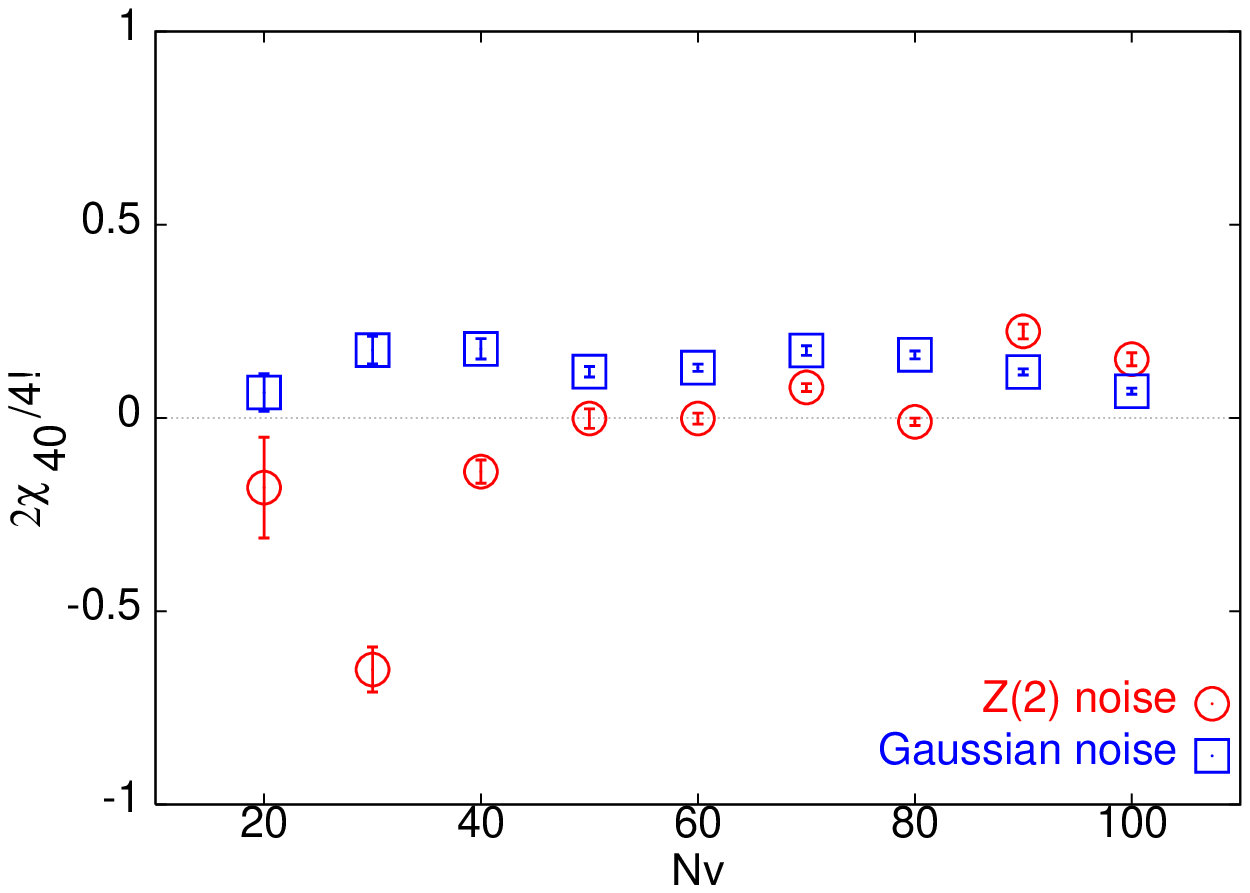}}
\scalebox{0.65}{\includegraphics{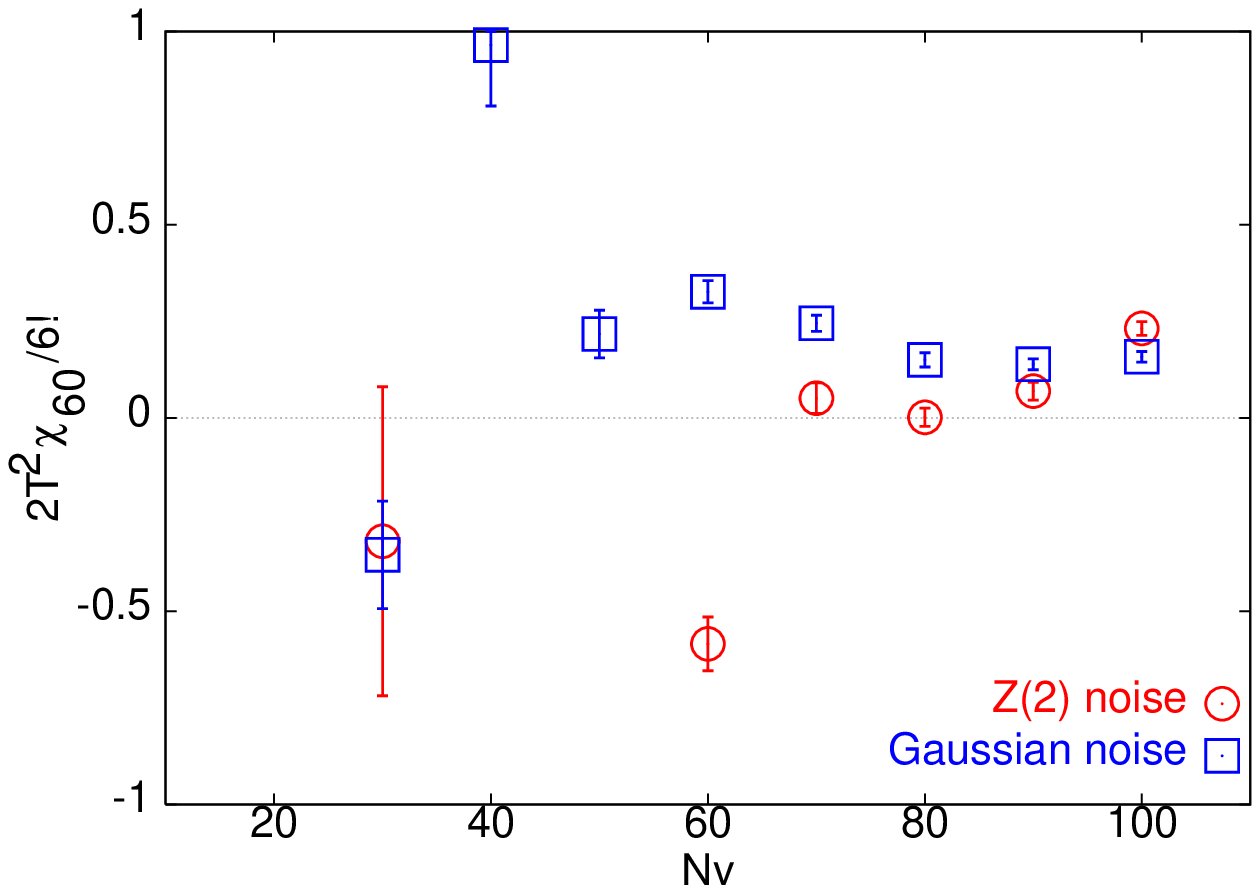}}
\end{center}
\caption{A comparison of Gaussian noise (boxes) and $Z_2$ noise (circles)
   for the evaluation of the NLS $\chi_{40}$ (left panel) and $\chi_{60}$
   (right panel) at $0.75T_c$ on a $4\times16^3$ lattice.
   Gaussian noise requires less vectors for stable evaluation of these
   quantities.}
\label{fg.z2}\end{figure}

\begin{figure}
\begin{center}
\scalebox{0.65}{\includegraphics{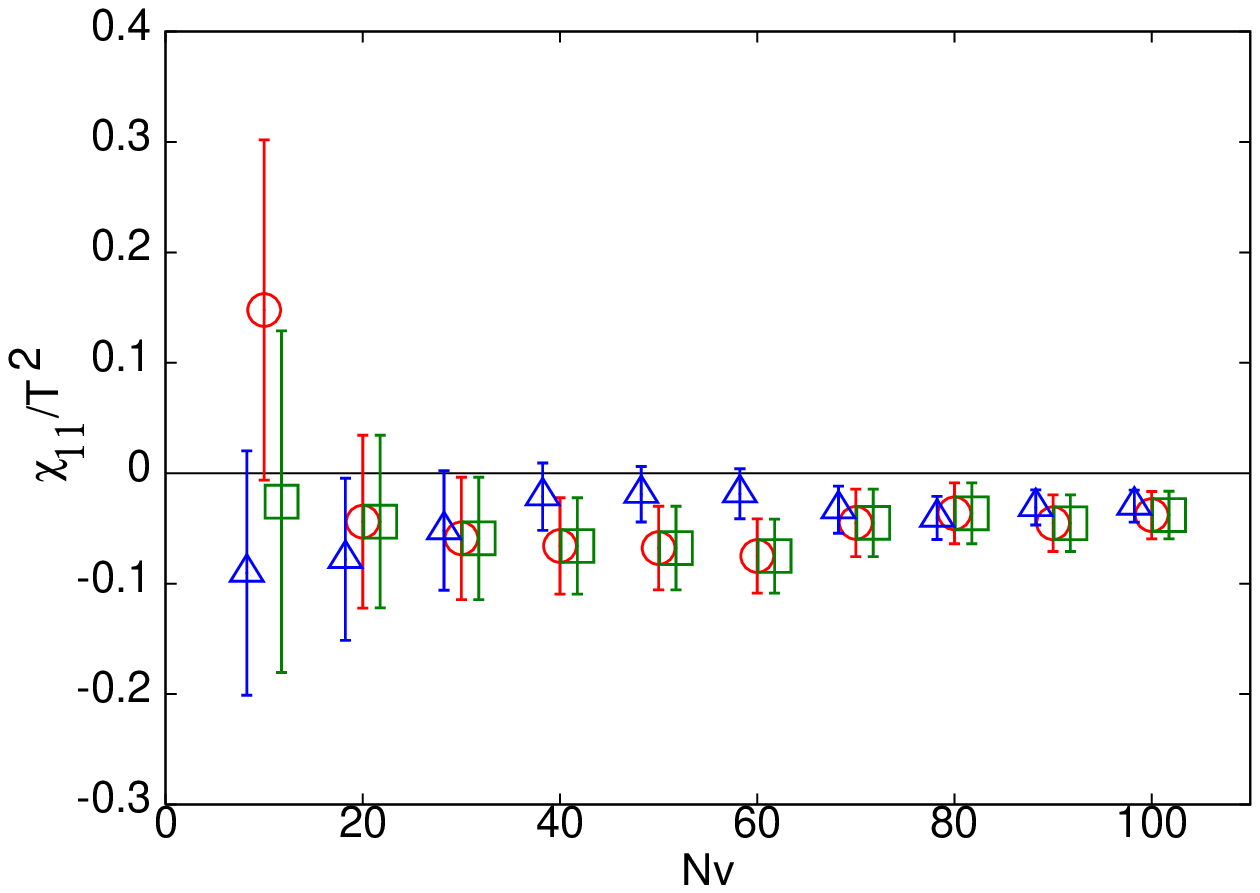}}
\scalebox{0.65}{\includegraphics{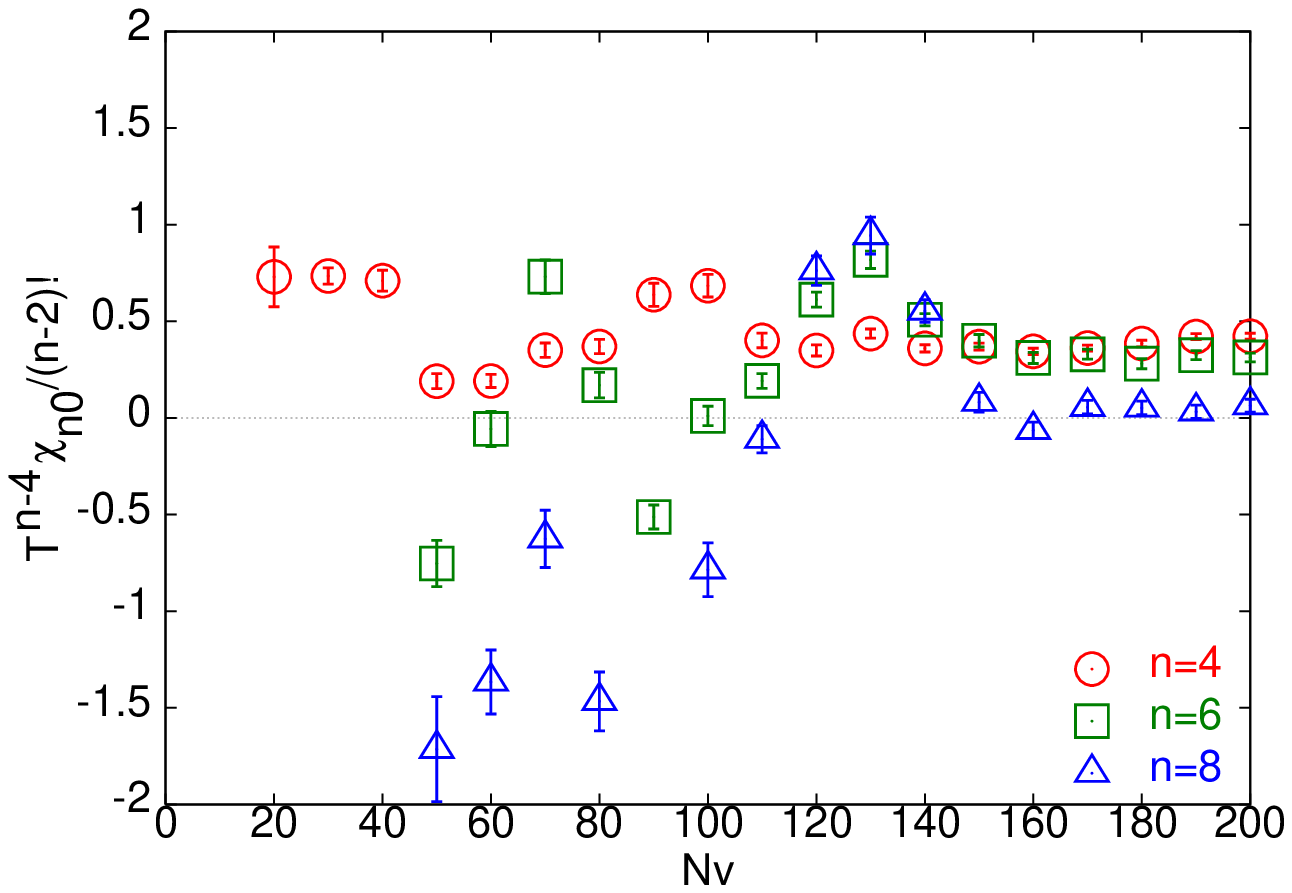}}
\end{center}
\caption{Estimates of $\chi_{11}/T^2$ as functions of the number of vectors
   $N_v$ for $0.75T_c$ on a $4\times24^3$ lattice. Shown are the estimates
   using Gaussian noise and CG stopping criterion of $10^{-4}$ (circles) and
   $10^{-6}$ (boxes) in Method I and $Z_2$ noise with CG stopping criterion
   of $10^{-4}$ in Method I (triangles). $N_v$ are chosen to be multiples of
   10, but some of the points are displaced slightly in the x-direction for
   improved visibility. The panel on the right shows $\chi_{n0}$ as a function
   of $N_v$ for $n=4$ (circles), 6 (boxes) and 8 (triangles). This data was
   collected at $0.75T_c$ on a $4\times10^3$ lattice with Gaussian noise and
   $\epsilon_{CG}=10^{-4}$.}
\label{fg.accuracyqns}\end{figure}

All the simulations reported in this paper were carried out at the lattice
spacing $a=1/4T$, corresponding to lattices with temporal extent $N_t=4$.
We used the R-algorithm in the simulations and chose the conjugate
gradient stopping criterion to be Method 1 of Section \ref{sc.traces}
with $\epsilon=10^{-5}$.  The finite temperature crossover at this
cutoff has been determined to be at the coupling $\beta=5.2875\pm0.0025$
\cite{gotttc}, which therefore corresponds to the cross over temperature
$T_c$ at zero $\mu$. We have improved the bound on the cross-over coupling,
as we report later in this section.

We have set the scale by using plaquette values to define a renormalized
coupling \cite{scale} in a series of exploratory computations on small
lattices ($16^4$) at $T=0$, using trajectory lengths of one molecular
dynamics (MD) unit. The typical statistics utilised for scale setting was
about 500 such trajectories, of which the initial 200 were discarded for
thermalization. During the scale setting the input bare quark mass was
varied until a self-consistent determination of the scale showed that
$m/T_c=0.1$ within statistical errors. A two-loop analysis of the scale
shows several sources of uncertainty. Statistical uncertainties can be
easily made small, and with our statistics they are of the order of 3--6
parts per thousand. Harder to control are the differences between various
renormalization schemes and the scatter due to errors in the estimate
of the critical coupling. The latter induces an error of slightly
over 1\% in the scales. Differences between different renormalization
schemes indicate that $a=1/4T_c$ is too coarse for 2-loop formul{\ae}
to work with high precision. This problem is also known in the quenched
theory, where it was seen that a higher order correction was needed to
achieve scaling \cite{scale}.  Here we quantify this uncertainty by
taking the scatter between the scale determination through the E, V and
$\overline{\rm MS}$ schemes. The uncertainty defined in this manner is
within 2\% of the value in the $\overline{\rm MS}$ scheme for $T$ within
25\% of $T_c$, but grows rapidly beyond, upto about 10\% at the value of
$\beta_c$ determined at $N_t=8$ for $m/T_c=0.1$. Since the critical end
point lies close to $T_c$, we decided not to do an extensive set
of $T=0$ simulations to control the scale further by extracting higher
order terms in the $\beta$-function. The combined statistical and
systematic errors in the determination of $T/T_c$ are shown in Table
\ref{tb.runs}.

Hadron masses have been determined earlier at the critical coupling
\cite{gottm}. It was found that $m_\pi/m_\rho=0.31\pm0.01$, which is
somewhat larger than in nature. $T_c$ could be given in physical units
by noting that $m_\rho/T_c=5.4\pm0.2$. Further evidence that lattice
spacings are still fairly coarse can be seen by noting that the
nucleon is too heavy, since $m_N/m_\rho=1.8\pm0.02$. Pushing towards
the continuum will be a necessary future step.

The simulation parameters and statistics are shown in Table \ref{tb.runs}.
The MD trajectory was chopped into time steps of length 0.01, and
the full trajectory was chosen to be 1 time unit for the $4\times8^3$
runs. As the spatial size, $N_s$ of the lattice changed, the trajectory
length was changed in proportion to $N_s$.  An earlier study had shown
such a tuning of the trajectory length to be useful in controlling the
growth of autocorrelations \cite{mdtau}.  We found that the number of
iterations in the conjugate gradient algorithm remained unchanged as
the spatial volume increased at fixed $T$.

The pseudo-critical point is usually identified by the Wilson line
susceptibility, $\chi_L$, the chiral susceptibility, $\chi_m$, which
is the second derivative of the free energy with respect to the quark
mass \cite{karsch}, or autocorrelations of various measurements. All of
these are expected to peak at $T_c$. If the point is critical, then one
expects these quantities to grow as a power of the volume, $V$. If, on
the other hand, there is a cross over, then these peak heights should
saturate at large enough $V$. Also, different quantities could peak at
somewhat different temperatures even for $V\to\infty$ \cite{ray}. 

In Figure \ref{fg.vdep} we show a measurement of $\chi_L$ at the couplings
we have used. The drop in the peak of $\chi_L$ on the largest $V$ is
clear indication that $\beta=5.2875$ is not the location of the peak in
the thermodynamic limit. In order to bracket the position of the peak we
have run separate simulations on $4\times12^3$ and $4\times16^3$ lattices
at nearby couplings with bare quark mass fixed at $ma=0.025$. Collecting
statistics from about 30 to 50 independent configurations at each
coupling, we find that the finite volume shift in the peak is less than
$\Delta\beta=0.00125$ for $N_s=16$. This corresponds to an error of 1\%
or less on $T_c$.

Such a shift is not of significant concern for our study since the
dominant systematic error in the scale setting comes from the fact that
the lattice spacings for $N_t=4$ in the temperature range of interest
are too coarse for 2-loop scaling to work--- improving the estimate of
$T_c$ beyond the present 1\% level of scale does not improve the total
systematic error by 1\%. Of course, as a consequence, these data cannot
yet be used to check whether there is a phase transition or a cross
over in the $\mu=0$ theory. This is an interesting but separate problem
\cite{edwin,digiacomo}. It might be useful in future to
explore a narrow region of $\beta$ near the peak in greater detail to
track the growth of $\chi_L$ with $V$ and give a definitive answer to
this question.

Several local gauge operators as well as quark operators such as
the chiral condensate were measured once per trajectory. The longest
autocorrelation, $\tau_{\rm max}$, in these measurements was used to
set the scale of autocorrelations. It was seen that $\tau_{\rm max}$,
when measured in MD time units, was roughly independent of $N_s$ at fixed
$T$, as expected from the scaling of the trajectory length \cite{mdtau}.
As a consequence of these scalings, the time taken for generating one
new configuration scaled approximately as $N_s^5$
at fixed $T$ in thermal equilibrium.

\begin{figure}
\begin{center}
\scalebox{0.65}{\includegraphics{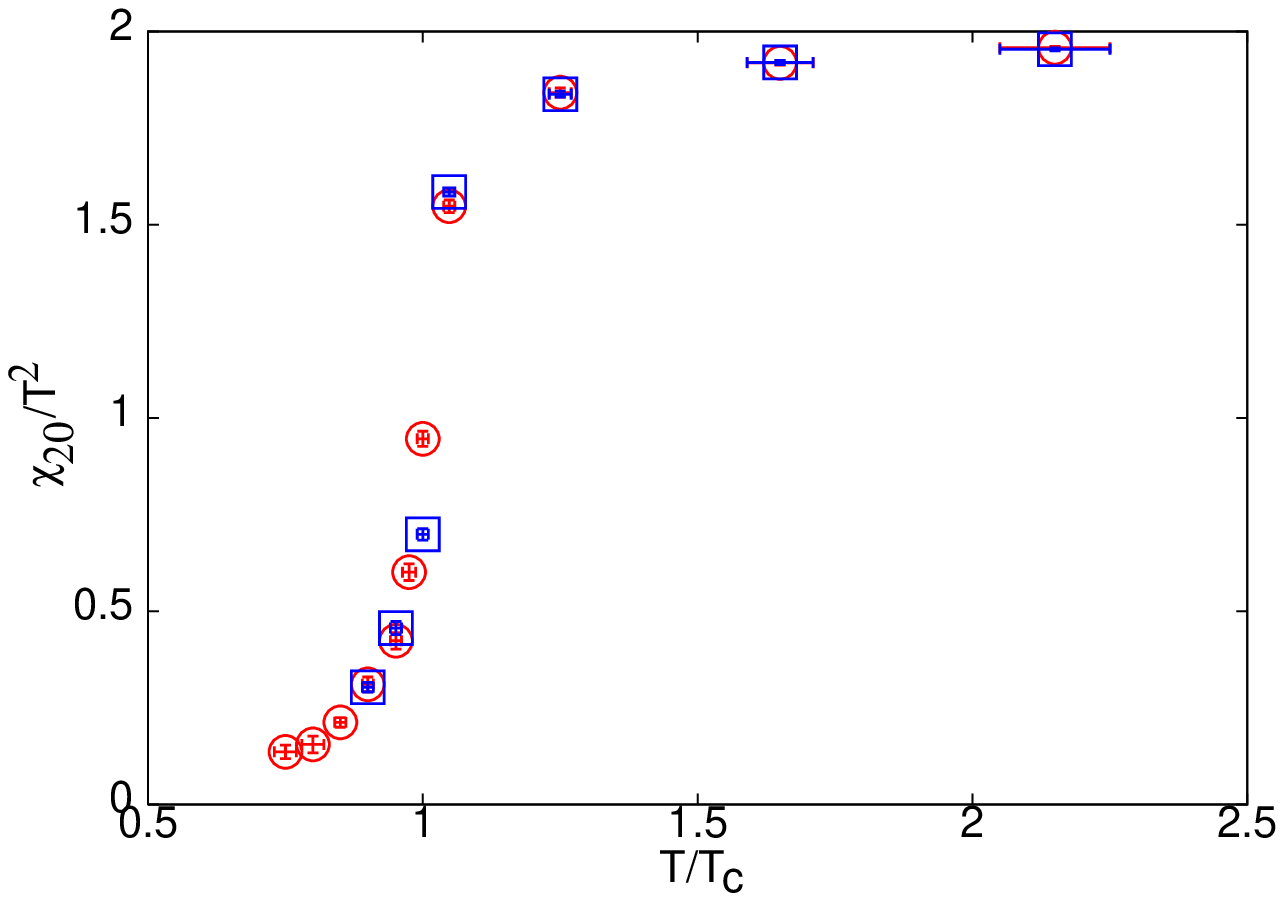}}
\scalebox{0.65}{\includegraphics{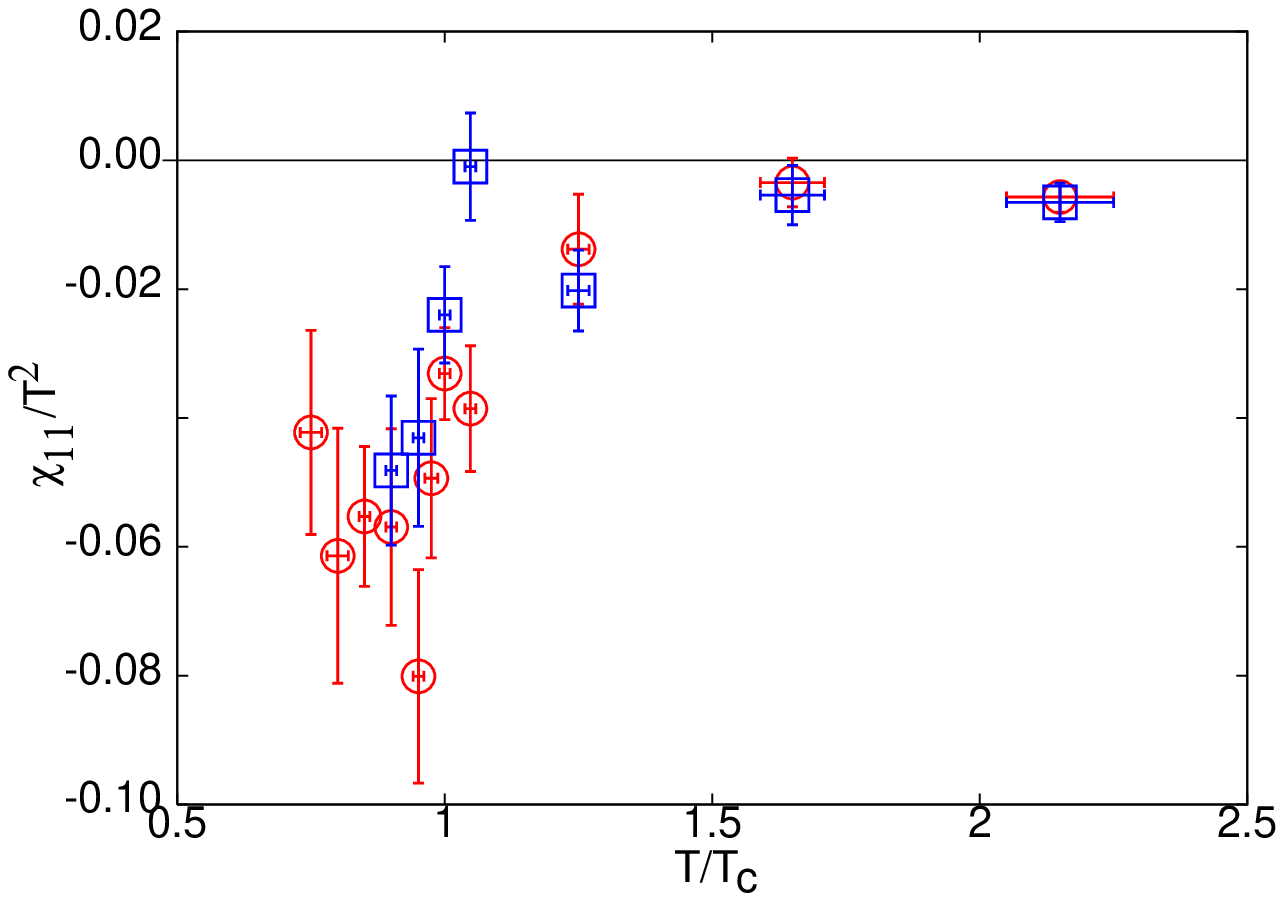}}
\scalebox{0.65}{\includegraphics{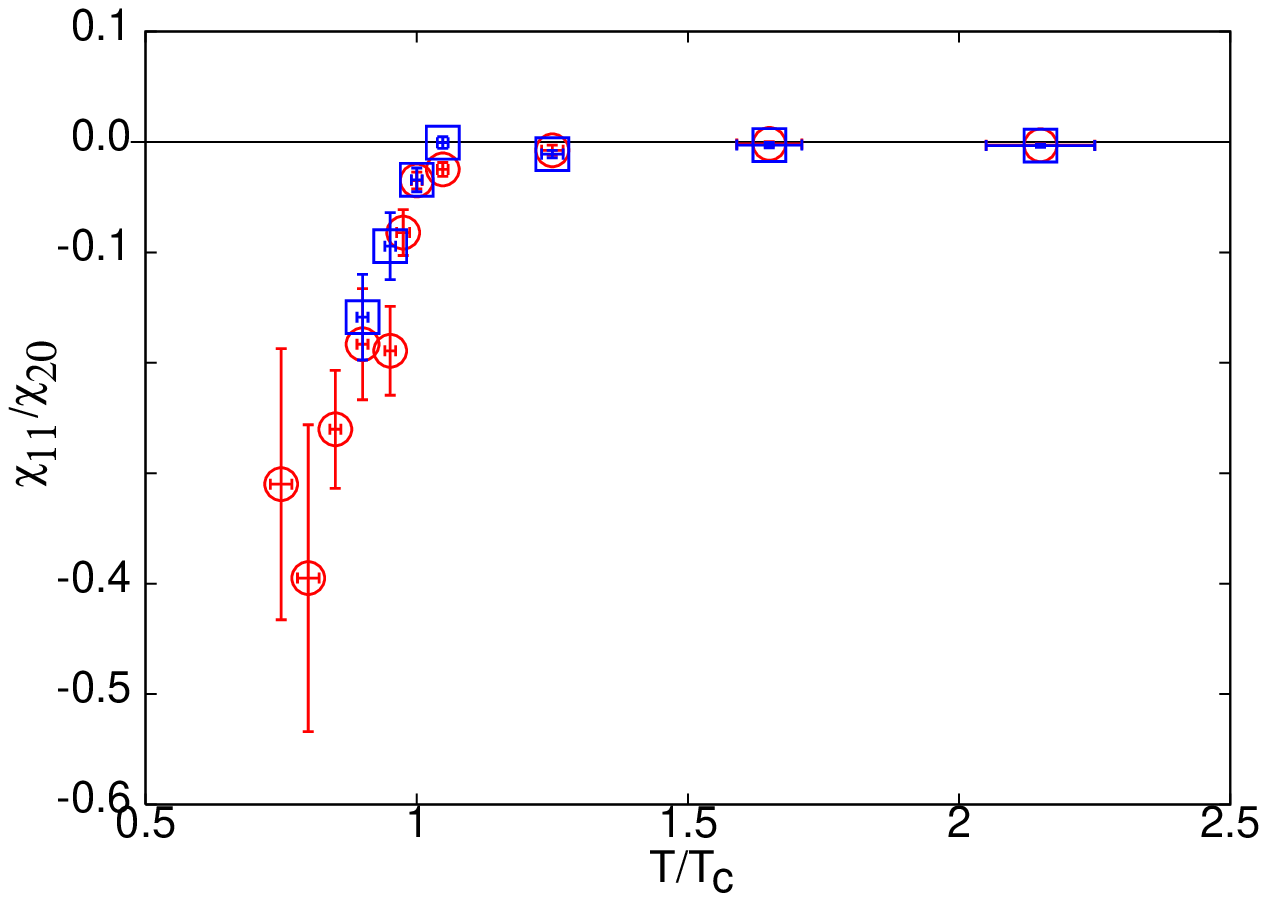}}
\end{center}
\caption{The quark number susceptibilities on lattice sizes $4\times16^3$ 
   (circles) and $4\times24^3$ (boxes). The first panel shows $\chi_2/T^2$,
   the second $\chi_{11}/T^2$, and the third the ratio.}
\label{fg.qns}\end{figure}

\begin{figure}
\begin{center}
\scalebox{0.65}{\includegraphics{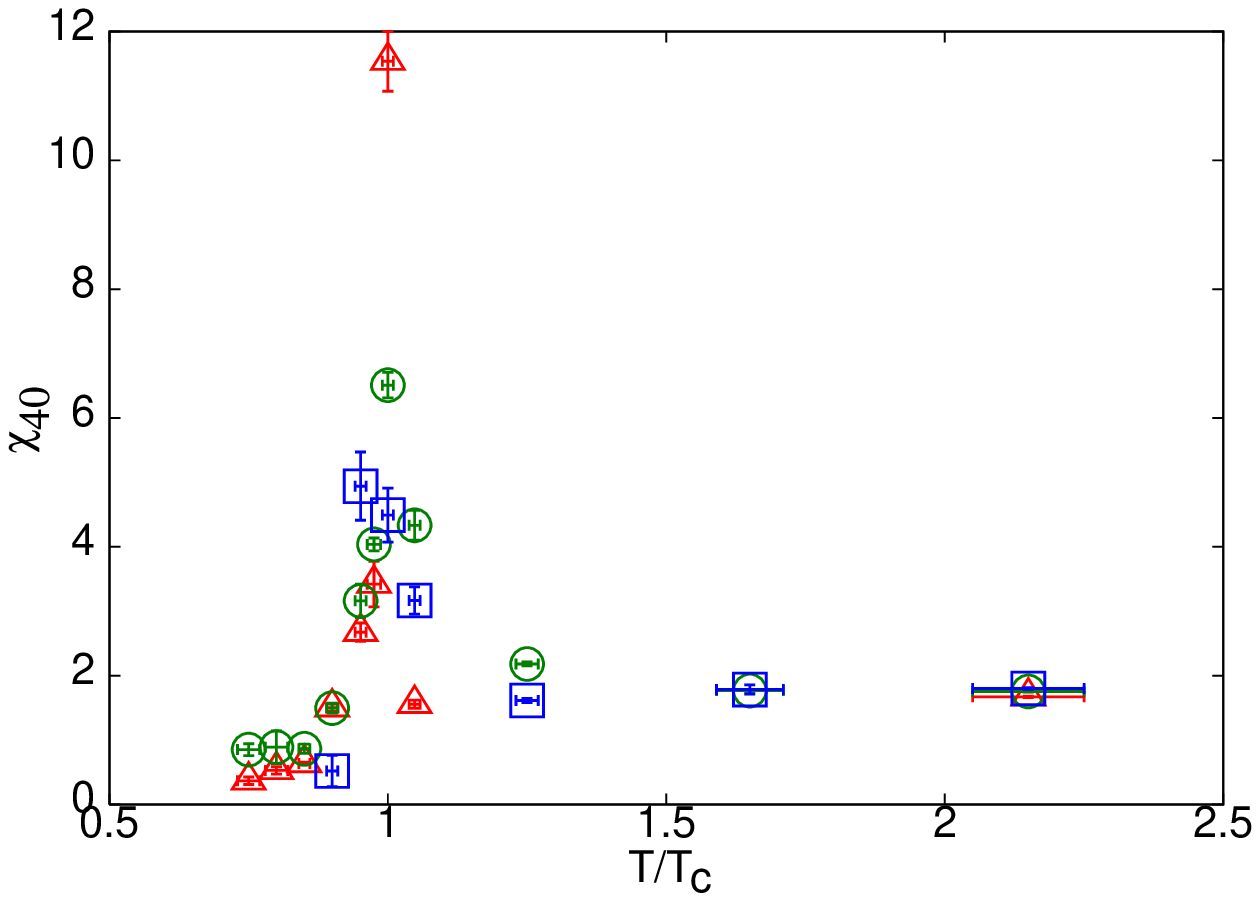}}
\scalebox{0.65}{\includegraphics{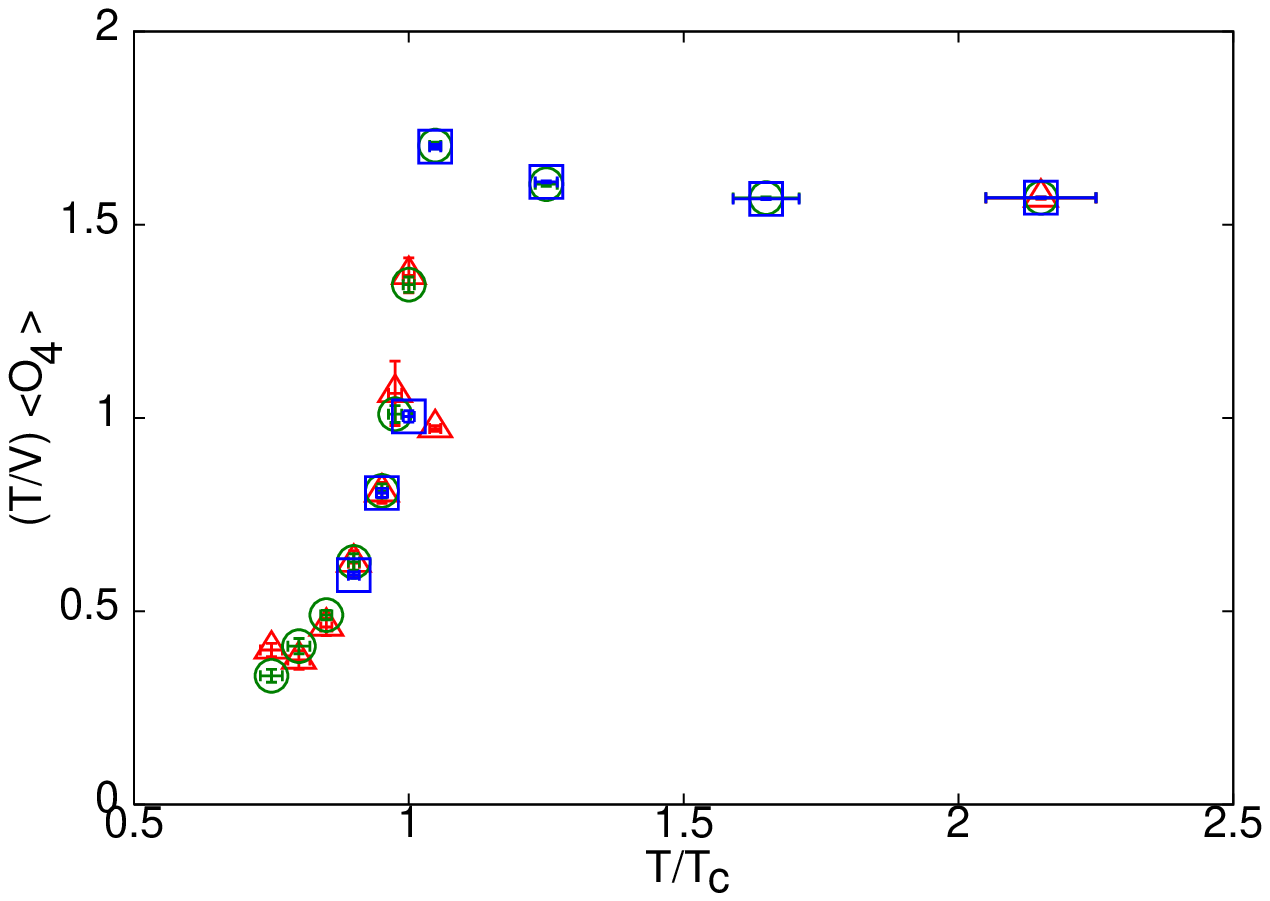}}
\end{center}
\caption{A fourth order QNS and some connected parts on lattice sizes
   $4\times12^3$ (triangles), $4\times16^3$ (circles) and $4\times24^3$
   (boxes). The first panel shows $\chi_{40}$ and the second $(T/V) \langle
   \O_4 \rangle$.}
\label{fg.ord4}\end{figure}

\begin{figure}
\begin{center}
\scalebox{0.65}{\includegraphics{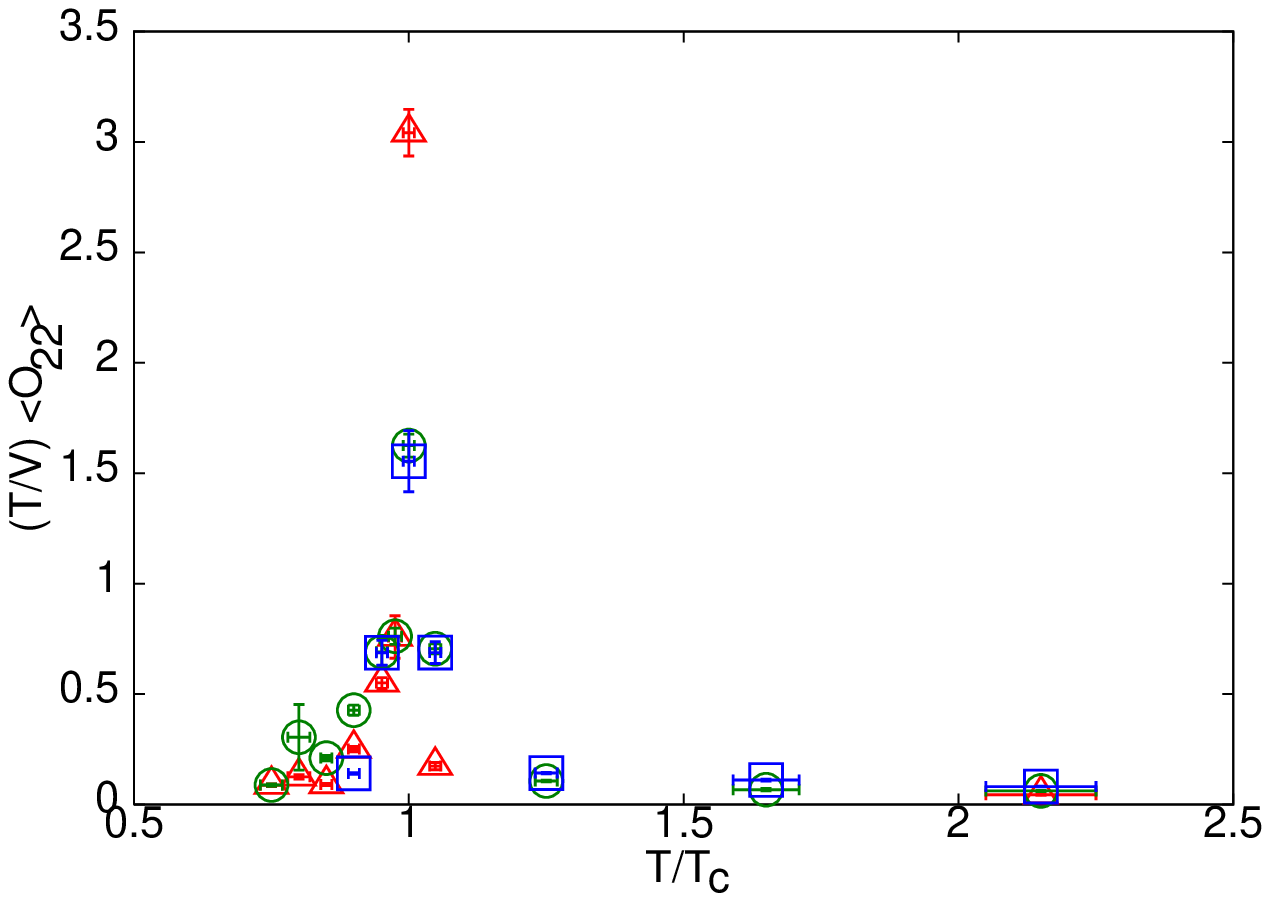}}
\scalebox{0.65}{\includegraphics{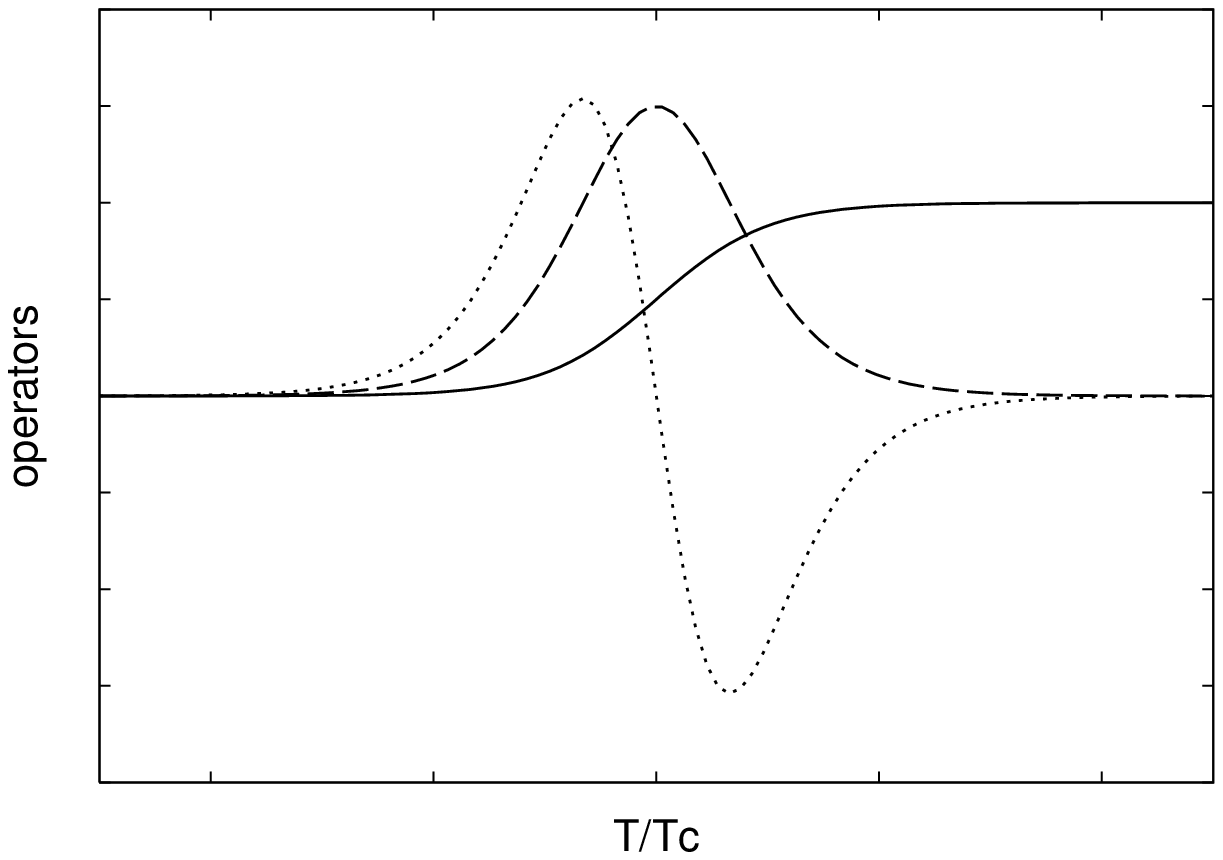}}
\scalebox{0.65}{\includegraphics{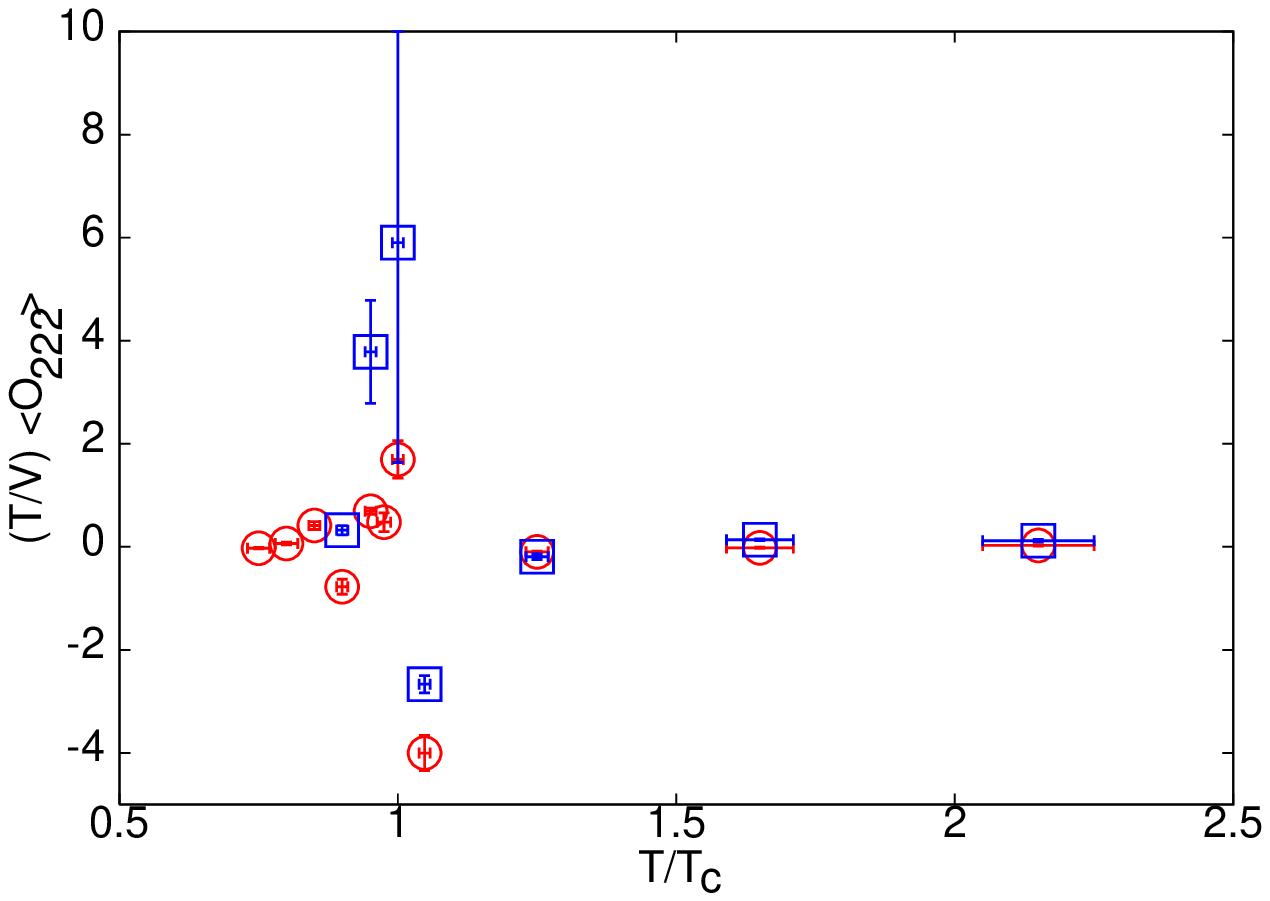}}
\end{center}
\caption{Some connected parts on lattice sizes $4\times16^3$ (circles)
   and $4\times24^3$ (boxes). The first panel shows $(T/V) \langle
   \O_{22} \rangle$. The second panel shows a model cross-over (full line)
   with first (dashed) and second (dotted) derivatives. The third panel
   shows $(T/V)\langle\O_{222}\rangle_c$. Note the similarities between
   data and model.}
\label{fg.nlin}\end{figure}

\subsection{Noise and measurement}\label{sc.noise}

Full control over errors in determining the non-linear susceptibilities is
crucial for the job of extrapolation to non-zero $\mu$. The major source
of numerical errors is the numerical cancellation of divergences which
go as powers of the volume in order to get a finite result. One part
of this program is the identification of primitive volume independent
quantities--- the connected parts of fermion-loop disconnected operators,
some of which were enumerated in Section \ref{sc.connected}.  In this
section we complete the program by showing that some choice of the
noise ensemble works better than others \cite{liu}, and that the number
of vectors is the crucial parameters, not the choice of the conjugate
gradient stopping criterion.

In Figure \ref{fg.volloss} we show the easiest part of this program---
the scaling with volume of the expectation values $(T/V) \langle
O_n\rangle$. These are fairly easy to handle because the operators
have no disconnected pieces and there are no divergences to cancel.
We find that the values and the error estimates are very stable against
the choice of the conjugate gradient stopping criterion, in method I
we can change $\epsilon$ from $10^{-4}$ to $10^{-6}$ without making a
change to these quantities. There may be a little volume dependence in
going from $N_s=8$ to $N_s=12$ lattices, but this saturates for $N_s$
between 12 and 24.  The errors become larger as $n$ increases.

We found that, in agreement with expectations, the disconnected pieces
of traces such as $\langle\O_{ij\cdots}\rangle$ grow with volume as a
power equal to the number of fermion-line disconnected pieces in the
operator. However, as shown in Figure \ref{fg.volloss}, the connected
part grows roughly linearly with volume. The residual volume dependence
is small, and the fact that $(T/V)\langle\O_{222}\rangle_c$ decreases
with $V$ shows that the cancellation of the leading divergent pieces
has been performed satisfactorily. These results have been obtained with
100 noise vectors drawn from a Gaussian ensemble.

We have made further tests of the effect of the choice of the noise
ensemble, as shown in Figure \ref{fg.z2}. It seems that it is preferable
to use Gaussian noise rather than $Z_2$ noise, since the former is more
stable against changes in the number of random vectors used. Since quite
the opposite conclusions have been presented in the literature, albeit
for different measurements \cite{liu}, we investigated this a little
further. It seems that indeed $Z_2$ noise performs better for single
traces, as in the measurements of $\langle\O_n\rangle$. However, the
measurement of products of traces such as $\langle\O_{ij\cdots}\rangle$
is cleaner with Gaussian noise.

\begin{figure}
\begin{center}
\scalebox{0.65}{\includegraphics{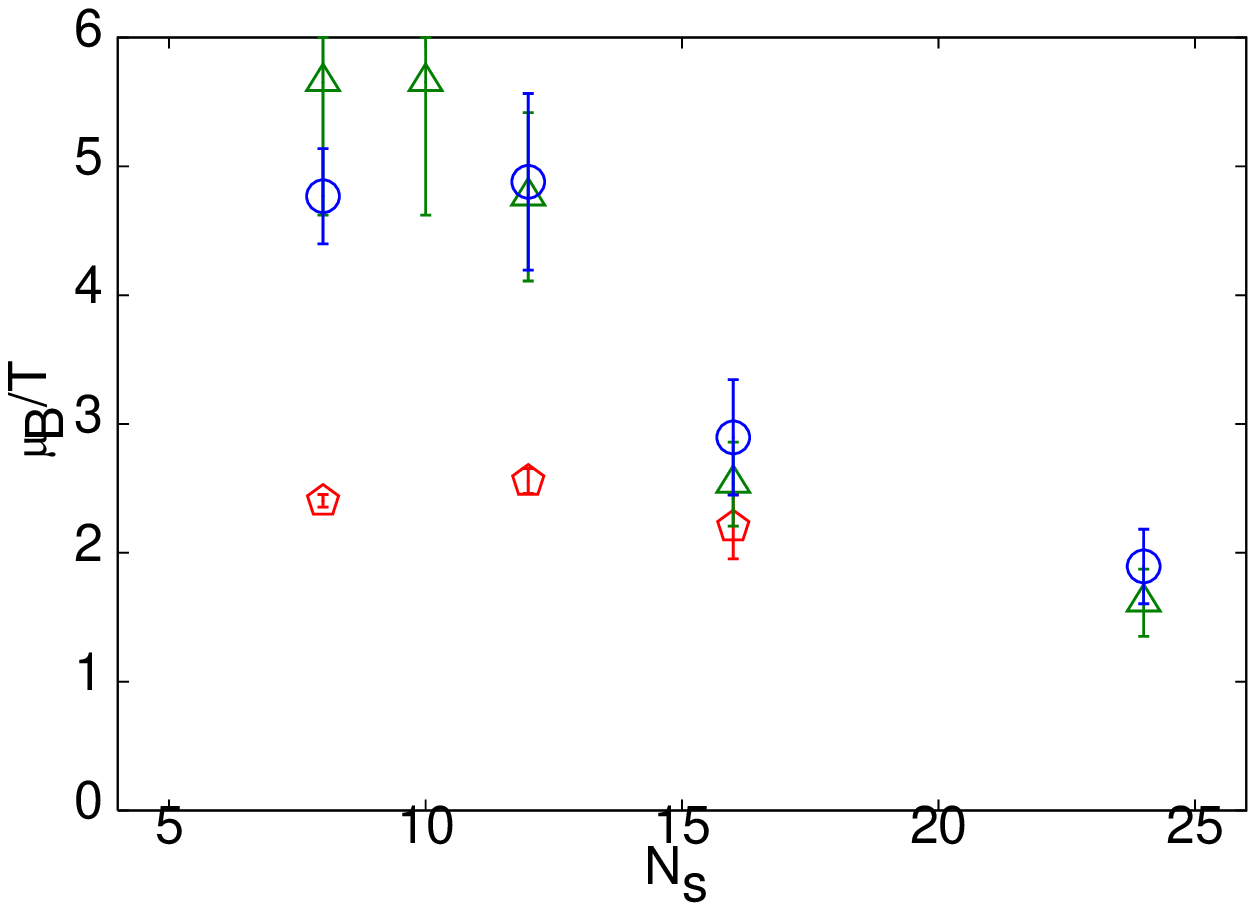}}
\scalebox{0.65}{\includegraphics{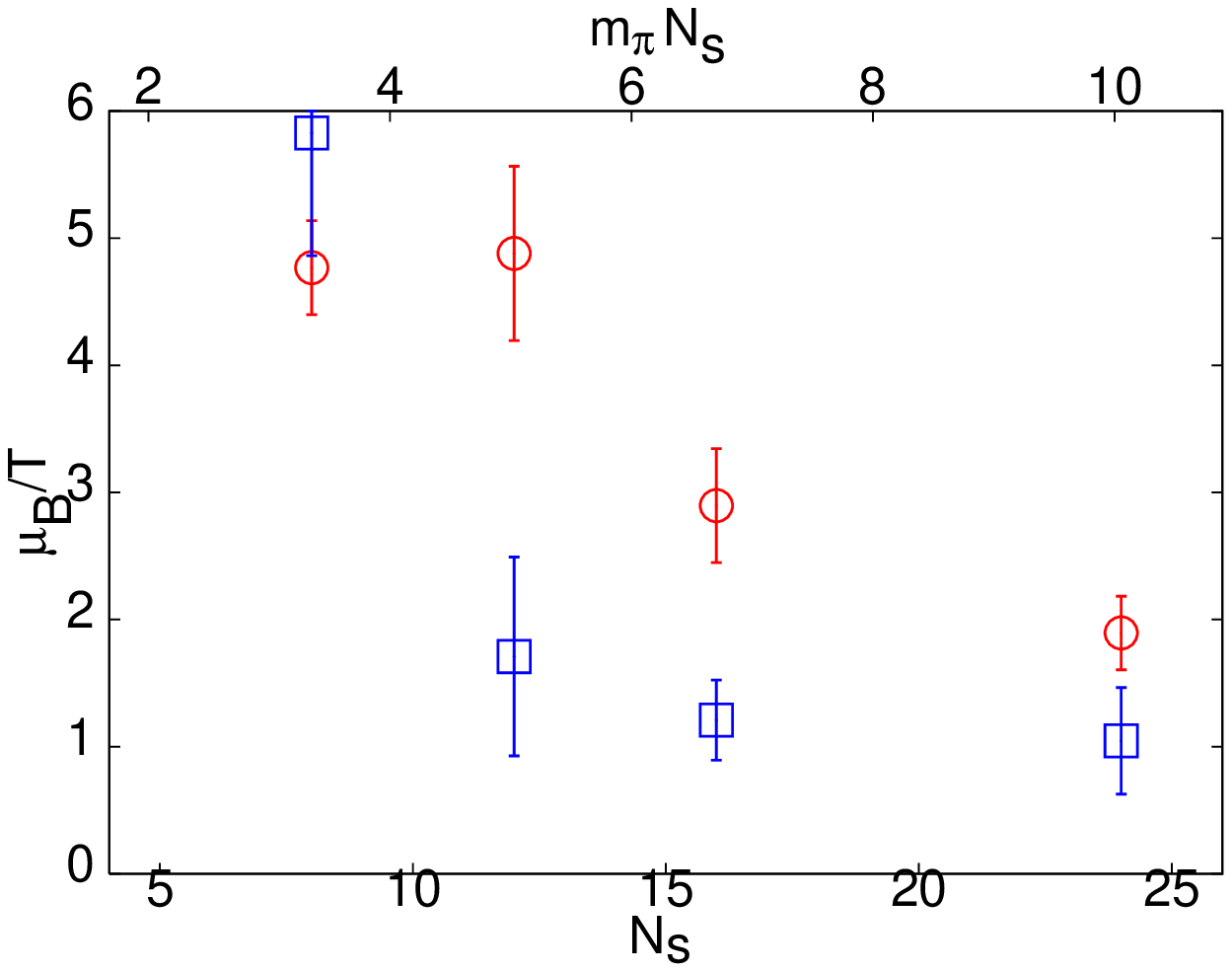}}
\end{center}
\caption{The radius of convergence of the series expansion for $\chi_{20}$.
   The first panel shows $r_6$ at $T/T_c=0.9$ (pentagons), 0.95 (circles)
   and 1 (triangles). The second panel shows results at $T/T_c=0.95$ for
   $r_4$ (circles) and $r_6$ (boxes).}
\label{fg.radii}\end{figure}

\begin{figure}
\begin{center}
\scalebox{1.00}{\includegraphics{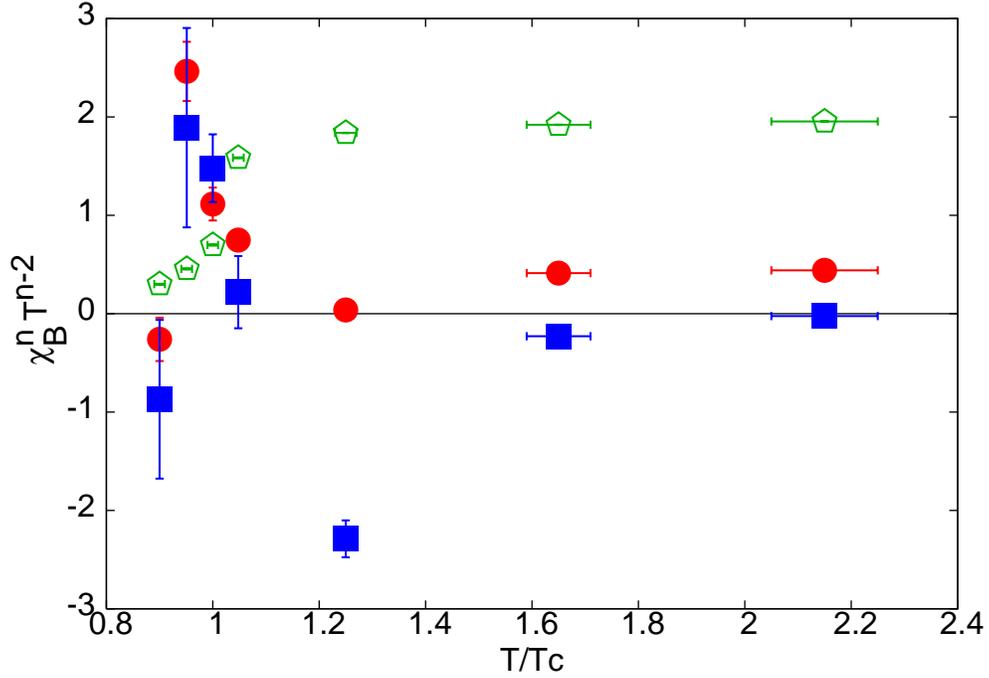}}
\end{center}
\caption{We show the dimensionless series coefficients, $\chi_B^n T^{n-2}$
   (eq.\ \ref{taylord}), obtained on $4\times24^3$ lattices, for $n=0$
   (pentagons), $n=2$ (circles) and $n=4$ (boxes). The $n=2$ coefficients
  have been divided by 2 and the $n=4$, by 12, in order to bring them all
  to roughly the same scale.}
\label{fg.expcoeff}\end{figure}

\begin{figure}
\begin{center}
\scalebox{1.0}{\includegraphics{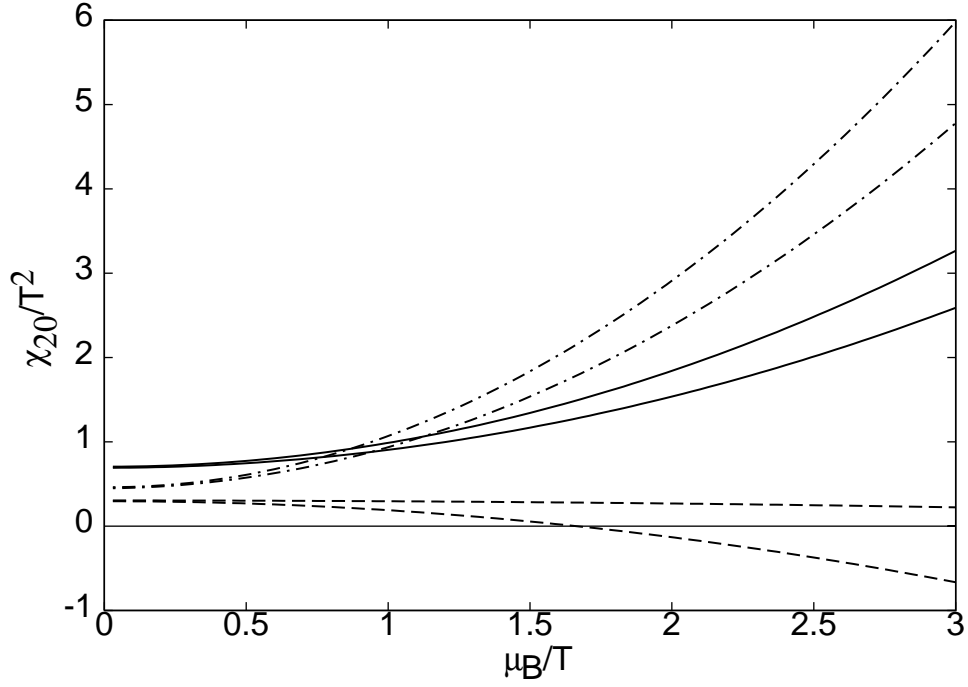}}
\end{center}
\caption{$\chi_{20}(\mu_B)/T^2$ obtained using merely two terms in the Taylor
   expansion already shows the qualitative change above $T=0.9T_c$. The 66\%
   confidence intervals for the extrapolation are shown for $T_c$ (full lines),
   $0.95T_c$ (dash-dotted) and $0.9T_c$ (dashed) from computations on a
   $4\times24^3$ lattice.}
\label{fg.chiexp}\end{figure}

\begin{figure}
\begin{center}
\scalebox{0.65}{\includegraphics{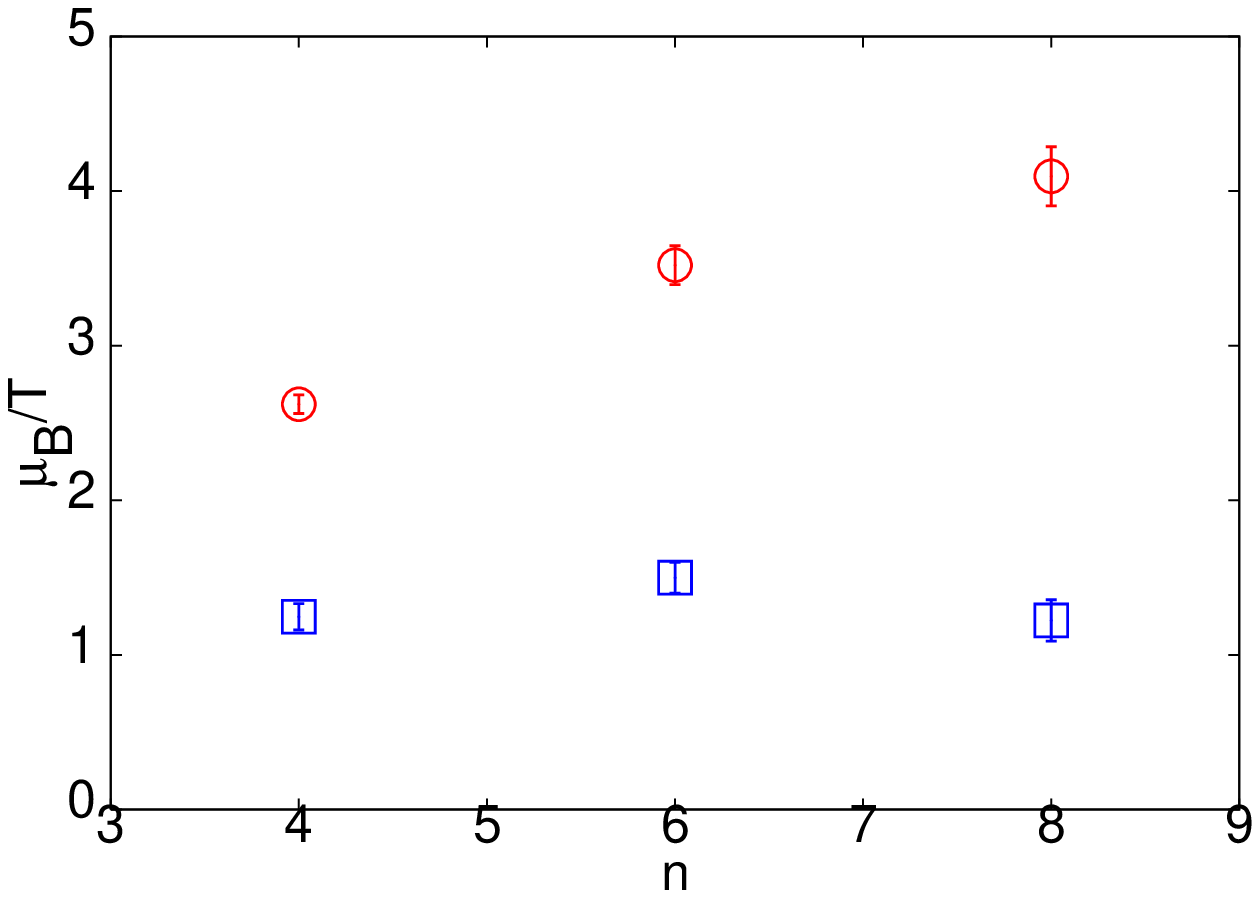}}
\scalebox{0.65}{\includegraphics{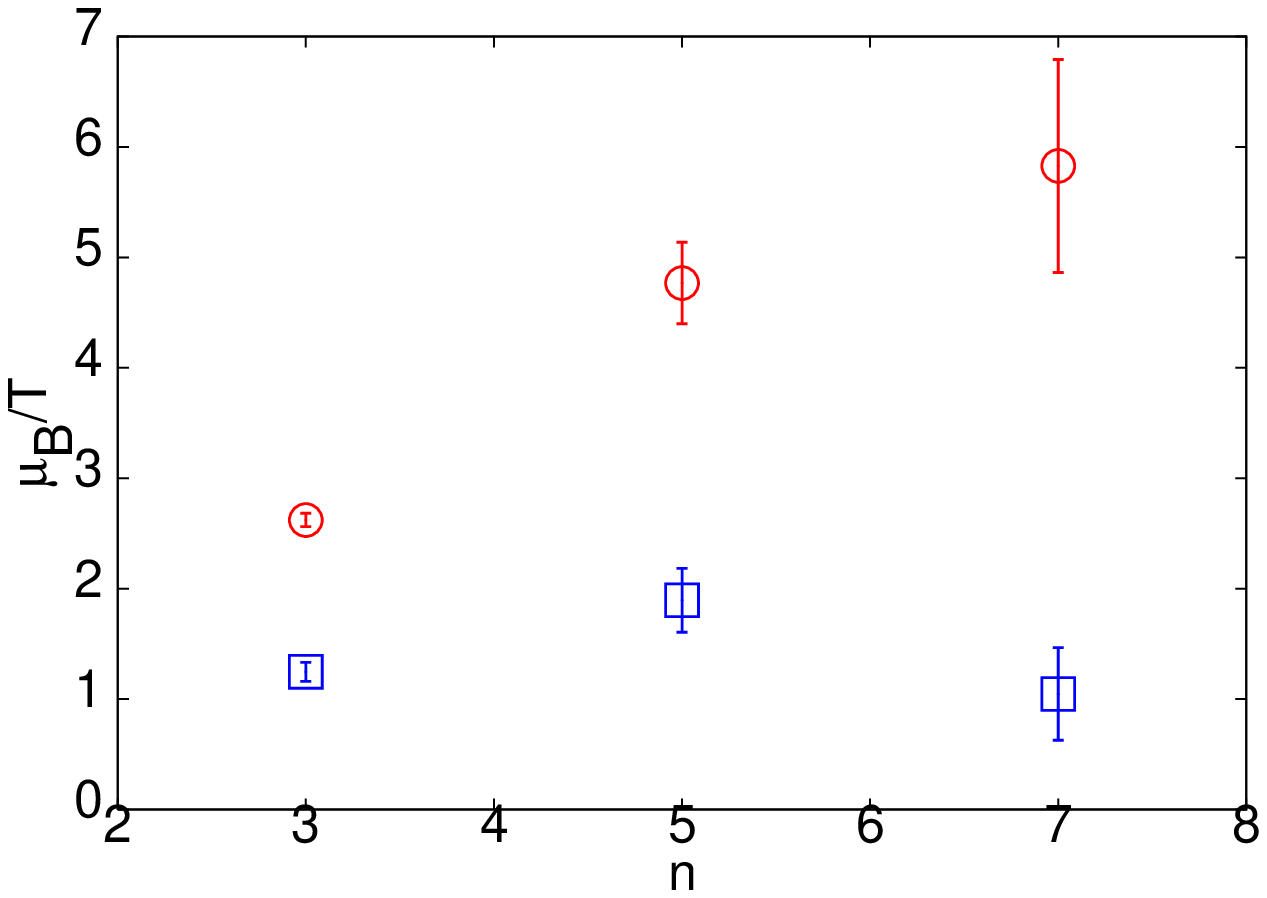}}
\end{center}
\caption{We show the radii of convergence as a function of the order of
   the expansion at $T=0.95 T_c$ on a $4\times8^3$ lattice (circles) and
   a $4\times24^3$ lattice (boxes). The panel on the left shows $\rho_n$
   and the one on the right is for $r_n$.}
\label{fg.orders}\end{figure}

The two final issues to be resolved are the role of the accuracy in the
conjugate gradient, and the number of vectors required for measurements.
In Figure \ref{fg.accuracyqns} we show that the choice of the CG stopping
criterion, $\epsilon_{CG}$, does not significantly affect the choice
of $N_v$. It is also clear from this figure that larger values of $N_v$
are required for the higher order measurements. In particular, $N_v=100$
seems to suffice for measurements up to the 4-th order, whereas $N_v=200$
is required beyond that for measurements up to the 8-th order, in the
hadronic phase. In fact, $N_v$ also depends on the lattice size. On
smaller lattices $N_v=200$ seem to suffices, but this grows with
lattice size and on the $4\times24^3$ lattices one needs $N_v=500$ below
$T_c$. Above $T_c$, $N_v=100$ seems to suffice for all the measurements
and lattice sizes.

\subsection{Susceptibilities and radii of convergence}\label{sc.results}

In this subsection we present results on the temperature dependence of
the linear and non-linear quark number susceptibilities.  Comparable
data on the linear QNS in the high temperature phase were presented in
\cite{pushan}. In figure \ref{fg.qns} we present data for $\chi_{20}/T^2$
and $\chi_{11}/T^2$. Some volume dependence is visible in the immediate
vicinity of $T_c$. The high temperature behaviour is compatible with
the predictions of \cite{bir,alexi}. The results are also completely
compatible with earlier results in \cite{pushan} after correcting for a
division by an extra factor of $(T/V)$ for $\chi_{11}/T^2$ reported there.
Comparison with the recent results of \cite{milc} are harder to perform
since the actions are different. As a result, such a comparison can be
meaningfully performed only after taking the continuum limit.

There is a clear qualitative difference in $\chi_{11}$ above and
below $T_c$. Above $T_c$ it is significantly easier to measure,
is small and compatible with zero. Below, and in the vicinity of
$T_c$, it is possibly larger, but much harder to measure because
of fluctuations.  The ratio $\chi_{11}/\chi_2$, taken at $\mu_B=0$,
quantifies the magnitude of the fermion sign problem at small $\mu_B$,
as discussed in Section \ref{sc.taylor}.  This ratio is also shown
in Figure \ref{fg.qns}.  It is clear that the fermion sign problem is
negligible at higher temperatures.  The success of modern extrapolation
methods at finite $\mu$ \cite{fk,pressure,biswa} is, at least partially,
due to this. Due to the rapid increase in phase fluctuations on lowering
$T$, visible in Figure \ref{fg.qns}, any extrapolation technique would
be hard to use significantly below the lowest temperatures we have
used. Since $\chi_{11}$ depends strongly on the pion mass, we expect
phase fluctuations to pose greater difficulties as one approaches the
physical pion mass. Conversely, the sign problem would be easier to deal
with in simulations at higher pion masses, such as \cite{biswa}.

We show the fourth order NLS, $\chi_{40}$, in Figure \ref{fg.ord4}
for our three largest lattice volumes. Away from the critical region
the volume dependence is seen to be negligible. In the critical region,
however, $\chi_{40}$ exhibits behaviour similar to that of $\chi_L$
(Figure \ref{fg.vdep}). The susceptibility peaks near $T_c$ and shows
pseudo-critical behaviour, since the near-critical region visibly shrinks
with increasing volume. Since $\chi_{40}$ decreases with $V$ at fixed
$\beta=5.2875$, we again have evidence that the cross-over coupling is
near this, but not exactly here.

Further insight into the nature of this  peak comes from examining
the other NLS at this order. It turns out that $\chi_{22}$ also has a
similar sharp peak, whereas $\chi_{31}$ is much smoother. These lead us
to examine the connected parts which contribute at this order (see eq.\
\ref{nf4inv}).  We found that $(T/V)\langle\O_4\rangle$, shown in Figure
\ref{fg.ord4}, seems to behave almost as a (pseudo) order parameter
for the crossover at $T_c$. In this it behaves like $\chi_{20}/T^2$
and, indeed, like all the single traces $(T/V)\langle\O_n\rangle$ that we have
examined. The peaking behaviour is essentially due to a similar peak
in $(T/V)\langle\O_{22} \rangle_c$, which we have displayed in Figure
\ref{fg.nlin}.

Note that $\O_2$ is a composite bosonic operator.  It may be
possible to write down effective long-distance theories in which
it is treated as a field operator whose expectation value shows the
correct cross over behaviour. In that case $\langle\O_{22}\rangle_c$
would be the susceptibility of this field, and being proportional
to the temperature derivative of $\langle\O_2\rangle$, would peak,
as observed. In fact, one can push this idea further, and examine the
expectation values $\langle\O_{222}\rangle_c$.  In such an effective
theory one would find this to be proportional to the next derivative
of $\langle\O_2\rangle$. Then the $T$-dependence of these quantities
at $\mu_f=0$ would have the shapes shown in the second panel of Figure
\ref{fg.nlin}. In fact, the data, shown in the final panel of Figure
\ref{fg.nlin}, does show the same qualitative behaviour. Thus, the
cross-over at $T_c$ can be probed by any of these connected parts.

There is significant volume dependence in the close vicinity of $T_c$,
although for $T\le0.95T_c$ one is clearly outside the pseudo-critical
region, and the volume dependence is weak.  Within the pseudo-critical
region the peak of $\langle\O_{22}\rangle_c$ is seen to have the
same qualitative behaviour as the peak of $\chi_L$ shown in Figure
\ref{fg.vdep}, including similar evidence for finite volume shift in
the cross over coupling. There are consequent effects on the higher
order connected parts, as shown in Figure \ref{fg.nlin}.

With the NLS at hand, we construct the Taylor expansion of the
diagonal QNS, $\chi_{20}(\mu_B)$, where the coefficients are given
in eq.\ (\ref{taylord}).  Radii of convergence can be obtained for
the series expansion $f(x)=\sum f_{2n} x^{2n}$ using two definitions---
\beq
   \rho_n=\left|\frac{f_0}{f_{2n}}\right|^{1/2n},
     \qquad{\rm and}\qquad
   r_n=\left|\frac{f_{2n}}{f_{2n+2}}\right|^{1/2},
\label{rads}\eeq
At large $n$ both definitions should be equivalent, although for
many series it is known that the latter gives somewhat smoother
approach to the limit of $n\to\infty$. Also, at a critical point,
one may extrapolate the latter to the limit $n\to\infty$ by a fit
to a $1/n$ behaviour \cite{gaunt}. We apply these definitions to
the series for the baryon number susceptibility (see above eq.\
\ref{taylord}).

In Figure \ref{fg.radii} we display the radii, $r_4$ and $r_6$, at
several different temperatures as a function of $N_s$: the volume
dependence is strong.  For $0.95\le T/T_c\le1$, the radii roughly agree
with the estimate of the critical end point in \cite{fk} for 
$N_s$ comparable to those used in that study. However,
as the volume increases, the radius of convergence drops. The magnitude
of the finite volume effects is parametrized by the pion Compton wavelength,
and for $m_\pi N_s\ge5$--6 one finds that these effects saturate. 
This is in accord with the discussion in Section \ref{sc.fss}.
Similar
saturation of finite size effects has been seen in other measurements
in the chiral sector at similar values of $m_\pi N_s$ \cite{ray}.

At temperatures below $0.95T_c$ the radius of convergence falls
quite dramatically. However, examination of the series coefficients
shows that in this lower range of temperatures the 6th order
coefficient, $\chi_B^6$, is negative (see, Figure \ref{fg.expcoeff}).
Thus the radius of convergence does not indicate a divergence.
Rather, it shows that at this temperature $\chi_{20}(\mu_B)<0$,
thus violating thermodynamic stability, and therefore indicating
that higher order terms in the expansion become necessary. This is
consistent with the increase in the ratio $\chi_{11}/\chi_{20}$ at
these $T$ (see Figure \ref{fg.qns}).  At temperatures between $T_c$
and $0.95T_c$, the character of the expansion is quite different,
with all the Taylor coefficients being positive. As discussed
earlier, when this is true the series diverges on the real
$\mu_B$ axis. The qualitative difference in these two expansions
is already visible at low order, and is illustrated in Figure
\ref{fg.chiexp}. Note that a figure like this illustrates whether
or not there is a divergence, but since it shows the sum of a finite
number of terms, it cannot show anything special happenning at the
radius of convergence.

In Figure \ref{fg.orders} we show the variation of the two different
definitions of the radius of convergence, $\rho_n$ and $r_n$, with
the order $n$. On our smallest lattice, $4\times8^3$, the successive estimates
diverge, and the two sets of estimates are compatible at the
2-$\sigma$ level. After the cross over to large volume behaviour,
the estimates stabilize with order, being nearly independent of
$n$. Also $\rho_n$ and $r_n$ are in close agreement with each other.
The radius of convergence extrapolated to all orders at $T^E=0.95T_c$
gives $\mu^E/T^E=1.1\pm0.2$.

If we assume that the end point is in the Ising universality class,
then the finite volume shift in the end point is given by
\beq
   \mu^E(V)=\mu^E(\infty) + \frac a{V^{1/\delta}},
   \qquad{i.e.}\qquad
   \mu^E(V)-\mu^E(V') = \frac a{V^{1/\delta}}-\frac a{{V'}^{1/\delta}},
\eeq
where $\delta=5$ is the 3-d Ising magnetic exponent and $V=N_s^3$.
Using the data exhibited in Figure \ref{fg.radii} for $N_s=16$ and
24, which are both above the crossover, it is easy to check that the
finite volume shift in $\mu^E$ is bounded by the statistical error
in the estimate for $N_s=24$. In view of this, we quote the estimate
and its error obtained for the $N_s=24$ lattice as the estimate in the thermodynamic limit.

\section{Summary and discussion}\label{sc.disc}

The main result we present in this paper is the strong volume
dependence of location of the critical region, leading to a
substantially smaller estimate of $\mu_B^E$ on large volumes than
before.  We began by
constructing the Taylor expansion for the quark number susceptibility,
$\chi_{20}(T,\mu_B)$, for $N_f=2$ QCD at for several large volumes.
The leading ($\mu_B$ independent) term of the series
is the quark number susceptibility (QNS) which has received extensive
attention recently \cite{pushan,valence,milc,bir,alexi,mustafa}. The
remaining Taylor coefficients involve the non-linear susceptibilities
(NLS) which were defined in \cite{pressure}. We have taken the series
up to the 6th order term, which involves 8th order NLS. A systematic
and efficient procedure for generating and computing the quark number
susceptibilities at any order was presented in Sections \ref{sc.taylor}
and \ref{sc.traces}.

\begin{table}\begin{center}\begin{tabular}{r|rr|cc|rrr}
\hline
    $m_\rho/T_c$ & $m_\pi/m_\rho$ & $m_N/m_\rho$ 
              & $N_s m_\pi$ & flavours 
              & $T^E/T_c$ & $\mu_B^E/T^E$ & reference
   \\
\hline
    5.372 (5) & 0.185 (2) & --- & 1.9--3.0 & 2+1 
                & 0.99 (2) & 2.2 (2) & \cite{fktwo} 
   \\
    5.12 (8)  & 0.307 (6) & ---  & 3.1--3.9 & 2+1
                & 0.93 (3) & 4.5 (2) & \cite{fk}
   \\
    5.4 (2)   & 0.31 (1)  & 1.8 (2) & 3.3--10.0 & 2
               & 0.95 (2) & 1.1 (2) & this work
   \\
    5.4 (2)   & 0.31 (1)  & 1.8 (2) & 3.3 & 2 & --- & --- & \cite{owe}
   \\
    5.5 (1)   & 0.70 (1)  & --- & 15.4 & 2
               & --- & --- & \cite{biswa} 
   \\
\hline
\end{tabular}\end{center}
\caption{Summary of critical end point estimates--- the lattice spacing
   is $a=1/4T$. $N_s$ is the spatial size of the lattice and $N_sm_\pi$ is
   the size in units of the pion Compton wavelength, evaluated for $T=\mu=0$.
   The ratio $m_\pi/m_K$ sets the scale of the strange quark mass. As the
   mass scales indicate, the lattice spacings and $u$ and $d$ quark masses
   of \protect\cite{fk} and this work are comparable. The numbers in brackets
   indicate the errors on the least significant digit. Since no estimates of
   the critical end point are quoted in \cite{biswa,owe}, the comparison is
   more complicated, and is described in the text.}
\label{tb.summ}\end{table}

The main thrust of this work is to approach the large volume limit.
From the point of view of thermodynamics, this limit has to be taken in
order to decide whether rapid changes in certain quantities and peaks in
others are due to critical behaviour, cross over or a first order phase
transition. For the $\mu_B=0$ cross over at $T_c$, good progress has
been made \cite{edwin}. However, due to questions such as the magnitude
of the shift in the critical coupling, and the absence of evidence for
peaking of different susceptibilities at slightly different points, an
unqualified answer is still not available. Note, however, the remarkably
high accuracy in the measurement of $T_c$ that is possible even without
answering this question. The situation is similar at critical points.
It is usually much easier to identify the critical point (assuming it is
critical) than it is to ``prove'' criticality through extensive analysis.
In this work we have shown that there is a minimum lattice size that
one should use in order to get a reliable estimate of the position
of the critical end point. Further work involved in ``proving''
criticality by studying detailed finite size effects, with substantially
larger lattice volumes and extracting critical indices, is beyond our
present computational capability and is left for the future.

\begin{figure}
\begin{center}
\scalebox{1.0}{\includegraphics{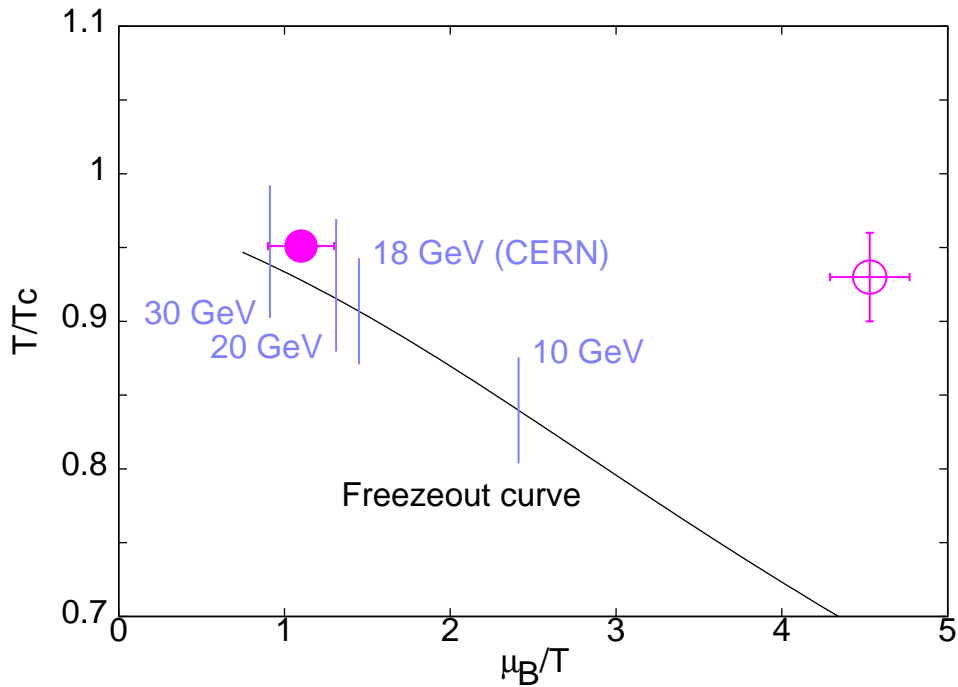}}
\end{center}
\caption{The phase diagram and the freeze-out curve superimposed. The
   filled circle denotes the estimate of the critical point which was
   obtained in this work. The open circle is an earlier estimate from
   \cite{fk} using smaller lattices and nearly the same quark mass
   as in this work (see the text for a discussion of their comparability).
   The freezeout curve \cite{pvt} has been converted for this figure
   using a value of $T_c$ appropriate to our computation.  A scale of
   the CM energy per nucleon, $\sqrt S$, has been marked on the freezeout
   curve.}
\label{fg.composite}\end{figure}

Taking the large volume limit is also necessary to resolve the small
eigenvalues of the Dirac operator, and thus control the chiral behaviour
required to see the critical end point. Since we use large lattices
as well as small quark masses, there are new technical problems to be
solved (see Section \ref{sc.noise}). Using the methods developed here,
the time required to compute the expansion to 8th order is between 1
to 3 times that required to generate an independent configuration well
away from the finite temperature cross over at $T_c$.  Closer to $T_c$
the time required for computing the Taylor expansion drops rapidly to
1/10 of the time required to generate a statistically independent
configuration. Since one never actually computes anything at the critical
point, there is no critical slowing down.

Away from $T_c$ we found little volume dependence for the QNS and
the NLS.  This extends our earlier results, which were confined to
$T>T_c$ \cite{pushan}.  The off-diagonal quark number susceptibilities,
$\chi_{11}$, are reasonably large (but noisy) below $T_c$ and become
small in the high temperature phase, where they are in rough agreement
with the results of a perturbative computation in \cite{bir}. The ratio
of these two susceptibilities is a direct estimate of the severity
of the sign problem (see eq.\ \ref{dexpn}).  We found that the sign
problem is very severe at small $T$, and acute even at $T\approx0.85T_c$
(see Figure \ref{fg.qns}). This manifests itself in the Taylor expansion
by requiring that many terms in the series be summed in order to get
physically sensible results (see Figure \ref{fg.chiexp} and the discussion
in Section \ref{sc.results}).  This aspect of the sign problem has been
investigated in \cite{cohen} and named the ``silver blaze problem''.

Near $T_c$, on the other hand, volume dependences were significant. This
is, of course, necessary for a system to become critical. However,
we presented additional evidence that there is no criticality
at $\mu_B=0$ by examining the volume dependence of the maximum
of $\chi_L$ (Figure \ref{fg.vdep}). This adds to the growing body of
evidence \cite{ray,edwin,digiacomo} that the critical point moves out to
finite $\mu_B$ when the pion mass (or equivalently, the quark mass) is
finite. Finally, the analysis of the series expansion of $\chi_{20}/T^2$
showed clear evidence of a breakdown and consequent criticality for
$T^E\approx0.95$ and $\mu_B^E/T^E\approx1.1$ on large lattices (see
Table \ref{tb.summ} for the error bars). We saw a crossover from small
to large volume behaviour for lattice sizes $N_sm_\pi\approx6$ (see
Figure \ref{fg.chiexp}). For the pion mass used here, that corresponds
to $N_s\approx14$. Similar finite size effects have been reported in
\cite{ray}, and ascribed to the lack of energy resolution for the small
eigenvalues of the Dirac operator when working at small volumes.

In Table \ref{tb.summ} we have collected all estimates of the critical
end point in recent lattice computations, along with other relevant
scales. It is interesting that the nucleon mass is still very large
compared with the real world. While this mass would change on tuning the
quark mass, the ratio $m_N/m_\rho$ will not decrease significantly towards
the continuum value except by changing the lattice spacing. Since the
nucleon mass is important for determining the value of the QNS $\chi_{20}$
below $T_c$, it is clear that going towards smaller lattice spacing is
equally important for getting a reliable estimate of the critical end point
of QCD. Unfortunately for computations with improved actions, there is
as yet no evidence that $m_N/m_\rho$ takes on its physical value already
at coarse lattice spacing. We emphasize that our method is capable of
handling the larger lattices required to take the continuum limit; the
only constraint is the usual polynomial growth in the required CPU time
as a function of $N_\tau$. From a comparison in Table \ref{tb.summ}
of our results with those obtained in \cite{fk,owe}, it appears that
the strange quark in \cite{fk} seems to play only a small role in the
determination of the scales.

Instead of an estimate of the critical end point, \cite{owe} quotes an
estimate of the pseudo-critical line $\beta(\mu)$, in the expectation that
the critical end point lies on this. The end point estimate of \cite{fk}
is indeed comparable.  On comparable lattice volumes we found that our
method of estimating the critical end point gives results which are
comparable to \cite{fk,owe}. Our true estimate of the end point, quoted in
Table \ref{tb.summ} however comes from computations on larger volumes,
as explained above.

The estimate of the end point in \cite{biswa} is indeed carried
out on large volumes in units of the pion Compton wavelength. However
the pion mass used there is twice as large as here, and hence far from
the physical value. The compatibility of the results of \cite{fk}
and \cite{biswa} can then be ascribed to a fortuitous cancellation
of volume dependence and the pion mass effects. Knowing that $m_\pi$ in
these two computations differed by a factor of two, this conclusion could have been
reached earlier by noting that the value of $\mu_B^E/T^E$ drops by a factor
of two when $m_\pi$ is decreased by a further factor of two \cite{fktwo}.

This is an appropriate place to list caveats. By identifying the radius of
convergence with the location of the critical point, we have found a fairly
broad region over which indications of criticality can be seen. Strictly,
the radius of convergence is just the lower bound for the location, and
higher order terms should be examined in future.
The extent of this critical region in the $\mu_B$ direction is marked out by the filled circle in
Figure \ref{fg.composite}. The fact that present computations, ours
included, are performed on coarse lattices with lattice spacing $a=1/4T$
has several implications.  The most important is the question of scale. At
such lattice spacings the ratio $m_N/m_\rho$ is too large. This implies
that the value of $T_c$ changes by a large factor if one sets the scale
by the nucleon mass rather than by the rho mass. In effect this means that
one has to go to finer lattice spacings to get the scale. We have set the
scale by the somewhat more stable method of using $\lamms$ identified from
a running coupling on the lattice \cite{scale,su3}. However, one needs
to go to smaller lattice spacing in order to stabilize this estimate
(see Section \ref{sc.setup} for details).  It has also been pointed out
in \cite{pressure} that the continuum limit has to be taken in order to
remove an ambiguity in the prescription for putting chemical potential
on the lattice.

We further note that direct proof of criticality through the extraction
of the critical indices at the end point would require significantly
longer series expansions than we have used here. It would be useful to
convert the algorithms for generating the coefficients into ``compilers''
so that this job can be done automatically. In the interesting region
near $T_c$ the measurements take a small fraction of the time needed to
generate the configurations. As a result it seems plausible that such
extensions of the present work can be undertaken in the near future.

Nevertheless, it is interesting to speculate on questions whose answers
are beyond our ability to compute at present. Since $\mu_B^E/T^E$ is
expected to decrease with $m_\pi$, our estimate of the end point
puts an upper bound to this quantity. Finite lattice spacing errors
were estimated in \cite{pressure} and can be controlled in computations in
the near future. This upper bound on the estimate of the critical end
point we obtain suggests that an experimental search for the end point
is feasible today.

We have illustrated this in Figure \ref{fg.composite} by plotting our
end point estimate along with a freezeout curve determined in
\cite{pvt}. This freezeout curve has been converted to $T/T_c$ units
using a value of $T_c$ appropriate to our computation.
Since this curve is
obtained by treating a resonance gas as ideal, we expect this freezeout
curve to lie below the line of transitions and the critical point,
as it indeed does. We have marked out on the freezeout curve values of
the CM energy, $\sqrt S$, per nucleon needed to reach that point on the
curve. The fact that the heavy-ion experiments at CERN with $\sqrt S=18$
GeV did not see any sign of a nearby end point is a confirmation that
the pion mass in the real world is lower than that used here.  An energy
scan at the RHIC should be able to locate the critical end point of QCD.

{\bf Acknowledgements:}
We would like to thank Simon Hands and Owe Phillipsen for discussions
during the program ``QCD at finite density: from lattices to stars'' of
INT Seattle.  SG would also like to acknowledge Jaikumar Radhakrishnan
for developing the connection between the problem of efficient trace
evaluation and the Steiner problem, Keh-Fei Liu for a discussion on the
properties of $Z_2$ noise, and Edwin Laermann for discussions during a
visit to the University of Bielefeld funded by the French-German Graduate
school funded by DFG under grant No. GRK 881/1-04.  RVG would like to
thank the Alexander von Humboldt Foundation for financial support and
the members of the Theoretical Physics Department of Bielefeld
university for their kind hospitality.  This computation was carried out
on the Indian Lattice Gauge Theory Initiative's CRAY X1 at the Tata
Institute of Fundamental Research. It is a pleasure to thank Ajay Salve
for his administrative support on the Cray.

\appendix
\section{The susceptibilities}
\subsection{The susceptibilities}

There are two steps to writing the NLS. First the derivatives
of the free energy (or $\log Z$) are expressed in terms of the derivatives of $Z$.
Second, the derivatives of $Z$ are expressed in terms of products of traces
(the quark operators).

The first step is to take the derivatives---
\beq
   \chi_{n_u,n_d} = 
   \frac{\partial^n P(T,\{\mu\})}{\partial\mu_i\partial\mu_j\cdots} =
    \left(\frac TV\right)
      \frac{\partial^{n_u+n_d} \log Z(T,\{\mu\})}{\partial\mu_u^{n_u}\partial\mu_d^{n_d}}.
\label{ders}\eeq
This is easily accomplished in any computer algebra system; for example, in
Mathematica using the program fragment---
{\tt chi[nu\_,nd\_] := D[Log[Z[muu,mud]],\{muu,nu\},\{mud,nd\}]}
augmented by substitution rules for setting $\chi_{n_u,n_d}=0$ for odd
$n_u+n_d$, and implementing the symmetry $n_u\leftrightarrow n_d$. The
results are written out below. Since $Z$ is a moment generating function,
$\log Z$ is a cumulant generating function. Statistics mavens will
therefore recognize the expressions below as the definitions of cumulants
in terms of moments. At the leading order we have
\beq
   \chi_{10} = \left(\frac TV\right)\frac{Z_{10}}Z.
\label{sus1}\eeq
Since this is zero, we shall use the relation $Z_{10}=0$ to simplify successive
derivatives. At the second order we find
\beq
   \chi_{20} = \left(\frac TV\right)\frac{Z_{20}}Z,\qquad
   \chi_{11} = \left(\frac TV\right)\frac{Z_{11}}Z.
\label{sus2}\eeq
At the third order we have we obtain the relations
$Z_{30}=Z_{21}=0$ which can be used to simplify the later derivatives. At fourth order
we have
\beqa
\nonumber
   \chi_{40} &=& \left(\frac TV\right)\left[\frac{Z_{40}}Z
                     -3\left(\frac{Z_{20}}Z\right)^2\right],
   \\
\nonumber
   \chi_{31} &=& \left(\frac TV\right)\left[\frac{Z_{31}}Z
                  -3\left(\frac{Z_{20}}Z\right)\left(\frac{Z_{11}}Z\right)\right],
   \\
   \chi_{22} &=& \left(\frac TV\right)\left[\frac{Z_{22}}Z
                  -\left(\frac{Z_{20}}Z\right)^2-2\left(\frac{Z_{11}}Z\right)^2\right].
\label{sus4}\eeqa
At the fifth order
CP symmetry allows us to write $Z_{50}=Z_{41}=Z_{32}=0$.
The 6th order susceptibilities are---
\beqa
\nonumber
    \chi_{60} &=& \left(\frac TV\right) \left[\frac{Z_{60}}Z
        -15\left(\frac{Z_{20}}Z\right)\left(\frac{Z_{40}}Z\right)
        +30\left(\frac{Z_{20}}Z\right)^3\right],\\
\nonumber
    \chi_{51} &=& \left(\frac TV\right) \left[\frac{Z_{51}}Z
        -10\left(\frac{Z_{20}}Z\right)\left(\frac{Z_{31}}Z\right)
        -5\left(\frac{Z_{11}}Z\right)\left(\frac{Z_{40}}Z\right)
        +30\left(\frac{Z_{11}}Z\right)\left(\frac{Z_{20}}Z\right)^2\right],\\
\nonumber
    \chi_{42} &=& \left(\frac TV\right) \left[\frac{Z_{42}}Z
        -\left(\frac{Z_{20}}Z\right)\left(\frac{Z_{40}}Z\right)
        -8\left(\frac{Z_{11}}Z\right)\left(\frac{Z_{31}}Z\right)
        -6\left(\frac{Z_{20}}Z\right)\left(\frac{Z_{22}}Z\right)
        +24\left(\frac{Z_{11}}Z\right)^2\left(\frac{Z_{20}}Z\right)
        +6\left(\frac{Z_{20}}Z\right)^3\right],\\
    \chi_{33} &=& \left(\frac TV\right) \left[\frac{Z_{33}}Z
        -6\left(\frac{Z_{20}}Z\right)\left(\frac{Z_{31}}Z\right)
        -9\left(\frac{Z_{11}}Z\right)\left(\frac{Z_{22}}Z\right)
        +18\left(\frac{Z_{11}}Z\right)\left(\frac{Z_{20}}Z\right)^2
        +12\left(\frac{Z_{11}}Z\right)^3\right].
\label{sus6}\eeqa
The 7th order susceptibilities give the further relations
$Z_{70}=Z_{61}=Z_{52}=Z_{43}=0$.
At the 8th order we obtain the susceptibilities
\beqa
\nonumber
    \chi_{80} &=& \left(\frac TV\right) \biggl[\frac{Z_{80}}Z
        -28\left(\frac{Z_{20}}Z\right)\left(\frac{Z_{60}}Z\right)
        -35\left(\frac{Z_{40}}Z\right)^2
        +420\left(\frac{Z_{20}}Z\right)^2 \left(\frac{Z_{40}}Z\right)
        -630\left(\frac{Z_{20}}Z\right)^4\biggr],\\
\nonumber
    \chi_{71} &=& \left(\frac TV\right) \biggl[\frac{Z_{71}}Z
        -7\left(\frac{Z_{11}}Z\right)\left(\frac{Z_{60}}Z\right)
        -35\left(\frac{Z_{31}}Z\right)\left(\frac{Z_{40}}Z\right)
        -21\left(\frac{Z_{20}}Z\right)\left(\frac{Z_{51}}Z\right)
        +210\left(\frac{Z_{11}}Z\right)\left(\frac{Z_{20}}Z\right)
            \left(\frac{Z_{40}}Z\right)\biggr.\\
    \nonumber &&\qquad\qquad\biggl.
        +210\left(\frac{Z_{20}}Z\right)^2 \left(\frac{Z_{31}}Z\right)
        -630\left(\frac{Z_{20}}Z\right)^3\left(\frac{Z_{11}}Z\right)\biggr],\\
\nonumber
    \chi_{62} &=& \left(\frac TV\right) \biggl[\frac{Z_{62}}Z
        -\left(\frac{Z_{20}}Z\right)\left(\frac{Z_{60}}Z\right)
        -12\left(\frac{Z_{11}}Z\right)\left(\frac{Z_{51}}Z\right)
        -15\left(\frac{Z_{22}}Z\right)\left(\frac{Z_{40}}Z\right)
        -15\left(\frac{Z_{20}}Z\right)\left(\frac{Z_{42}}Z\right)\biggr.\\
    \nonumber &&\qquad\qquad\biggl.
        +30\left(\frac{Z_{20}}Z\right)^2\left(\frac{Z_{40}}Z\right)
        +60\left(\frac{Z_{11}}Z\right)^2\left(\frac{Z_{40}}Z\right)
        -20\left(\frac{Z_{31}}Z\right)^2
        +240\left(\frac{Z_{11}}Z\right)\left(\frac{Z_{20}}Z\right)
           \left(\frac{Z_{31}}Z\right)\biggr.\\
    \nonumber &&\qquad\qquad\biggl.
        -90\left(\frac{Z_{20}}Z\right)^4
        +90\left(\frac{Z_{20}}Z\right)^2 \left(\frac{Z_{22}}Z\right)
        -540\left(\frac{Z_{11}}Z\right)^2\left(\frac{Z_{20}}Z\right)^2\biggr],\\
\nonumber
    \chi_{53} &=& \left(\frac TV\right) \biggl[\frac{Z_{53}}Z
        -3\left(\frac{Z_{20}}Z\right)\left(\frac{Z_{51}}Z\right)
        -5\left(\frac{Z_{31}}Z\right)\left(\frac{Z_{40}}Z\right)
        -15\left(\frac{Z_{11}}Z\right)\left(\frac{Z_{42}}Z\right)
        +30\left(\frac{Z_{11}}Z\right)\left(\frac{Z_{20}}Z\right)
           \left(\frac{Z_{40}}Z\right)\biggr.\\
    \nonumber &&\qquad\qquad\biggl.
        -30\left(\frac{Z_{22}}Z\right)\left(\frac{Z_{31}}Z\right)
        -10\left(\frac{Z_{20}}Z\right)\left(\frac{Z_{33}}Z\right)
        +180\left(\frac{Z_{11}}Z\right)\left(\frac{Z_{20}}Z\right)
           \left(\frac{Z_{22}}Z\right)
        +90\left(\frac{Z_{20}}Z\right)^2\left(\frac{Z_{31}}Z\right)\biggr.\\
    \nonumber &&\qquad\qquad\biggl.
        +120\left(\frac{Z_{11}}Z\right)^2\left(\frac{Z_{31}}Z\right)
        -270\left(\frac{Z_{11}}Z\right)\left(\frac{Z_{20}}Z\right)^3
        -360\left(\frac{Z_{11}}Z\right)^3\left(\frac{Z_{20}}Z\right)\biggr]\\
\nonumber
    \chi_{44} &=& \left(\frac TV\right) \biggl[\frac{Z_{44}}Z
        -12\left(\frac{Z_{20}}Z\right)\left(\frac{Z_{42}}Z\right)
        +12\left(\frac{Z_{20}}Z\right)^2\left(\frac{Z_{40}}Z\right)
        -\left(\frac{Z_{40}}Z\right)^2
        -54\left(\frac{Z_{20}}Z\right)^4
        -16\left(\frac{Z_{31}}Z\right)^2\biggr.\\
    \nonumber &&\qquad\qquad\biggl.
        -16\left(\frac{Z_{11}}Z\right)\left(\frac{Z_{33}}Z\right)
        +192\left(\frac{Z_{11}}Z\right)\left(\frac{Z_{20}}Z\right)
            \left(\frac{Z_{31}}Z\right)
        +144\left(\frac{Z_{11}}Z\right)^2\left(\frac{Z_{22}}Z\right)
        -18\left(\frac{Z_{22}}Z\right)^2\biggr.\\
    &&\qquad\qquad\biggl.
        +72\left(\frac{Z_{20}}Z\right)^2\left(\frac{Z_{22}}Z\right)
        -432\left(\frac{Z_{11}}Z\right)^2\left(\frac{Z_{20}}Z\right)^2
        -144\left(\frac{Z_{11}}Z\right)^4\biggr].
\label{sus8}\eeqa

The second step is to write the derivatives of $Z$ in terms of
products of traces. A diagrammatic method for this has been given
before \cite{dgm}. We put down a second method suited for symbolic
manipulations, which, however, is recursive. The two basic identities
are $Z_{10}=Z\langle\O_1\rangle$ and $\O_n'=\O_{n+1}$. These give first
the low order derivatives---
\beq
    Z_{20} = Z\biggl\langle\O_{11}+\O_2\biggr\rangle,\qquad\qquad
    Z_{11} = Z\biggl\langle\O_{11}\biggr\rangle.
\label{ord23}\eeq
Here the notation $\O_{ij\cdots l}$ stands
for the product, $\O_i\O_j\cdots\O_l$.
At the fourth order we get---
\beqa
\nonumber
    Z_{40} &=& Z\biggl\langle\O_{1111}+6\O_{112}+4\O_{13}+3\O_{22}+\O_4\biggr\rangle,\\
\nonumber
    Z_{31} &=& Z\biggl\langle\O_{1111}+3\O_{112}+\O_{13}\biggr\rangle,\\
    Z_{22} &=& Z\biggl\langle\O_{1111}+2\O_{112}+\O_{22}\biggr\rangle.
\label{ord4}\eeqa
At the 6th order the derivatives are
\beqa
\nonumber
    Z_{60} &=& Z\biggl\langle \O_{111111} + 15\O_{11112} + 20\O_{1113}
       + 45\O_{1122} + 15\O_{114} + 60\O_{123} 
      \biggr.\\ \nonumber &&\qquad\biggl.
       + 6\O_{15} + 15\O_{222} + 15\O_{24} + 10\O_{33} + \O_6 \biggr\rangle.\\
\nonumber
    Z_{51} &=& Z\biggl\langle \O_{111111} + 10\O_{11112} + 10\O_{1113}
       + 15\O_{1122} + 5\O_{114} + 10\O_{123} + \O_{15} \biggr\rangle,\\
\nonumber
    Z_{42} &=& Z\biggl\langle \O_{111111} + 7\O_{11112} + 4\O_{1113}
       + 9\O_{1122} + \O_{114} + 4\O_{123} 
       + 3\O_{222} + \O_{24} \biggr\rangle,\\
    Z_{33} &=& Z\biggl\langle \O_{111111} + 6\O_{11112} + 2\O_{1113}
       + 9\O_{1122} + 6\O_{123} + \O_{33} \biggr\rangle.
\label{zder6}\eeqa
Finally we write down some of the eighth order derivatives---
\beqa
\nonumber
    Z_{80} &=& Z\biggl\langle \O_{11111111} + 28\O_{1111112} + 56\O_{111113}
       +70\O_{11114}+210\O_{111122}
       +56\O_{1115}+560\O_{11123}+28\O_{116}
     \biggr.\\ \nonumber &&\qquad\biggl.
       +420\O_{11222}+420\O_{1124}
       +280\O_{1133}+8\O_{17}+840\O_{1223}+168\O_{125}+280\O_{134}
       +105\O_{2222}
     \biggr.\\ \nonumber &&\qquad\biggl.
       +210\O_{224}+280\O_{233}+28\O_{26}
       +56\O_{35}+35\O_{44}+\O_8\biggr\rangle.\\
\nonumber
    Z_{71} &=& Z\biggl\langle \O_{11111111} + 21\O_{1111112} + 35\O_{111113}
       +105\O_{111122}+35\O_{11114}
       +210\O_{11123}+21\O_{1115}+105\O_{11222}
     \biggr.\\ \nonumber &&\qquad\biggl.
       +70\O_{1133} +105\O_{1124}
       +7\O_{116}+105\O_{1223}+35\O_{134}+21\O_{125}+\O_{17}\biggr\rangle,\\
\nonumber
    Z_{62} &=& Z\biggl\langle \O_{11111111} + 16\O_{1111112} + 20\O_{111113}
       +60\O_{111122}+15\O_{11114} +80\O_{11123}+6\O_{1115}+60\O_{11222}+10\O_{1133}
     \biggr.\\ \nonumber &&\qquad\biggl.
       +30\O_{1124} +\O_{116}+6\O_{125}+60\O_{1223}+15\O_{2222}+15\O_{224}
       +10\O_{233}+\O_{26}\biggr\rangle,\\
\nonumber
    Z_{53} &=& Z\biggl\langle \O_{11111111} + 13\O_{1111112} + 11\O_{111113}
       +45\O_{111122}+5\O_{11114} +50\O_{11123}+\O_{1115}+45\O_{11222}+10\O_{1133}
     \biggr.\\ \nonumber &&\qquad\biggl.
       +15\O_{1124} +45\O_{1223}+3\O_{125}+5\O_{134}+10\O_{233}+\O_{35}\biggr\rangle,\\
\nonumber
    Z_{44} &=& Z\biggl\langle \O_{11111111} + 12\O_{1111112} + 8\O_{111113}
       +42\O_{111122}+2\O_{11114}
       +48\O_{11123}+36\O_{11222}+16\O_{1133}+12\O_{1124}
     \biggr.\\ &&\qquad\biggl.
       +24\O_{1223}+8\O_{134}+9\O_{2222}+6\O_{224}+\O_{44}\biggr\rangle.
\label{zder8}\eeqa
Combinatorial rules for generating the terms have been given before \cite{dgm}. They can
be used as checks. As an example, we evaluate the coefficient of the term
$\O_{1122}$ in $Z_{60}$. This is the number of ways of partitioning 6
objects into groups of 2 ones and 2 twos---
\beq
   \left\{\frac12 {6\choose1} \, {5\choose1} \right\} \times
   \left\{\frac12 {4\choose2} \right\}
   = \frac{6.5.4.3}{2^3} = 45.
\label{1122}\eeq

\subsection{Notation for traces}

Before writing out the traces we introduce the compact notation
\beq
   \lb n_1\cdot p_1\osum n_2\cdot p_2\osum \cdots\rb
      = \tr\left[\bigg(M^{-1}M^{(p_1)}\bigg)^{n_1}
                 \bigg(M^{-1}M^{(p_2)}\bigg)^{n_2}\cdots\right],
\eeq
where $M^{(p)}$ is the $p$-th derivative of $M$. Also, $\lb 1\cdot p\rb$
is written as $\lb p\rb$. The `addition', $\oplus$, is not commutative,
but all cyclic permutations of terms inside the brackets, $\lb\cdots\rb$,
are allowed. `Multiplication', (denoted by the dot) is distributive over
addition, subject to restrictions due to non-commutativity. Thus $\lb
n\cdot p\osum m\cdot p\rb = \lb (n+m)\cdot p\rb$, but no simplification is
possible for $\lb n\cdot p\osum m\cdot p'\osum l\cdot p\osum\cdots\rb$.
Traces can be added, \eg, $a \lb n\cdot p\rb + b \lb n\cdot p\rb =
(a+b) \lb n\cdot p\rb$.  Derivatives are easy to write---
\beq
   \lb n\cdot p\rb' = -n \lb 1\osum n\cdot p\rb
                      +n \lb (n-1)\cdot p\osum (p+1)\rb.
\label{der}\eeq
The operation of taking derivatives is linear over the `addition' in
$\lb n_1\cdot p_1\osum n_2\cdot p_2\osum \cdots\rb$, which is just the
rule for taking derivatives of products.

\subsection{Operators}
We have the lowest orders
\beq
   \O_1 = \lb 1\rb,\qquad\O_2 = -\lb 2\cdot1\rb+\lb 2\rb.
\label{o12}\eeq
Then, the remaining known ones are obtained simply by applying the
rules again. Since $\lb 2\cdot1\rb'=-2\lb 3\cdot1\rb+2\lb 1\osum 2\rb$
and $\lb 2\rb'=-\lb 1\osum 2\rb+\lb 3\rb$, we first obtain
\beq
   \O_3 = 2\lb 3\cdot1\rb-3\lb 1\osum 2\rb+\lb 3\rb.
\label{o3}\eeq
Beyond the second order the results depend on the prescription for
putting chemical potential on the lattice.  In the Hasenfratz-Karsch
(HK) prescription since all the odd derivatives are equal to each other,
and so are the even derivatives, we can rewrite the above as
\beq
   \O_3 = 2\lb 3\cdot1\rb-3\lb 1\osum 2\rb+\lb 1\rb\qquad({\rm HK}).
\label{o3hk}\eeq

\subsubsection{4th order}
At the 4th order we have 
\beqa
\nonumber && \lb 3\cdot1\rb'=-3\lb 4\cdot1\rb+3\lb 2\cdot1\osum 2\rb,\\
\nonumber && \lb 1\osum 2\rb'=-2\lb 2\cdot1\osum 2\rb+\lb 2\cdot2\rb+\lb 1\osum 3\rb,\\
          && \lb 3\rb'=-\lb 1\osum 3\rb+\lb 4\rb,
\label{der4}\eeqa
using the rules of derivatives. Note that the coefficients in each line
sum up to zero. This is a consequence of the rule for derivatives in eq.\
(\ref{der}). Also note that each operator, $\lb \cdots\osum n_i\cdot
p_i\osum \cdots\rb$, which contributes to $\O_n$ must satisfy the
constraint $\sum n_ip_i=n$.

The expressions in eq.\ (\ref{der4}) give 
\beq
   \O_4 =
          -6\lb 4\cdot1\rb+12\lb 2\cdot1\osum 2\rb-3\lb 2\cdot2\rb
          -4\lb 1\osum 3\rb+\lb 4\rb
        = -6\lb 4\cdot1\rb+12\lb 2\cdot1\osum 2\rb-3\lb 2\cdot2\rb
          -4\lb 2\cdot1\rb+\lb 2\rb,
\label{o4}\eeq
where the second expression holds only in the HK prescription. Note
that a further consequence of the rule for derivatives is that the sum
of the coefficients is zero for each $\O_n$ for $n\ge2$.

\subsubsection{5th order}
At the 5th order we use
\beqa
\nonumber &&\lb 4\cdot1\rb'  = -4\lb 5\cdot1\rb+4\lb 3\cdot1\osum 2\rb,\\
\nonumber &&\lb 2\cdot1\osum 2\rb'  = -3\lb 3\cdot1\osum 2\rb+2\lb 1\osum 2\cdot2\rb+\lb 2\cdot1\osum 3\rb,\\
\nonumber &&\lb 2\cdot2\rb'  = -2\lb 1\osum 2\cdot2\rb+2\lb 2\osum 3\rb,\\
\nonumber &&\lb 1\osum 3\rb'  = -2\lb 2\cdot1\osum 3\rb+\lb 2\osum 3\rb+\lb 1\osum 4\rb,\\
          &&\lb 4\rb'  = -\lb 1\osum 4\rb+\lb 5\rb,
\label{der5}\eeqa
to get the following expression in the HK scheme
\beq
   \O_5 
        = 24\lb 5\cdot1\rb-60\lb 3\cdot1\osum 2\rb+30\lb 1\osum 2\cdot2\rb
        +20\lb 3\cdot1\rb-15\lb 1\osum 2\rb+\lb 1\rb.
\label{o5}\eeq

\subsubsection{6th order}
At the 6th order we use
\beqa
\nonumber &&\lb 5\cdot1\rb'  = -5\lb 6\cdot1\rb+5\lb 4\cdot1\osum 2\rb,\\
\nonumber &&\lb 3\cdot1\osum 2\rb'  = -4\lb 4\cdot1\osum 2\rb+2\lb 2\cdot1\osum 2\cdot2\rb
               +\lb 1\osum 2\osum 1\osum 2\rb+\lb 3\cdot1\osum 3\rb,\\
\nonumber &&\lb 1\osum 2\cdot2\rb'  = -2\lb 2\cdot1\osum 2\cdot2\rb-\lb 1\osum 2\osum 1\osum 2\rb
               +\lb 3\cdot2\rb+\lb 1\osum 2\osum 3\rb+\lb 1\osum 3\osum 2\rb,\\
\nonumber &&\lb 2\cdot1\osum 3\rb'  = -3\lb 3\cdot1\osum 3\rb+\lb 1\osum 2\osum 3\rb+\lb 1\osum 3\osum 2\rb+\lb 2\cdot1\osum 4\rb,\\
\nonumber &&\lb 2\osum 3\rb'  = -\lb 1\osum 2\osum 3\rb+\lb 2\cdot3\rb-\lb 1\osum 3\osum 2\rb+\lb 2\osum 4\rb,\\
\nonumber &&\lb 1\osum 4\rb'  = -2\lb 2\cdot1\osum 4\rb+\lb 2\osum 4\rb+\lb 1\osum 5\rb,\\
          &&\lb 5\rb'  = -\lb 1\osum 5\rb+\lb 6\rb,
\label{der6}\eeqa
where we see the consequences of non-commutativity of `addition' for the first
time. In the HK scheme this gives the result
\beqa
\nonumber
   \O_6 
        &=& -120\lb 6\cdot1\rb-120\lb 4\cdot1\rb+360\lb 4\cdot1\osum 2\rb-16\lb 2\cdot1\rb+150\lb 2\cdot1\osum 2\rb\\
        &&\qquad-180\lb 2\cdot1\osum 2\cdot2\rb-90\lb 1\osum 2\osum 1\osum 2\rb
        +30\lb 3\cdot2\rb-15\lb 2\cdot2\rb+\lb 2\rb.
\label{o6}\eeqa
An earlier error in the derivative $\lb 1\osum 2\cdot2\rb'$ gave
errors in the coefficients of $\lb 2\cdot1\osum 2\cdot2\rb$ ($-210$
instead of $-180$) and $\lb 1\osum 2\osum 1\osum 2\rb$ ($-60$ instead of
$-90$). These erroneous expressions were used in \cite{lat03}.  However,
correcting this error has little numerical consequence.

\subsubsection{7th order}
At the 7th order we need---
\beqa
\nonumber &&\lb 6\cdot1\rb'  = -6\lb 7\cdot1\rb+6\lb 5\cdot1\osum 2\rb,\\
\nonumber &&\lb 4\cdot1\osum 2\rb'  = -5\lb 5\cdot1\osum 2\rb+2\lb 3\cdot1\osum 2\cdot2\rb
            +2\lb 1\osum 2\osum 2\cdot1\osum 2\rb+\lb 4\cdot1\osum 3\rb,\\
\nonumber &&\lb 2\cdot1\osum 2\cdot2\rb'  = -3\lb 3\cdot1\osum 2\cdot2\rb-\lb 1\osum 2\osum 2\cdot1\osum 2\rb+2\lb 1\osum 3\cdot2\rb
             +\lb 2\cdot1\osum 2\osum 3\rb+\lb 2\cdot1\osum 3\osum 2\rb,\\
\nonumber &&\lb 1\osum 2\osum 1\osum 2\rb'  = -4\lb 1\osum 2\osum 2\cdot1\osum 2\rb+2\lb 1\osum 2\osum 1\osum 3\rb+2\lb 1\osum 3\cdot2\rb,\\
\nonumber &&\lb 3\cdot1\osum 3\rb'  = -4\lb 4\cdot1\osum 3\rb+\lb 2\cdot1\osum 2\osum 3\rb+\lb 2\cdot1\osum 3\osum 2\rb
             +\lb 1\osum 2\osum 1\osum 3\rb+\lb 3\cdot1\osum 4\rb,\\
\nonumber &&\lb 3\cdot2\rb'  = -3\lb 1\osum 3\cdot2\rb+3\lb 2\cdot2\osum 3\rb,\\
\nonumber &&\lb 1\osum 2\osum 3\rb'  = -2\lb 2\cdot1\osum 2\osum 3\rb-\lb 1\osum 2\osum 1\osum 3\rb
             +\lb 2\cdot2\osum 3\rb+\lb 1\osum 2\cdot3\rb+\lb 1\osum 2\osum 4\rb,\\
\nonumber &&\lb 1\osum 3\osum 2\rb'  = -2\lb 2\cdot1\osum 3\osum 2\rb-\lb 1\osum 2\osum 1\osum 3\rb
             +\lb 2\cdot2\osum 3\rb+\lb 1\osum 2\cdot3\rb+\lb 1\osum 4\osum 2\rb,\\
\nonumber &&\lb 2\cdot1\osum 4\rb'  = -3\lb 3\cdot1\osum 4\rb+\lb 1\osum 2\osum 4\rb+\lb 1\osum 4\osum 2\rb+\lb 2\cdot1\osum 5\rb,\\
\nonumber &&\lb 2\cdot3\rb'  = -2\lb 1\osum 2\cdot3\rb+2\lb 3\osum 4\rb,\\
\nonumber &&\lb 2\osum 4\rb'  = -\lb 1\osum 2\osum 4\rb-\lb 1\osum 4\osum 2\rb+\lb 3\osum 4\rb+\lb 2\osum 5\rb,\\
\nonumber &&\lb 1\osum 5\rb'  = -2\lb 2\cdot1\osum 5\rb+\lb 2\osum 5\rb+\lb 1\osum 6\rb,\\
          &&\lb 6\rb'  = -\lb 1\osum 6\rb+\lb 7\rb.
\label{der7}\eeqa
In the HK scheme this yields
\beqa
\nonumber
   \O_7 
        &=& 720\lb 7\cdot1\rb
       -2520\lb 5\cdot1\osum 2\rb
       +1260\lb 3\cdot1\osum 2\cdot2\rb
       -1470\lb 3\cdot1\osum2\rb
       +1260\lb 1\osum 2\osum 2\cdot1\osum 2\rb
       +840\lb 5\cdot1\rb
\\ &&\qquad\qquad
       -630\lb 1\osum 3\cdot2\rb
       +420\lb 1\osum 2\cdot2\rb
       +182\lb 3\cdot1\rb
       -63\lb 1\osum 2\rb
       +\lb 1\rb.
\label{o7}\eeqa

\subsubsection{8th order}
At the 8th order we need---
\beqa
\nonumber &&\lb 7\cdot1\rb'  = -7\lb 8\cdot1\rb+7\lb 6\cdot1\osum 2\rb,\\
\nonumber &&\lb 5\cdot1\osum 2\rb'  = -6\lb 6\cdot1\osum 2\rb+2\lb 4\cdot1\osum 2\cdot2\rb+\lb 2\cdot1\osum 2\osum 2\cdot1\osum 2\rb
             +2\lb 1\osum 2\osum 3\cdot1\osum 2\rb+\lb 5\cdot1\osum 3\rb,\\
\nonumber &&\lb 3\cdot1\osum 2\cdot2\rb'  = -4\lb 4\cdot1\osum 2\cdot2\rb
             -\lb 3\cdot1\osum 2\osum 1\osum 2\rb+2\lb 2\cdot1\osum 3\cdot2\rb
             +\lb 1\osum 2\osum 1\osum 2\cdot2\rb
+\\ \nonumber &&\qquad\qquad\qquad\qquad\qquad
             +\lb 3\cdot1\osum 2\osum 3\rb+\lb 3\cdot1\osum 3\osum 2\rb,\\
\nonumber &&\lb 2\cdot1\osum 2\osum 1\osum 2\rb'  = 
             -2\lb 2\cdot1\osum 2\osum 2\cdot1\osum 2\rb
             -3\lb 3\cdot1\osum 2\osum 1\osum 2\rb
             +\lb 2\cdot1\osum 2\osum 1\osum 3\rb+\lb 2\cdot1\osum 3\cdot2\rb
+\\ \nonumber&&\qquad\qquad\qquad\qquad\qquad
             +2\lb 1\osum 2\cdot2\osum 1\osum 2\rb
             +\lb 2\cdot1\osum 3\osum 1\osum 2\rb,\\
\nonumber &&\lb 4\cdot1\osum 3\rb'  = -5\lb 5\cdot1\osum 3\rb
             +\lb 3\cdot1\osum 2\osum 3\rb+\lb 3\cdot1\osum 3\osum 2\rb
             +\lb 2\cdot1\osum 2\osum 1\osum 3\rb
             +\lb 2\cdot1\osum 3\osum 1\osum 2\rb+\lb 4\cdot1\osum 4\rb,\\
\nonumber &&\lb 1\osum 3\cdot2\rb'  = -2\lb 2\cdot1\osum 3\cdot2\rb
             -2\lb 1\osum 2\cdot2\osum 1\osum 2\rb+\lb 4\cdot2\rb
             +\lb 1\osum 3\osum 2\cdot2\rb+\lb 1\osum 2\cdot2\osum 3\rb
             +\lb 1\osum 2\osum 3\osum 2\rb,\\
\nonumber &&\lb 2\cdot1\osum 2\osum 3\rb'  = 
             -3\lb 3\cdot1\osum 2\osum 3\rb-\lb 2\cdot1\osum 2\osum 1\osum 3\rb
             +\lb 1\osum 2\osum 3\osum 2\rb+\lb 1\osum 2\cdot2\osum 3\rb
+\\ \nonumber &&\qquad\qquad\qquad\qquad\qquad
             +\lb 2\cdot1\osum 2\cdot3\rb
             +\lb 2\cdot1\osum 2\osum 4\rb,\\
\nonumber &&\lb 2\cdot1\osum 3\osum 2\rb'  = 
             -3\lb 3\cdot1\osum 3\osum 2\rb-\lb 2\cdot1\osum 3\osum 1\osum 2\rb
             +\lb 1\osum 2\osum 3\osum 2\rb+\lb 1\osum 3\osum 2\cdot2\rb
+\\ \nonumber &&\qquad\qquad\qquad\qquad\qquad
             +\lb 2\cdot1\osum 2\cdot3\rb
             +\lb 2\cdot1\osum 4\osum 2\rb,\\
\nonumber &&\lb 1\osum 2\osum 1\osum 3\rb'  = 
             -2\lb 2\cdot1\osum 2\osum 1\osum 3\rb
             -2\lb 2\cdot1\osum 3\osum 1\osum 2\rb
             +\lb 1\osum 3\osum 1\osum 3\rb+\lb 1\osum 3\osum 2\cdot2\rb
+\\ \nonumber &&\qquad\qquad\qquad\qquad\qquad
             +\lb 1\osum 2\cdot2\osum 3\rb
             +\lb 1\osum 2\osum 1\osum 4\rb,\\
\nonumber &&\lb 3\cdot1\osum 4\rb'= -4\lb 4\cdot1\osum 4\rb+\lb 2\cdot1\osum 2\osum 4\rb+\lb 2\cdot1\osum 4\osum 2\rb
             +\lb 1\osum 2\osum 1\osum 4\rb+\lb 3\cdot1\osum 5\rb,\\
\nonumber &&\lb 2\cdot2\osum 3\rb'= -\lb 1\osum 2\cdot2\osum 3\rb+\lb 1\osum 2\osum 3\osum 2\rb+\lb 1\osum 3\osum 2\cdot2\rb
             +2\lb 2\osum 2\cdot3\rb+\lb 2\cdot2\osum 4\rb,\\
\nonumber &&\lb 1\osum 2\cdot3\rb'= -3\lb 2\cdot1\osum 2\cdot3\rb+\lb 2\osum2\cdot3\rb+\lb 1\osum 4\osum 3\rb+\lb 1\osum 3\osum 4\rb,\\
\nonumber &&\lb 1\osum 2\osum 4\rb'=-2\lb 2\cdot1\osum 2\osum 4\rb-\lb 1\osum2\osum 1\osum 4\rb+\lb 2\cdot2\osum4\rb
             +\lb 1\osum 3\osum 4\rb+\lb 1\osum 2\osum 5\rb,\\
\nonumber &&\lb 1\osum 4\osum 2\rb'=-2\lb 2\cdot1\osum 4\osum 2\rb-\lb 1\osum2\osum 1\osum 4\rb+\lb 2\cdot2\osum4\rb
             +\lb 1\osum 4\osum 3\rb+\lb 1\osum 5\osum 2\rb,\\
\nonumber &&\lb 2\cdot1\osum 5\rb'=-3\lb 3\cdot1\osum 5\rb+\lb 1\osum 2\osum 5\rb+\lb 1\osum 5\osum 2\rb+\lb 2\cdot1\osum 6\rb,\\
\nonumber &&\lb 3\osum 4\rb'  = -\lb 1\osum 3\osum 4\rb-\lb 1\osum 4\osum 3\rb+\lb 2\cdot4\rb+\lb 3\osum 5\rb,\\
\nonumber &&\lb 2\osum 5\rb'  = -\lb 1\osum 2\osum 5\rb-\lb 1\osum 5\osum 2\rb+\lb 3\osum 5\rb+\lb 2\osum 6\rb,\\
\nonumber &&\lb 1\osum 6\rb'  = -2\lb 2\cdot1\osum 6\rb+\lb 2\osum 6\rb+\lb 1\osum 7\rb,\\
          &&\lb 7\rb'  = -\lb 1\osum 7\rb+\lb 8\rb.
\label{der8}\eeqa
Putting this together yields, in the HK prescription, the expression
\beqa
\nonumber
   \O_8 &=& 
 -5040\lb 8\cdot1\rb
 +20160\lb 6\cdot1\osum2\rb
 -10080\lb 4\cdot1\osum2\cdot2\rb
 -10080\lb 3\cdot1\osum2\osum1\osum2\rb
 -6720\lb 6\cdot1\rb
\\ \nonumber&&\qquad
 -5040\lb 2\cdot1\osum2\osum2\cdot1\osum2\rb
 +5040\lb 2\cdot1\osum3\cdot2\rb
 +5040\lb 1\osum2\cdot2\osum1\osum2\rb
 +15120\lb 4\cdot1\osum2\rb
\\ \nonumber&&\qquad
 -630\lb 4\cdot2\rb
 -5040\lb 2\cdot1\osum2\cdot2\rb
 -2520\lb 1\osum2\osum1\osum2\rb
 -2016\lb 4\cdot1\rb
 +1512\lb 2\cdot1\osum2\rb
\\ &&\qquad\qquad
 +420\lb 3\cdot2\rb
 -63\lb 2\cdot2\rb
 -64\lb 2\cdot1\rb
 +\lb 2\rb
\label{o8}\eeqa


\begin{thebibliography}{99}
\bibitem{pettini}
  A.\ Barducci \etal, \pl, B 231 (1989) 463.
\bibitem{kr}
  J.\ Berges and K.\ Rajagopal, \np, B 538 (1999) 215.
\bibitem{ms}
  M.\ A.\ Halasz \etal, \pr, D 58 (1998) 096007.
\bibitem{shuryak}
  M.\ A.\ Stephanov, K.\ Rajagopal and E.\ V.\ Shuryak, \pr, D 60 (1999) 114028;\\
  M.\ A.\ Stephanov, K.\ Rajagopal and E.\ Shuryak, \prl, 81 (1998) 4816.
\bibitem{pressure}
  R.\ V.\ Gavai and S.\ Gupta, \pr, D 68 (2003) 034506.
\bibitem{fk}
  Z.\ Fodor and S.\ Katz, \jhep, 0203 (2002) 014.
\bibitem{biswaold}
  C.\ R.\ Allton \etal, \pr, D 66 (2002) 074507.
\bibitem{maria}
  M.-P.\ Lombardo and M.\ d'Elia, \pr, D 67 (2003) 014505.
\bibitem{owe}
  Ph.\ de Forcrand and O.\ Philipsen, \np, B 642 (2002) 290.  
\bibitem{biswa}
  C.\ R.\ Allton, \etal, \pr, D 68 (2003) 014507.
\bibitem{biel}
  A.\ Peikert, Ph.\ D.\ thesis, University of Bielefeld, Germany, May 2000.
\bibitem{fktwo}
  Z.\ Fodor and S.\ Katz, \jhep, 0404 (2004) 050.
\bibitem{qcdpax}
  C.\ Bernard \etal, \pr, D 55 (1997) 6861;\\
  J.\ Engels \etal, \pl, B 396 (1997) 210;\\
  F.\ Karsch, E.\ Laermann and A.\ Peikert, \pl, B 478 (2000) 447;\\
  A.\ Ali Khan \etal, \pr, D 64 (2001) 074510.
\bibitem{gott}
  S.\ Gottlieb \etal, \prl, 59 (1987) 2247.
\bibitem{first}
  R.\ V.\ Gavai and S.\ Gupta, \pr, D 64 (2001) 074506.
\bibitem{gaunt}
  D.\ S.\ Gaunt and A.\ J.\ Guttmann, p.\ 181, {\sl Phase Transitions
  and Critical Phenomena\/}, Vol. 3, eds.\ C.\ Domb and M.\ S.\ Green,
  Academic Press, London, 1974.
\bibitem{zuber}
  J.-M.\ Drouffe and J.-B.\ Zuber, {\sl Phys.\ Rep.\/}, 102 (1983) 1.
\bibitem{ray}
  S.\ Gupta and R.\ Ray, \pr, D 70 (2004) 114015.
\bibitem{steiner}
  M.\ Charikar {\sl et al.\/}, {\sl Approximation Algorithms for
  Directed Steiner Tree Problems\/}, technical report STAN-CS-TN-97-56,
  Dept.\ of Computer Science, Stanford University, March 1997.
\bibitem{saad}
  Y.\ Saad \etal, {\sl SIAM J.\ Sci.\ Comput.\/}, 21 (2000) 1909.
\bibitem{gotttc}
  S.\ Gottlieb \etal, \prl, 59 (1987) 1513.
\bibitem{scale}
   S.\ Gupta, \pr, D 64 (2001) 034507.
\bibitem{gottm}
  S.\ Gottlieb \etal, \pr, D 38 (1988) 2245.
\bibitem{mdtau}
  S.\ Gupta, \np, B 370 (1992) 741.
\bibitem{karsch}
  F.\ Karsch and E.\ Laermann, \pr, D 50 (1994) 6954.
\bibitem{edwin}
  E.\ Laermann, {\sl Nucl.\ Phys.\ Proc.\ Suppl.\/}, 63 (1998) 114.
\bibitem{digiacomo}
  M.\ D'Elia, A.\ Di Giacomo and C.\ Pica, hep-lat/0408008.
\bibitem{liu}
  S.-J.\ Dong and K.-F.\ Liu \pl, B 328 (1994) 130.
\bibitem{pushan}
  R.\ V.\ Gavai, S.\ Gupta and P.\ Majumdar, \pr, D 65 (2002) 054506.
\bibitem{bir}
  J.-P.\ Blaizot, E.\ Iancu and A.\ Rebhan, \pl, B 523 (2001) 143.
\bibitem{alexi}
  A.\ Vuorinen, \pr, D 68 (2003) 054017.
\bibitem{milc}
  C.\ Bernard \etal, hep-lat/0405029.
\bibitem{valence}
  R.\ V.\ Gavai and S.\ Gupta, \pr, D 67 (2003) 034501.
\bibitem{mustafa}
  P.\ Chakraborty, M.\ G.\ Mustafa and M.\ H.\ Thoma, \pr, D 68 (2003) 085012.
\bibitem{pvt} 
  J.\ Cleymans, private communication.
\bibitem{cohen}
   T.\ D.\ Cohen, \prl, 91 (2003) 222001.  
\bibitem{su3}
  R.\ V.\ Gavai and S.\ Gupta, \pr, D 66 (2002) 094510.
\bibitem{dgm}
  S.\ Gupta, {\sl Acta Phys.\ Polon.\/}, B 33 (2002) 4259.
\bibitem{lat03}
  R.\ V.\ Gavai and S.\ Gupta, {\sl Nucl.\ Phys.\ Proc.\ Suppl.\/}, 129 (2004) 524.
\end{thebibliography}
\end{document}